\documentclass[fleqn,usenatbib]{mnras}

\usepackage[T1]{fontenc}

\DeclareRobustCommand{\VAN}[3]{#2}
\let\VANthebibliography\thebibliography
\def\thebibliography{\DeclareRobustCommand{\VAN}[3]{##3}\VANthebibliography}

\usepackage{graphicx}	
\usepackage{amsmath}	
\usepackage{amssymb}	
\usepackage{newtxtext,newtxmath}

\title[Resolving Disks \& Mergers in HRQs]{Resolving Disks \& Mergers in $z\sim2$ Heavily Reddened Quasars and their Companion Galaxies with ALMA}

\author[M. Banerji et al.]{
Manda Banerji,$^{1,2,3}$\thanks{E-mail: M.Banerji-Wright@soton.ac.uk}
Gareth C. Jones$^{4,3}$, 
Stefano Carniani$^{5}$,  
Colin DeGraf$^{2,3}$, 
Jeff Wagg$^6$
\\
$^{1}$School of Physics \& Astronomy, University of Southampton, Highfield Campus, Southampton SO17 1BJ\\
$^{2}$Institute of Astronomy, University of Cambridge, Madingley Road, Cambridge CB3 0HA, UK\\
$^3$Kavli Institute for Cosmology, University of Cambridge, Madingley Road, Cambridge CB3 0HA, UK\\
$^4$Cavendish Laboratory, University of Cambridge, 19 J. J. Thomson Ave., Cambridge CB3 0HE, UK\\
$^5$Scuola Normale Superiore, Piazza dei Cavalieri 7, I-56126 Pisa, Italy\\
$^6$SKA Organization, Lower Withington Macclesfield, Cheshire SK11 9DL, UK\\
}

\date{Accepted XXX. Received YYY; in original form ZZZ}

\pubyear{2021}

\begin{document}
\label{firstpage}
\pagerange{\pageref{firstpage}--\pageref{lastpage}}
\maketitle

\begin{abstract}
We present sub-arcsecond resolution ALMA imaging of the CO(3-2) emission in two $z\sim2.5$ heavily reddened quasars (HRQs) - ULASJ1234+0907 and ULASJ2315+0143 – and their companion galaxies. Dynamical modeling of the resolved velocity fields enables us to constrain the molecular gas morphologies and host galaxy masses. Combining the new data with extensive multi-wavelength observations, we are able to study the relative kinematics of different molecular emission lines, the molecular gas fractions and the locations of the quasars on the M$_{\rm{BH}}$-M$_{\rm{gal}}$ relation. Despite having similar black-hole properties, the two HRQs display markedly different host galaxy properties and local environments. J1234 has a very massive host – M$_{\rm{dyn}} \sim 5 \times 10^{11}$M$_\odot$ and two companion galaxies that are similarly massive located within 200 kpc of the quasar. The molecular gas fraction is low ($\sim$6\%). The significant ongoing star formation in the host galaxy is entirely obscured at rest-frame UV and optical wavelengths. J2315 is resolved into a close-separation major-merger ($\Delta$r=15 kpc; $\Delta$v=170 km/s) with a $\sim$1:2 mass ratio. The total dynamical mass is estimated to be $\lesssim$10$^{11}$M$_\odot$ and the molecular gas fraction is high ($>$45\%). A new HSC image of the galaxy shows unobscured UV-luminous star-forming regions co-incident with the extended reservoir of cold molecular gas in the merger. We use the outputs from the Illustris simulations to track the growth of such massive black holes from $z\sim6$ to the present day. While J1234 is consistent with the simulated $z\sim2$ relation, J2315 has a black hole that is over-massive relative to its host galaxy. 

\end{abstract}




\section{Introduction}

In our current paradigm of galaxy formation, accreting supermassive black-holes play a fundamental role in shaping the evolution of the most massive galaxies. Gas-rich major mergers have for some time been invoked in theoretical models to fuel both star formation and accretion onto the black hole in massive galaxies (e.g. \citealt{DiMatteo:05, Springel:05, Hopkins:08, Narayanan:10, Gabor:16, Steinborn:18}). Idealised as well as cosmological simulations of merging galaxies now have the requisite spatial and temporal resolution to be able to demonstrate that the merger leads to enhanced star formation (e.g. \citealt{Moreno:15}), triggering of an active galactic nucleus (AGN; e.g. \citealt{Capelo:15, Angles-Alcazar:20}), and an alteration in the interstellar medium properties of the galaxy (e.g. \citealt{Moreno:19}). Energy injection from the quasar during a feedback-dominated phase \citep{Fabian:12} is then thought to be responsible for quenching star formation and curtailing the growth of these massive galaxies. Cosmological simulations routinely implement such feedback in order to explain the observed distributions of massive galaxies (e.g. \citealt{Sijacki:15, Beckmann:17}).  A major aim of current observational research is to find compelling evidence in favour of this standard paradigm. Recently several aspects of this simple picture have been confronted by new observational data. For example, the existence of massive gas-disks in the high-redshift Universe (e.g. \citealt{Tacconi:13, Forster-Schreiber:18}) provides evidence that processes such as smooth accretion via cold gas flows \citep{Dekel:09} or minor mergers may play a more prominent role in assembling galaxies even at the high-mass end of the galaxy mass function, therefore challenging the major-merger paradigm. The effects of AGN feedback too can be more nuanced than previously thought. For example, theoretical models predict that compression of gas in galactic outflows can indeed trigger rather than inhibit star formation (e.g. \citealt{Ishibashi:17}). There have been some claims of observations of such positive feedback effects (e.g. \citealt{Cresci:15, Carniani:16}) with evidence now emerging that stars can in fact be formed \textit{in-situ} within a galactic outflow \citep{Maiolino:17, Gallagher:19}. 

In the context of the merger-driven evolutionary scenario for the formation and co-evolution of massive galaxies and their supermassive black-holes, populations of obscured quasars could represent an important phase as a galaxy is transitioning from a dust-obscured starburst to an ultraviolet luminous quasar (e.g. \citealt{Hickox:18}). The idea has its origins in the seminal work of \citet{Sanders:88}. With the advent of wide-area surveys in the X-ray, near and mid infrared wavelengths, population studies of the luminous, obscured quasar population are now possible (e.g. \citealt{Banerji:12, Banerji:13, Banerji:15, Glikman:12, Assef:15, Hamann:17, LaMassa:17, Temple:19}). The most luminous obscured quasars detected in these wide-area surveys are often seen to be hosted by merging galaxies (e.g. \citealt{Urrutia:08, Glikman:15, Fan:16, Bischetti:18, Diaz-Santos:18}) with significant amounts of ongoing star-formation (e.g. \citealt{Banerji:14, Banerji:17, Urrutia:12, Wethers:18}). Powerful AGN-driven outflows have also been detected in several of these systems (e.g. \citealt{Urrutia:09, Perna:15, Cresci:15, Zakamska:16, Temple:19, Kakkad:20}), with the observations suggesting that these outflows could have both a positive and negative feedback effect on star formation. 

Luminous quasars were among the first high-redshift ($z\gtrsim2$), extragalactic sources to be targeted in molecular gas emission (e.g. \citealt{Solomon:05} and references therein) and submillimeter and millimeter observations provide us with an opportunity to study host galaxy emission from these quasars over a wavelength range where the emission from the quasar itself is sub-dominant. Molecular gas has been detected in quasars out into the Epoch of Reionisation (e.g. \citealt{Walter:03, Walter:04, Maiolino:05, Wang:10, Wang:11, Venemans:17}) and the new capabilities of the Atacama Large Millimeter Array (ALMA) are now making it possible to spatially resolve both the dust and gas emission in high-redshift quasar hosts (e.g. \citealt{Carilli:13b, Wagg:14, Salome:12, Carniani:13, DeBreuck:14, Venemans:17, Trakhtenbrot:17, Shao:17, Brusa:18, Fogasy:20}). Nevertheless, the number of high-redshift quasars with such resolved measurements of the molecular gas emission are still relatively small. Resolved measurements of the host galaxies in the millimeter wavelengths enable unique constraints on the dynamical masses, gas morphologies and gas fractions in such systems. 

Observational evidence is mounting that many of the major physical processes thought to govern massive galaxy formation - e.g. mergers, starbursts, high-accretion black-hole growth and outflows - are synchronously occurring during the dust-obscured quasar phase. How are these processes triggered and what effect do they have on the quasar host galaxies? We address these questions here via detailed, spatially resolved observations of the CO(3-2) gas emission in the quasars ULASJ1234$+$0907 ($z\sim2.50$) and ULASJ2315$+$0143 ($z\sim2.56$). The two quasars are very similar in terms of the AGN properties. Both quasars are extremely luminous (L$_{\rm{6\mu m}}\sim10^{47}$ erg/s; L$_{\rm{X}}\sim10^{45}$ erg/s - see \citealt{Lansbury:20}) with massive black-holes (M$_{\rm{BH}}>10^9$M$_\odot$) obscured by significant columns of dust and gas (N$_{\rm{H}}\sim10^{22}$cm$^{-2}$; A$_{\rm{V}}\sim5-6$ mag;  \citealt{Banerji:12, Banerji:15, Banerji:17, Wethers:18, Lansbury:20}). Both quasars are known to reside in highly star-forming host galaxies \citep{Banerji:14, Wethers:18, Wethers:20}. The first detections of molecular gas via the CO(3-2) emission in the two quasars was presented in \citet{Banerji:17} (B17 hereafter). In \citet{Banerji:18} (Paper I hereafter) we combined the CO(3-2) detections with detections of CO(1-0), CO(7-6) and CI(2-1) to better constrain the interstellar medium properties of the quasar host galaxies. While Paper I focused on the integrated line emission properties of our quasar host galaxies, in this paper we concentrate on studying the resolved emission to gain a better understanding of the physical processes that govern the formation of these galaxies. Throughout the paper we assume a flat, $\Lambda$CDM cosmology with H$_0$=70 km s$^{-1}$ Mpc$^{-1}$, $\Omega_{\rm{M}}$=0.3 and $\Omega_{\Lambda}$=0.7.

\section{ALMA Band 3 Observations}

\label{sec:data}

Our previous work (B17) presented detections of the CO(3-2) line in a sample of four heavily reddened quasars (HRQs) at $z\sim2.5$ that were initially discovered in \citet{Banerji:12, Banerji:15}. These four quasars are among the reddest quasars found in our near-infrared search for luminous, obscured quasars that are not present in wide-field optically selected spectroscopic quasar samples such as those from the Sloan Digital Sky Survey (SDSS) and BOSS. Briefly, the quasars have extremely red near infra-red colours of $(H-K)_{\rm{Vega}} > 2$, corresponding to dust extinctions towards the quasar continuum of $E(B-V)\gtrsim1$ (A$_{\rm{V}}\gtrsim3$ mag for R$_{\rm{V}}=3.1$ mag). The bolometric luminosities and black-hole masses are comparable to the most luminous quasars selected in optical surveys such as the Sloan Digital Sky Survey. As a detection experiment, the Cycle 3 ALMA observations in B17 made use of the most compact available configuration of the telescope leading to an angular resolution of $\sim$2.5-3$\arcsec$. Two of the quasars from the sample - ULASJ1234+0907 ($z\sim2.50$; J1234 hereafter) and ULASJ2315+0143 ($z\sim2.56$; J2315 hereafter) - were marginally resolved in the Cycle 3 data and were therefore re-observed during ALMA Cycle 4 at $\sim$3$\times$ better angular resolution with the aim of spatially resolving the gas emission in both sources and constraining the gas dynamics. J1234 has two nearby gas-rich galaxies that were first discovered via their CO(3-2) detections in B17. Both companion galaxies (named G1234N and G1234S) were also imaged at higher resolution as part of the observations presented in this work. Further details of the new ALMA Cycle 4 observations can be found in Table \ref{tab:obs}. 

The spectroscopic observations of both quasar systems made use of a correlator configuration with four dual polarisation bands, each of bandwidth 1.875GHz. In the case of J1234, observations made use of the Time Division Mode (TDM) of the correlator, providing a channel width of 15.625 MHz. In the case of J2315, the Frequency Division Mode (FDM) of the correlator was instead used, providing a much finer channel width of $\sim$488 kHz across the entire bandwidth, which we then binned up to 15.625 MHz resolution. 

The raw data were processed using \textsc{casa} (v4.7.0), by executing the appropriate calibration scripts for each set of observations as provided by the ALMA Observatory. Line cubes were produced after u-v plane continuum subtraction employing the line-free channels. We then used the \textit{clean} algorithm with a \textit{natural} weighting of the visibilities to transform to the image plane. Continuum images were also produced by running \textit{clean} in \textit{mfs} mode on the line-free spectral windows, and once again using a \textit{natural} weighting. All images have had a primary beam correction applied to them using the \textsc{casa} task \textit{pbcor}. Images have been generated with a pixel-scale of 0.1$\arcsec$ per pixel. The beam sizes as well as RMS noise levels reached by the data for each of the sources is summarised in Table \ref{tab:obs}. The spectra have all had Hanning smoothing applied to them as part of the pipeline processing. 

\begin{table*}
\begin{center}
\caption{Details of the ALMA Cycle 4 Band 3 observations analysed in this work.}
\label{tab:obs}
\begin{tabular}{lcc}
& ULASJ1234$+$0907 & ULASJ2315$+$0143 \\
\hline
Dates Observed  & 2016-11-06 &  2016-11-07, 2017-05-22 \\
Number of Antennae & 40 & 43 \\
Exposure Time  & 1hr55m  & 2h42m\\
Beam Size (line) / $\arcsec$ & 0.96$\times$0.79 & 0.82$\times$0.79 \\
Beam Size (continuum) / $\arcsec$ & 1.01$\times$0.90 & 0.80$\times$0.75 \\
Channel R.M.S in 15.625 MHz channels / mJy beam$^{-1}$  &  0.10 & 0.14 \\
Continuum R.M.S / $\mu$Jy beam$^{-1}$ & 11.5 & 9.3 \\
\hline 
\end{tabular}
\end{center}
\end{table*}

\section{OBSERVED MOLECULAR GAS AND DUST CONTINUUM PROPERTIES}

\label{sec:results}

\begin{figure*}
\begin{center}
\begin{tabular}{cc}
\includegraphics[scale=0.45]{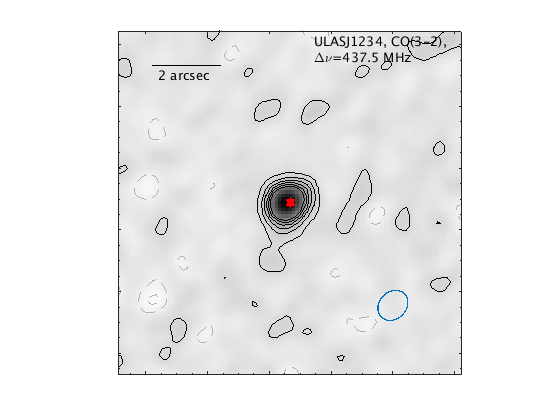} & \hspace{-1.0cm} \includegraphics[scale=0.45]{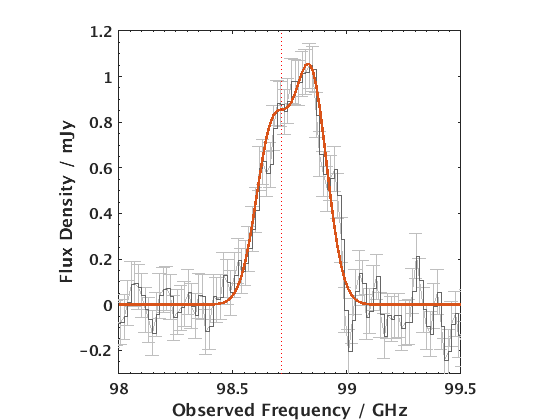} \\
\includegraphics[scale=0.45]{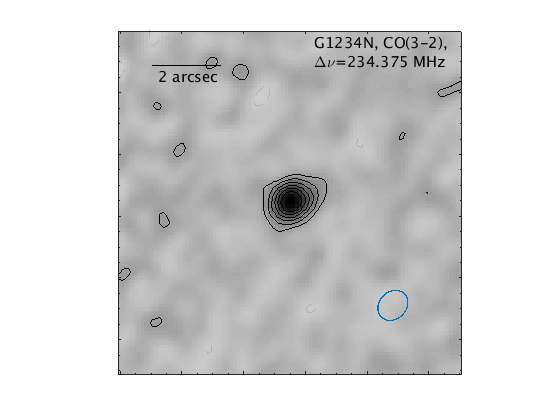} & \hspace{-1.0cm} \includegraphics[scale=0.45]{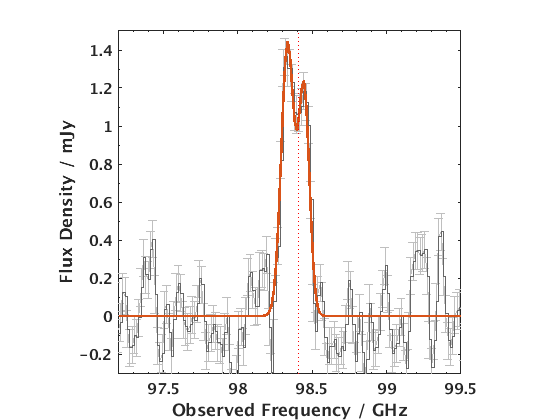} \\
\includegraphics[scale=0.45]{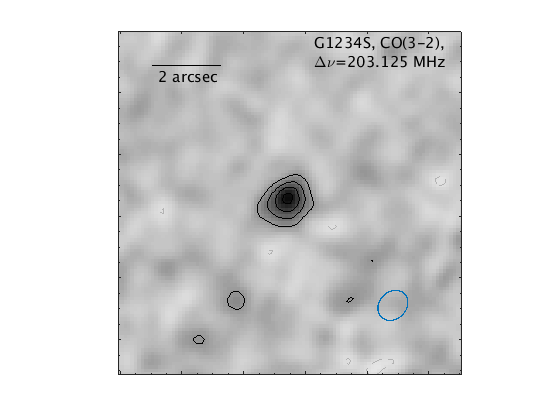} & \hspace{-1.0cm} \includegraphics[scale=0.45]{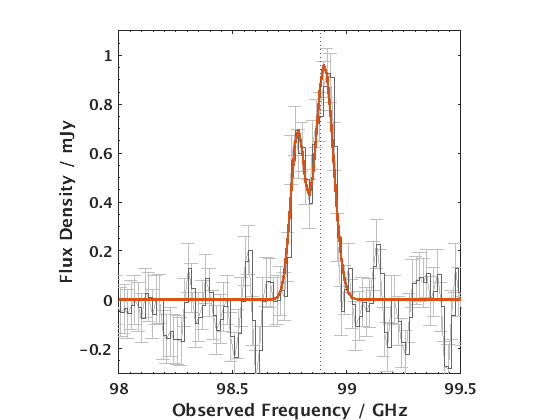} \\
\end{tabular}
\caption{Maps of the CO(3-2) emission in the three galaxies making up the J1234 system (left). North is up and East is to the right. The quasar position is marked with the star in the top-left image and the synthesized beam size and position angle are indicated by the blue ellipse at the bottom right of each image. The right-hand panels show the integrated CO(3-2) spectra for the three galaxies together with the best-fit double Gaussian fit to these line profiles. Double-peaked profiles are evident for all three galaxies. Zero velocity, indicated by the dotted vertical line corresponds to the systemic redshifts of these galaxies obtained from our ALMA Cycle 3 observations in B17.}
\label{fig:CO_J1234}
\end{center}
\end{figure*}

\begin{figure*}
\begin{center}
\begin{tabular}{c}
\includegraphics[scale=0.75]{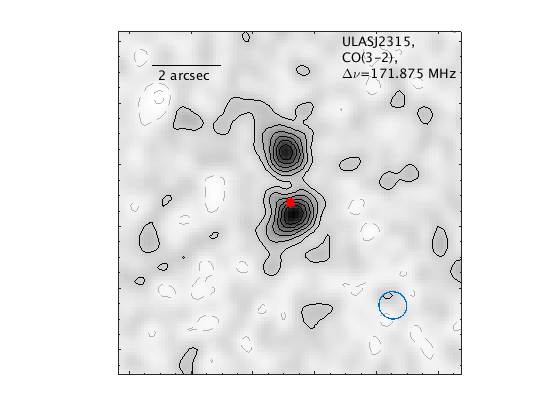} \\
\end{tabular}
\begin{tabular}{ccc}
\includegraphics[scale=0.4]{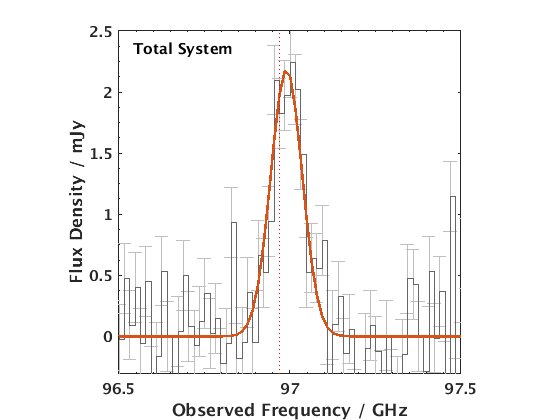} & \hspace{-1.2cm} \includegraphics[scale=0.4]{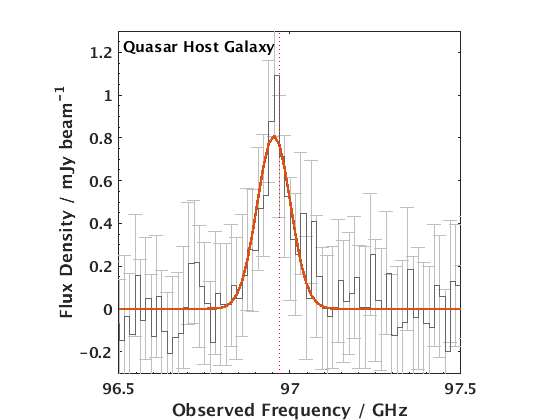} & \hspace{-1.2cm} \includegraphics[scale=0.4]{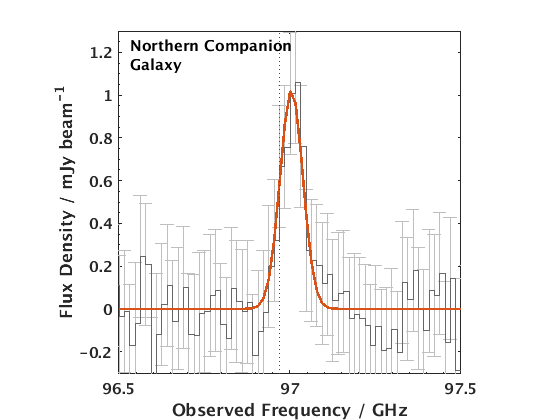} \\
\end{tabular}
\caption{\textit{Top:} Map of the CO(3-2) emission in the J2315 system. North is up and East is to the right. The source is clearly resolved into two spatial components and the quasar position is indicated by the star. The synthesized beam size is indicated by the blue ellipse at the bottom-right of the image. \textit{Bottom:} CO(3-2) spectra extracted over the entire system (left), at the quasar position (middle) and at the position of the northern companion galaxy (right) together with the best-fit single Gaussian used to fit these line profiles. Zero velocity, indicated by the dotted vertical line corresponds to the systemic redshift obtained from our ALMA Cycle 3 observations in B17.}
\label{fig:CO_J2315}
\end{center}
\end{figure*}

In Fig. \ref{fig:CO_J1234} and \ref{fig:CO_J2315} we present the CO(3-2) images as well as integrated spectra for all galaxies in the J1234 and J2315 systems respectively obtained via our ALMA Cycle 4 program. We infer initial source sizes by fitting a 2-dimensional Gaussian to the zeroth moment image using \textsc{casa}. The spectra have been extracted using apertures of variable sizes (diameter 1-2$\arcsec$) centred on the positions of each galaxy stated in Table \ref{tab:properties}. The aperture sizes chosen depend on the size of the CO emitting region for each galaxy, which is also quoted in Table \ref{tab:properties} and discussed further in Sections \ref{sec:J1234} and \ref{sec:J2315}. These new observations provide higher signal-to-noise ratio (S/N) spectra for all sources compared to the Cycle 3 data presented in B17, as well as higher S/N measurements of the continuum flux densities at observed frequencies of $\sim$86-103 GHz. These measurements are summarised in Table \ref{tab:properties}. 

\begin{table*}
\begin{center}
\caption{A summary of the positions, redshifts and line and continuum flux measurements for our $z\sim2.5$ quasars and their companion galaxies.}
\label{tab:properties}
\begin{tabular}{lccccc}
& ULASJ1234 & G1234N & G1234S & ULASJ2315 (QSO) & ULASJ2315 (Gal) \\
\hline
RA & 12:34:27.51 & 12:34:26.38 & 12:34:27.30 & 23:15:56.24 & 23:15:56.22 \\
Dec & 09:07:54.2 & 09:08:06.3 & 09:07:43.2 & 01:43:50.1 & 01:43:52.0 \\
Redshift & 2.503 & 2.541 & 2.497 & 2.566 & 2.566 \\
Continuum Observed Frequency / GHz & 86.2308 & 91.4662 & 91.4662 & 102.6042 & 102.6042 \\
Continuum Flux Density / $\mu$Jy beam$^{-1}$ & 71$\pm$7 & 31$\pm$2 & 33$\pm$6 & 160$\pm$8 & 45$\pm$7 \\
CO(3-2) $\Delta$FWHM / kms$^{-1}$ & 780$\pm$80 & 540$\pm$70 & 570$\pm$80 & 300$\pm$60 & 190$\pm$40 \\
CO(3-2) Line Intensity / Jy kms$^{-1}$ & 0.93$\pm$0.02 & 0.46$\pm$0.02 & 0.36$\pm$0.02 & 0.27$\pm$0.02 & 0.28$\pm$0.01 \\
CO(3-2) Line Luminosity / $\times$10$^7$L$_\odot$ & 3.97$\pm$0.09 & 1.98$\pm$0.07 & 1.56$\pm$0.07 & 1.19$\pm$0.07 & 1.26$\pm$0.07 \\
CO(3-2) Size$^{\ast}$ / arcsec & 0.48$\pm$0.14 & 0.57$\pm$0.18 & 0.53$\pm$0.24 & 1.0$\pm$0.2 & 1.4$\pm$0.2 \\
\hline 
\end{tabular}
\end{center}
\flushleft $^{\ast}$FWHM of major axis from a Gaussian fit to the CO moment 0 image using \textsc{CASA} \textit{imfit} and after deconvolving from the beam.
\end{table*}

All three galaxies in the J1234 system are spatially resolved in our data based on the fact that \textsc{CASA} is able to deconvolve the intrinsic emission from the convolved image. We have also checked that the radial profiles of all three galaxies are more extended than would be expected for a point source. The spectra exhibit clear double-peaked CO(3-2) emission line profiles. Such clear double-peaked profiles were not evident in the original Cycle 3 spectra in B17, and indicate that we have now resolved multiple kinematic components in the gas emission. The J2315 system, which was already spatially resolved in CO emission in B17, now breaks up into two distinct galaxies in a close-separation major merger in our new data. We are therefore able to extract CO(3-2) spectra for both the quasar host and the companion galaxy lying to the North of the quasar. Table \ref{tab:properties} presents the line and continuum flux densities for both our quasars and their companion galaxies. These measurements have been analysed in more detail in Paper I, where we combine them with observations of the CO(1-0), CO(7-6) and CI(2-1) lines as well as the far infra-red to radio continuum in order to infer the properties of the interstellar medium in these galaxies. In Paper I we find that the quasar J1234 as well as its companion galaxy G1234S are comprised of highly excited, dense gas consistent with what is seen in luminous AGN. J2315 on the other hand has excitation properties that are more similar to that observed in star-forming galaxies. Both the quasars J1234 and J2315 as well as their companion galaxies were inferred to be highly star-forming based on their dust continuum detections in Paper I. 

The focus of the current paper is on a detailed kinematic analysis of the CO(3-2) emission and inferred constraints on the gas dynamics and gas morphologies in our two quasar systems. Before turning to such an analysis we first present a description of the dynamical models we will employ.

\section{DYNAMICAL MODELLING}

\label{sec:dymod}

\subsection{$^{\rm 3D}$Barolo}

We model the CO(3-2) traced gas dynamics in our quasars and their companion galaxies using the tilted ring model fitting code $^{\rm 3D}$Barolo - 3D-Based Analysis of Rotating Objects from Line Observations \citep{DiTeodoro:15}. The approach is similar to our previous work (B17) employing the GIPSY \textit{rotcur} task, but with the key improvement that $^{\rm 3D}$Barolo fits models to the entire data cube rather than just the velocity field. By convolving the model with the restoring beam and comparing data and model on a channel-by-channel basis, the effects of beam smearing are reduced, and additional dynamical aspects may be examined. Such an approach has been used previously to study the rotation curves of nearby galaxies (e.g. \citealt{deBlok:08, Iorio:17, Shelest:20}) and is now being routinely applied at high-redshifts (e.g. \citealt{Jones:17, Xue:18, Tadaki:20}).

The three-dimensional model fitting considers some of the same variables that are included in two-dimensional fits: inclination, \textit{i}, position angle, \textit{PA}, rotational velocity, $v_{\rm{rot}}$, systemic redshift, $v_{\rm{sys}}$ and the spatial centroid, [$x0$, $y0$]. By expanding to the spectral dimension, we can also consider velocity dispersion $\sigma_v$, radial brightness profile $I(r)$, and scale height $Z_o$ as free parameters. While $^{\rm 3D}$Barolo has been well-tested on low-S/N and low-resolution datasets (e.g., \citealt{DiTeodoro:15}), the best-fit results are strongly dependent on both the model geometry (i.e., number of rings, ring width) and initial estimates for each parameter. To reduce any implicit bias, we automate the fitting process by using signal isolation and moment maps.

We first utilize the $^{\rm 3D}$Barolo \textit{SEARCH} utility (based on DUCHAMP; \citealt{Whiting:12}) to create a signal mask. This utility automatically finds the RMS noise level of an input data cube and identifies signal peaks above a user-provided signal-to-noise threshold (S/N$_{\rm{upper}}=3.5$). A search is then conducted around these peaks for emission above a second user-provided S/N$_{\rm{lower}}=3.0$, creating a three-dimensional mask of significant emission. This approach includes fewer non-signal pixels than a 2-D spaxel mask or a mask based on a single S/N threshold.

Next, we use the CASA task \textit{immoments} to create moment 0 (integrated intensity) maps using all channels containing significant line emission. By fitting this map with a two-dimensional Gaussian (CASA task \textit{imfit}), we find a best-fit beam-deconvolved FWHM of the major and minor axes, position angle, and morphological central position. We also apply \textit{immoments} to all pixels identified as signal to create moment 1 (line of sight velocity) and moment 2 (velocity dispersion) maps. These maps can be seen in the left most panels of Fig. \ref{fig:maps_J1234}, \ref{fig:maps_G1234N}, \ref{fig:maps_G1234S} and \ref{fig:maps_J2315}. 

With this mask and these maps in hand, we are able to provide physically motivated estimates for most model parameters. We base the maximum model radius on the 2-D Gaussian fit to the moment zero map: $1.0\times$(the deconvolved FWHM of the major axis). In order to not over-resolve the data, we adopt a minimum ring width of (FWHM of the minor axis of the restoring beam)/2.5 - see also \citet{Talia:18,Fan:19}. The number of model rings is then found by dividing the maximum model radius by the minimum ring width and rounding down to the nearest integer, while the ring width is taken to be the maximum model radius divided by this number. For all galaxies modelled in this work, this process returns models with three rings. The central position of the model is fixed to be the centroid of the 2-D Gaussian fit to the moment zero map. Since each emission line is well-detected, the systemic velocity is expected to be $0$\,km\,s$^{-1}$, with respect to the redshifts listed in Table \ref{tab:dynmod}. $^{\rm 3D}$Barolo provides an estimate of the rotational velocity based on the moment 1 map, while we provide the value of the moment 2 map at the morphological center as an estimate of the velocity dispersion. The inclination estimate is set to $45^{\circ}$, and allowed to vary between $10-80^{\circ}$, while the position angle is provided by the user. We assume a thin disk with a scale height of  $0.01''\sim100$\,pc.

Using this set of initial guesses for each of these variables and a list of ring radii and widths, $^{\rm 3D}$Barolo first constructs a model of the first ring by randomly populating a space with discrete emitting gas clouds. These clouds are then given velocities to replicate the estimated rotation curve and velocity dispersion. The resulting data cube is convolved with the synthesized beam of the observation, and the summed intensity in each pixel along the velocity axis of the model cube is rescaled to equal that of the data cube. The pixels of the data and model cubes are then compared, and each parameter is adjusted. When the residual is minimized, the fitting stops, and the spatial parameters (\textit{i, PA, $v_{\rm{sys}}$}) are averaged across all rings. The fitting is then repeated for the remaining parameters. When this reaches a minimum, the next ring is analyzed. The final outputs for this model fitting process are a model data cube and best-fit morphological (i.e., $i$, $PA$) and kinematic (i.e., rotational velocity curve, velocity dispersion curve, systemic redshift) parameters. The output model cube encodes information about the rotation curve - i.e. $v_{\rm{rot}}$ as a function of radius, $r$, from which we can derive the dynamical mass of the galaxy as follows: 

\begin{equation}
M_{\rm{dyn}}=\frac{v^2_{\rm{rot, max}} r_{\rm{max}}}{G}
\label{eq:mdyn_tr}
\end{equation}

The uncertainties on the dynamical mass only include the output rotational model uncertainties from $^{\rm 3D}$Barolo ($\delta v_{\rm{rot, max}}$) and the width of the model ring ($\delta r_{\rm{max}}$) and the significant uncertainty on the inclination is not included. Since the fitting process is three-dimensional and multi-staged, the inclusion of the inclination uncertainty in the error budget is non-trivial and we simply note that the formal uncertainties on the dynamical masses from $^{\rm 3D}$Barolo are likely to be under-estimated. 

\subsection{Exponential Disk Modelling}

\label{sec:expdisk}

$^{\rm 3D}$Barolo makes no assumption regarding the underlying density distribution and mass profile of the galaxy and simply fits the CO emitting material across the extent over which it is seen in our ALMA data. We therefore use an independent method to check the mass estimates by approximating the kinematics of the galaxies as a thin rotating disk \citep{Carniani:13, Pensabene:20}. Such a model assumes that the surface brightness of the disk has an exponential profile $\Sigma \propto e^{-r/R_d}$, where $R_d$ is the scale radius that corresponds to a half-light radius of $R_{1/2}=1.67 \times R_d$. By assuming that the mass surface density has the same profile as the gas emission, the circular velocity associated with the exponential profile can be calculated following equation 2.157 of \citet{Binney:08}. Both intensity and velocity model maps are then matched to the spatial resolution of the observations by taking into account the observed synthesized beam. The free parameters of this dynamical model are the spatial coordinates of the disk centre, the half-light radius (or scale radius), inclination, position angle, dynamical mass, and systemic velocity. The dynamical mass is the disk mass enclosed within a radius of 10 kpc. Because this radius is much larger than the typical disk scale radius at $z\sim2$ \citep{Cresci:09}, this mass corresponds to the total baryonic mass in the disk. Note also that this radius is much larger than the radius over which the CO emission is visible, and therefore the radius over which $^{\rm 3D}$Barolo fits are conducted. The full parameter space is explored by using the Markov Chain Monte Carlo (MCMC) algorithm \textit{emcee} \citep{Foreman-Mackey:13}), which allows one to obtain a probability distribution of all free-parameters along with marginalised uncertainties. The disk centre coordinates and half-light radius are derived from modelling the intensity maps, while the probability distributions of the other parameters are driven from the fitting of the velocity maps. 

\section{The J1234 System: A Quasar with Two Massive Companion Galaxies}

\label{sec:J1234}

\subsection{Dynamical Masses, Sizes \& Gas Fractions}

\begin{figure*}
\begin{center}
\begin{tabular}{ccc}
\includegraphics[scale=0.5]{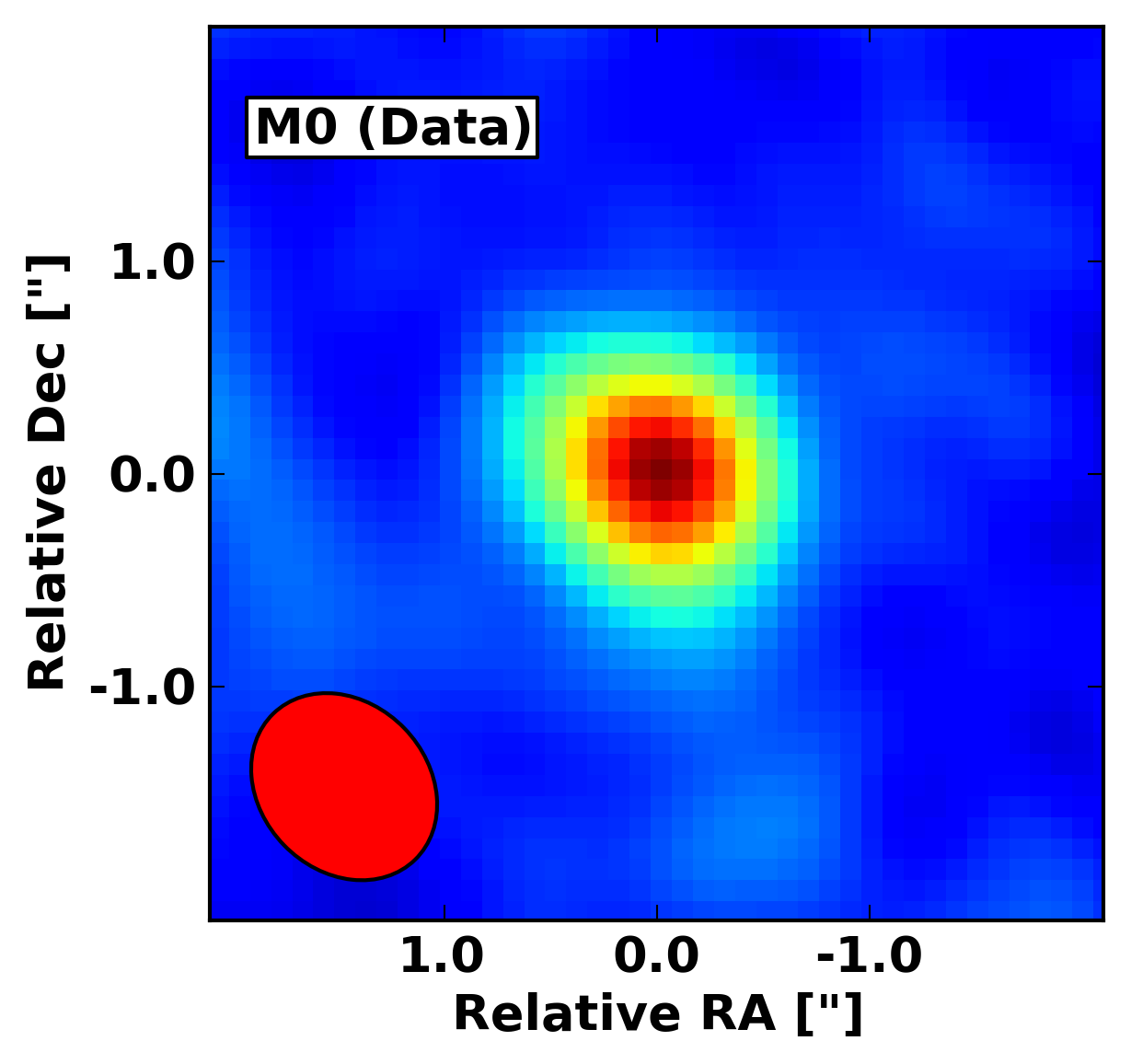} & \hspace{-0.5cm} \includegraphics[scale=0.5]{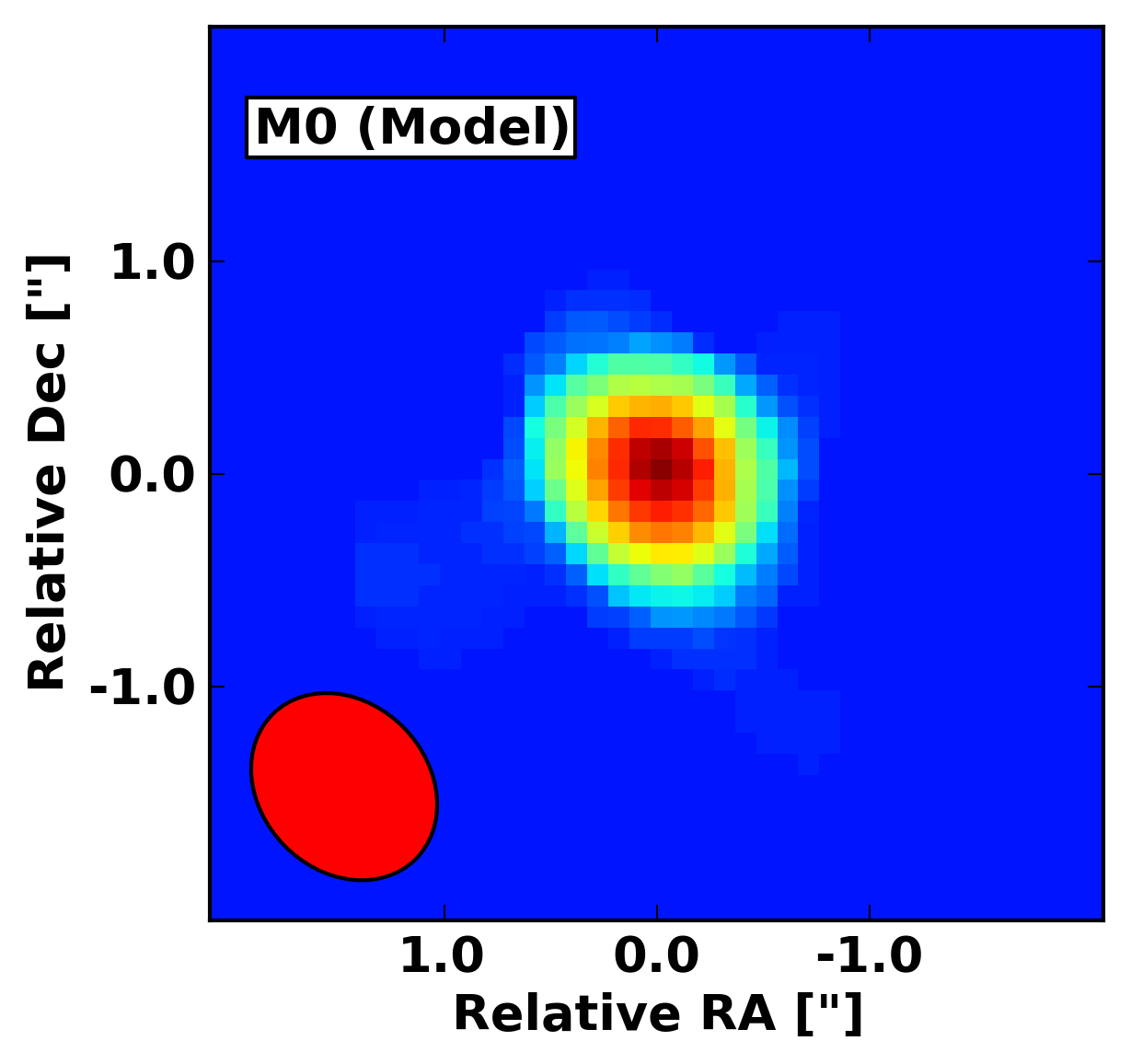} & \hspace{-0.5cm}
\includegraphics[scale=0.5]{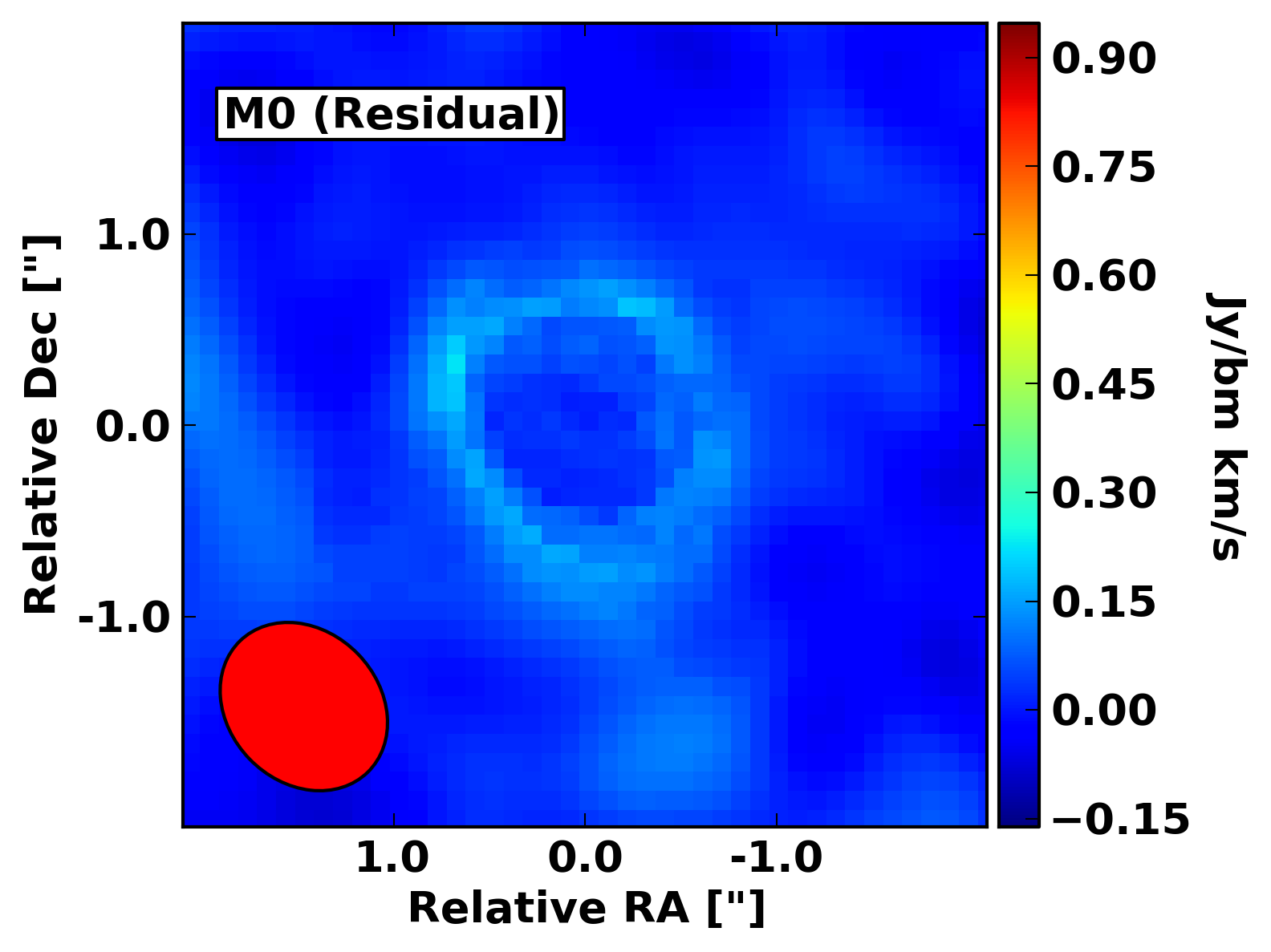} \\
\includegraphics[scale=0.5]{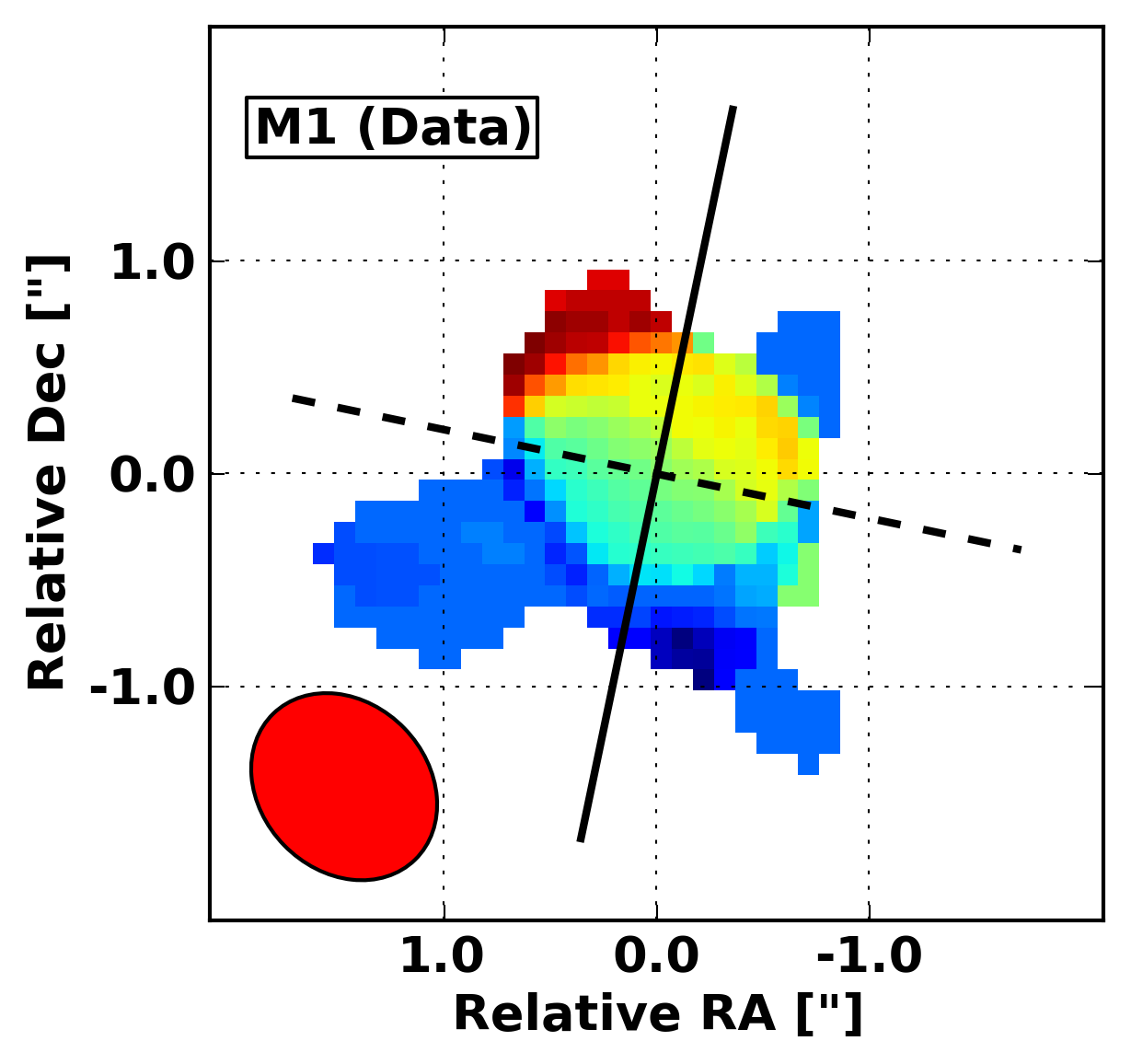} & \hspace{-0.5cm} \includegraphics[scale=0.5]{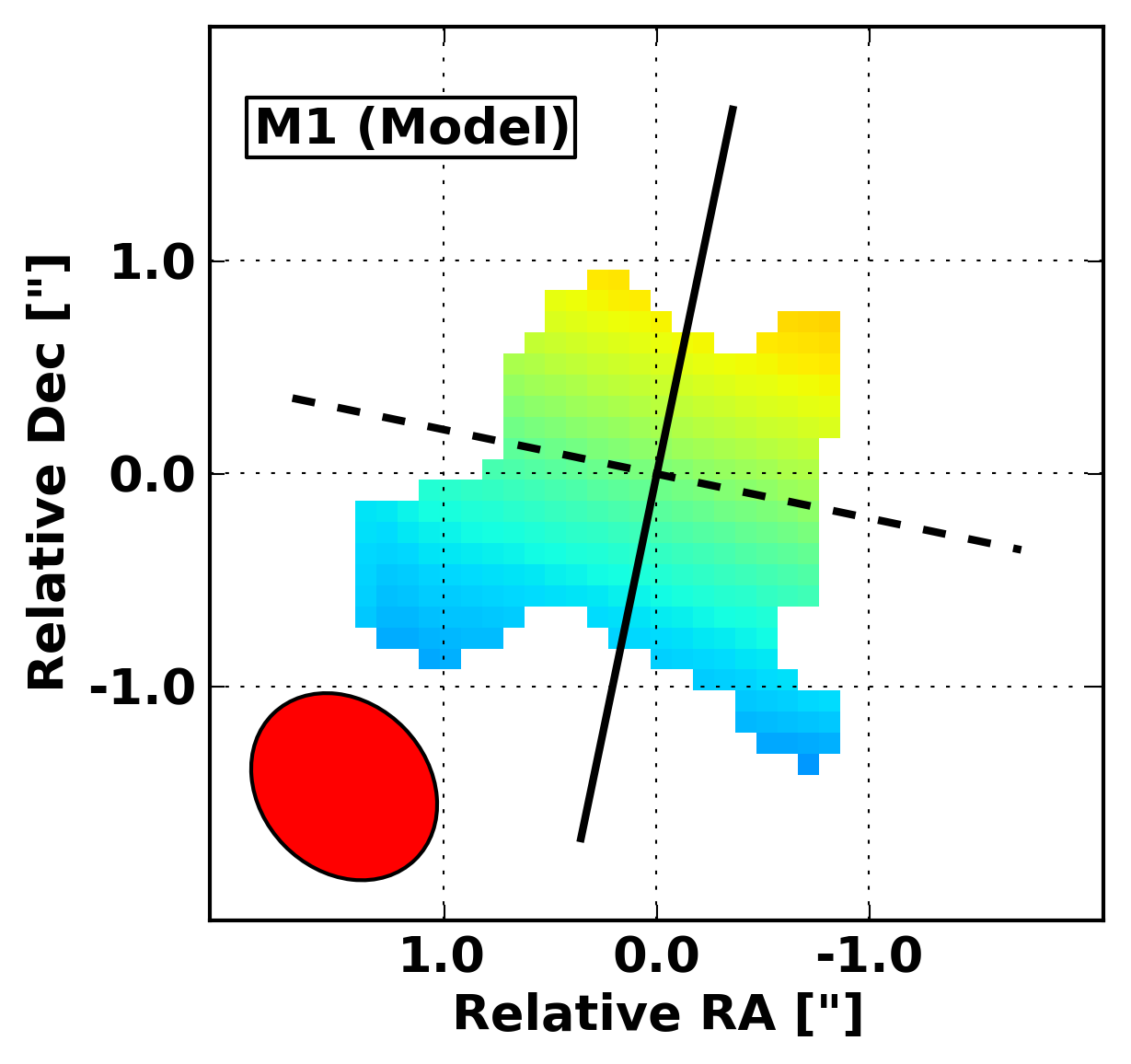} & \hspace{-0.5cm}
\includegraphics[scale=0.5]{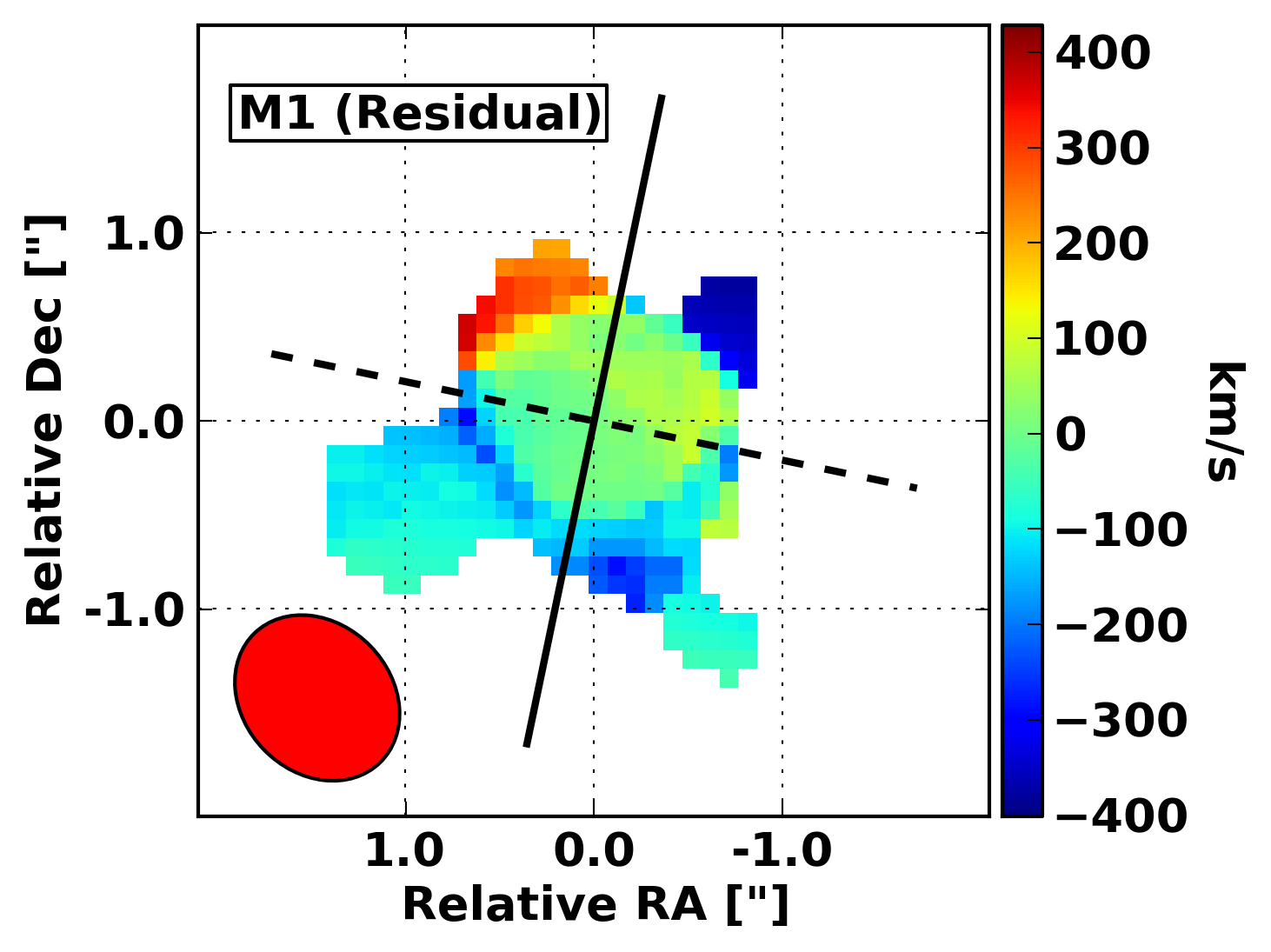} \\
\includegraphics[scale=0.5]{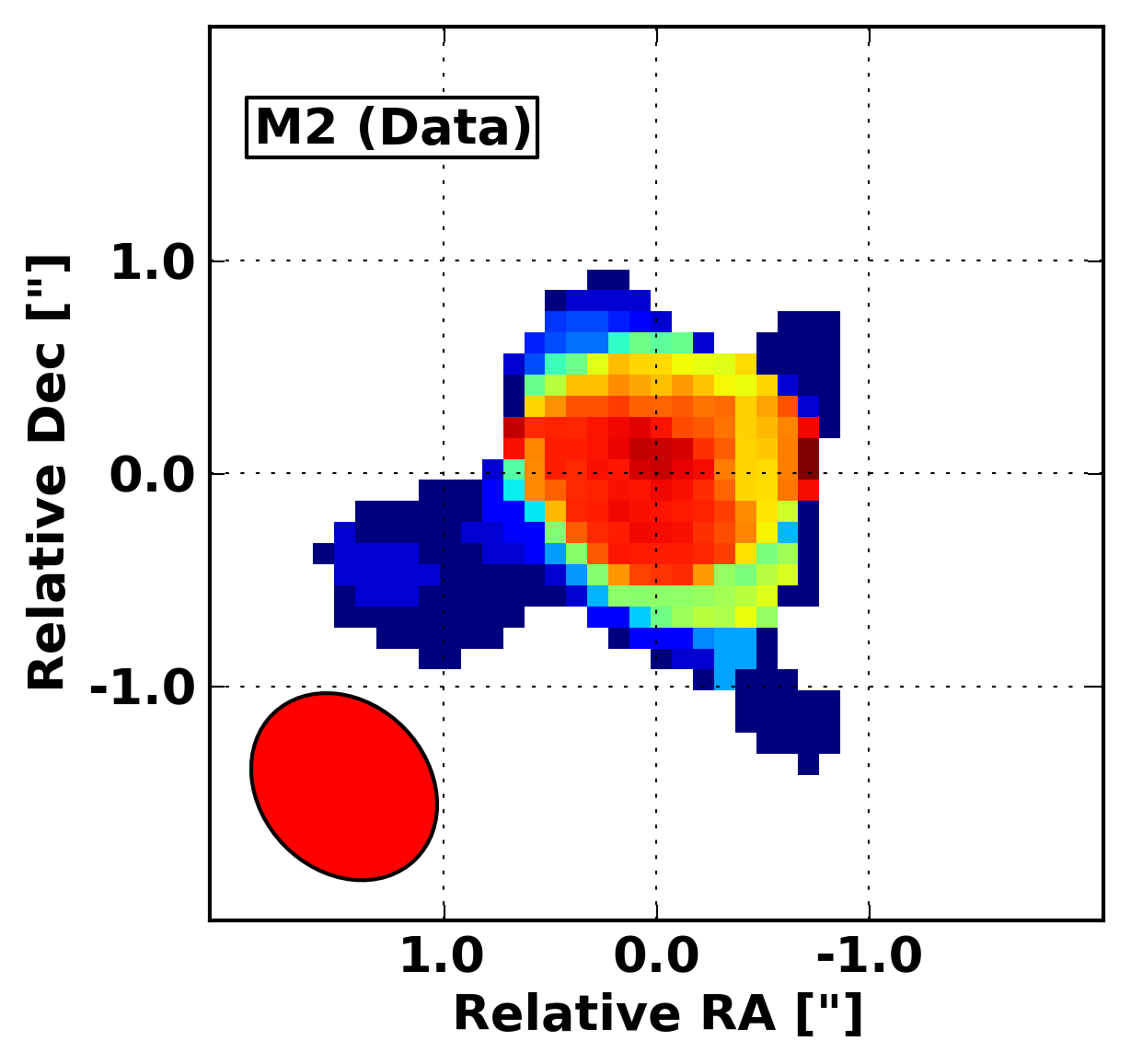} & \hspace{-0.5cm} \includegraphics[scale=0.5]{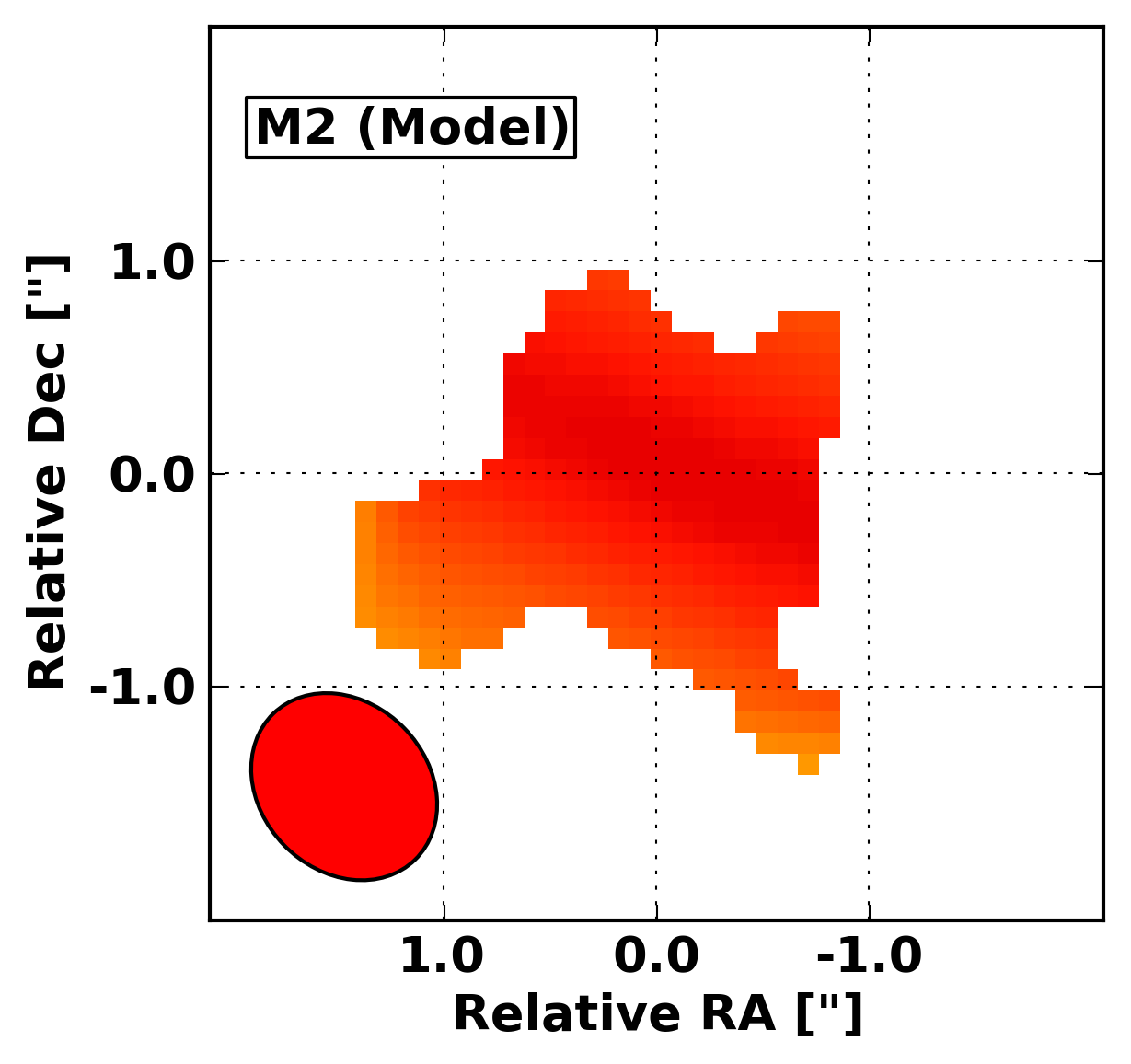} & \hspace{-0.5cm}
\includegraphics[scale=0.5]{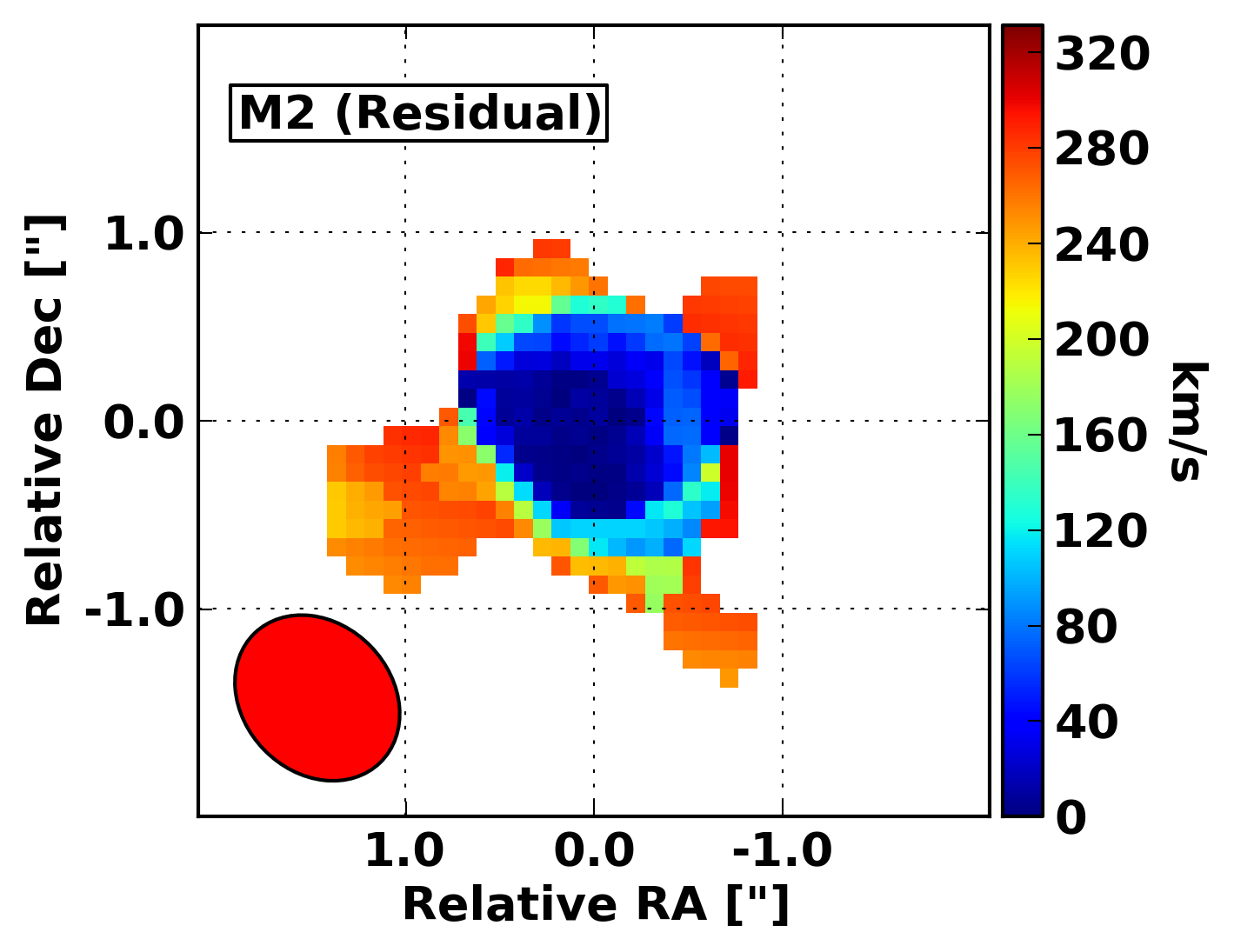} \\
\end{tabular}
\caption{Line intensity (top; M0), velocity (middle; M1) and velocity dispersion (bottom; M2) maps for ULASJ1234. The observed data are shown in the left most panels and the best-fit models from $^{\rm 3D}$Barolo together with the fit residuals are shown in the middle and right panels respectively.}
\label{fig:maps_J1234}
\end{center}
\end{figure*}

The quasar host galaxy J1234 ($z=2.501$) and its two CO-luminous companions - G1234N ($z=2.514$; $\sim$170 kpc projected distance from the quasar) and G1234S ($z=2.497$; $\sim$90 kpc projected distance from the quasar) - are from hereon referred to as the J1234 system. The CO images shown in Fig. \ref{fig:CO_J1234} have been constructed by collapsing all the spectral channels containing the line in each source. Initial source sizes for the gas emission were derived by fitting a single 2-dimensional Gaussian to the CO images using the \textsc{casa} task \textit{imfit}. Using these constraints as an initial guess, the fitting was then re-done in the u-v plane using \textit{uvmodelfit}. The results from the two fitting approaches are consistent for all the galaxies and the full-width-half-maximum (FWHM) of the major axis is (0.48$\pm$0.14)$\arcsec$, (0.57$\pm$0.18)$\arcsec$ and (0.53$\pm$0.24)$\arcsec$ for ULASJ1234, G1234N and G1234S respectively, after deconvolving from the beam. The FWHM values can be converted to a half-light radius using the equation, R$_{1/2}$=0.5$\times$FWHM for a Gaussian profile, and correspond to physical scales of $\sim$2 kpc at the redshifts of these galaxies. We summarise the half-light radius constraints for all three galaxies in Table \ref{tab:dynmod}. A second estimate of the half-light radius can be estimated from fitting the exponential disk models (Section \ref{sec:expdisk}) to the intensity maps. The half-light radii derived from the exponential disk models are also specified in Table \ref{tab:dynmod} and are generally larger than those derived from the Gaussian fits - $\sim$3-4 kpc at the redshifts of these galaxies. This is to be expected given the 2-D Gaussian profiles are not able to reproduce the full spatial extent of the CO emission seen from these galaxies.


We have already noted the double peaked line profiles for all three sources seen in Fig. \ref{fig:CO_J1234}, which provide evidence for kinematically distinct components of the gas emission. Two Gaussian components are required to adequately model the 1-dimensional spectral line profiles and these best spectral-fit models are over-plotted on the observed spectra in Fig. \ref{fig:CO_J1234}. The velocity separation between the two Gaussian components is 510$\pm$160 km/s, 340$\pm$50 km/s and 350$\pm$40 km/s for the quasar, G1234N and G1234S respectively. Such double-peaked profiles could arise from the approaching and receding sides of a rotating gas disk, a coalescing merger of two distinct galaxies, or large-scale inflows and outflows affecting the molecular gas. 

\begin{figure*}
\begin{center}
\begin{tabular}{ccc}
\includegraphics[scale=0.5]{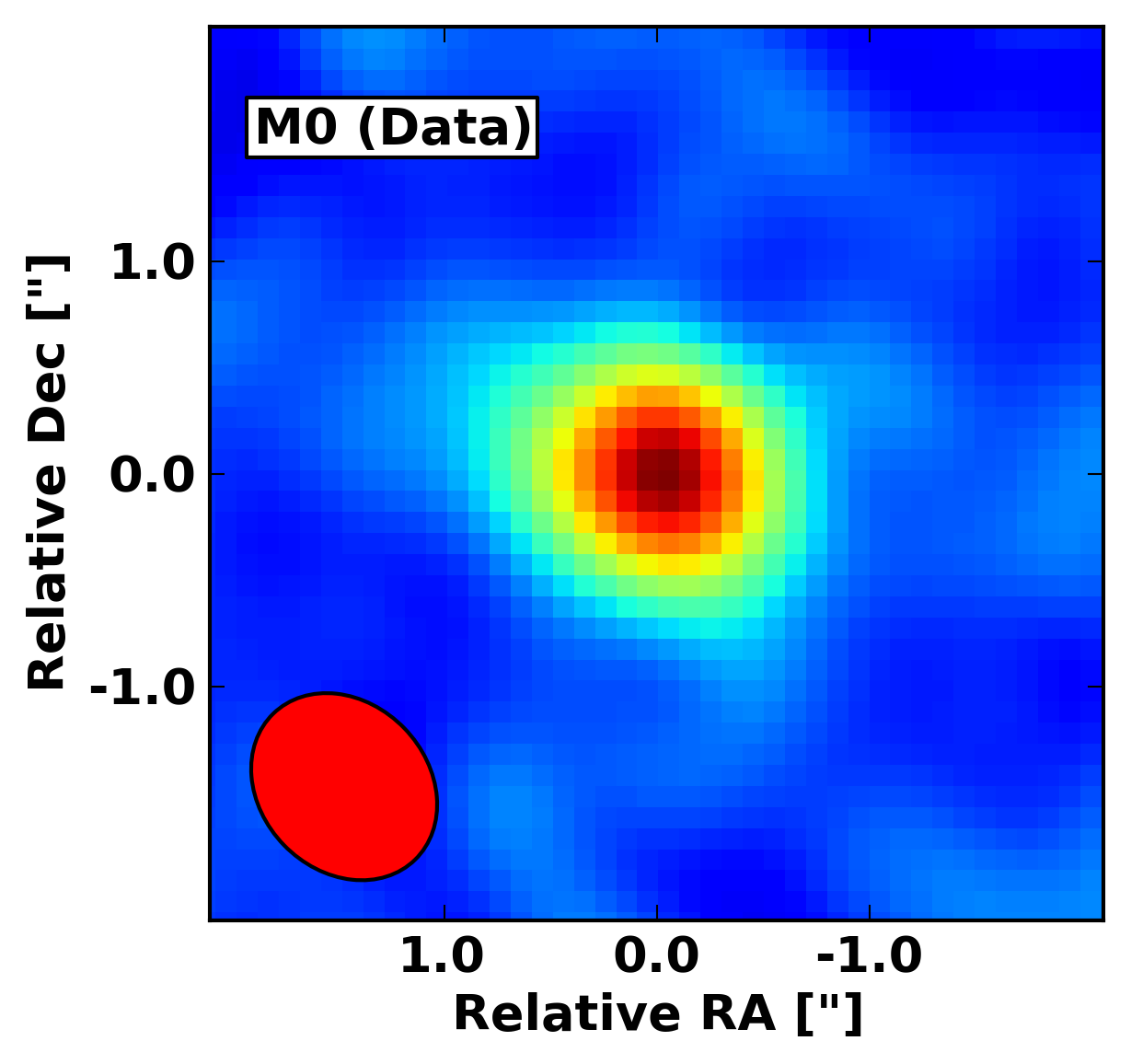} & \hspace{-0.5cm} \includegraphics[scale=0.5]{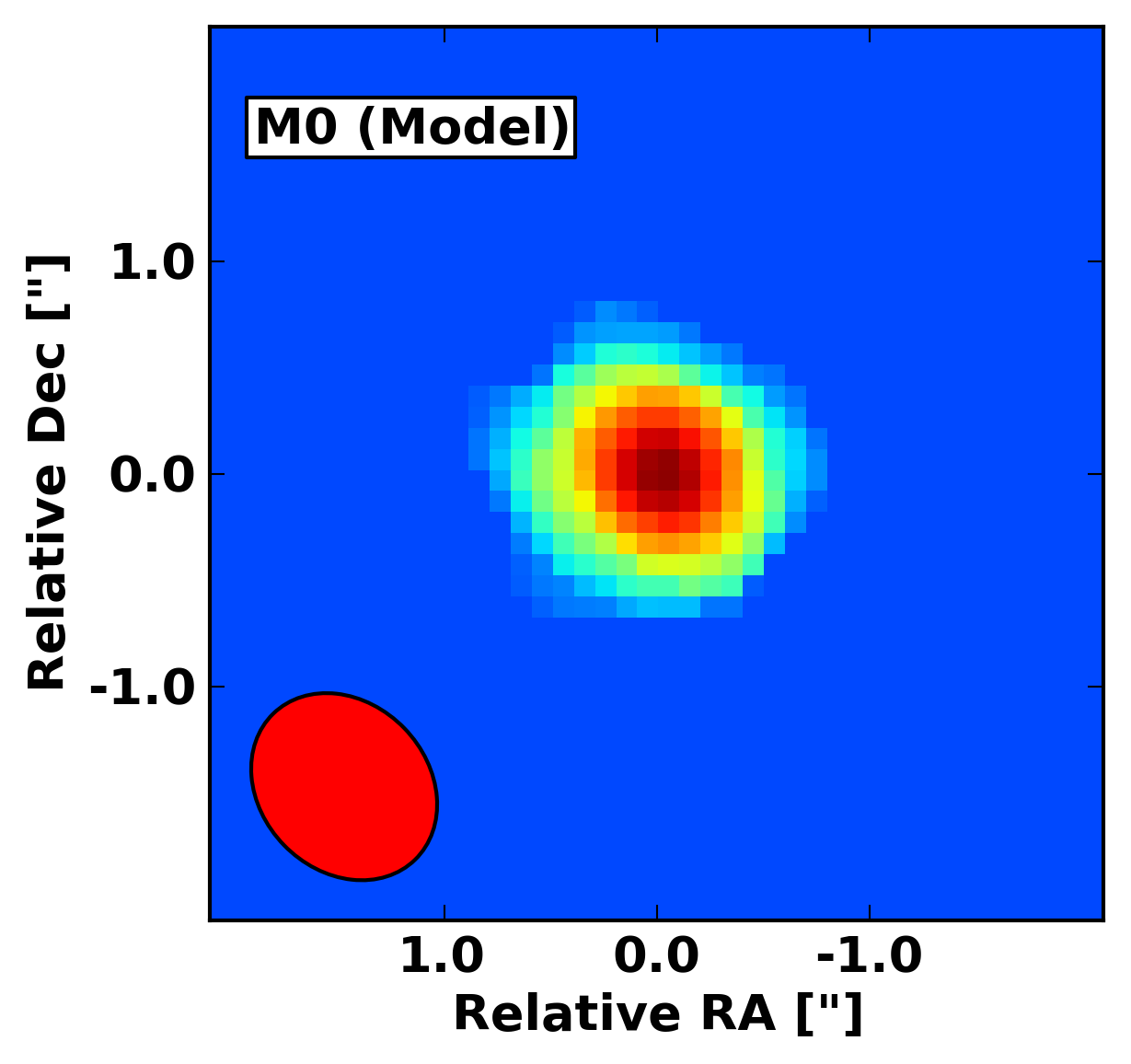} & \hspace{-0.5cm}
\includegraphics[scale=0.5]{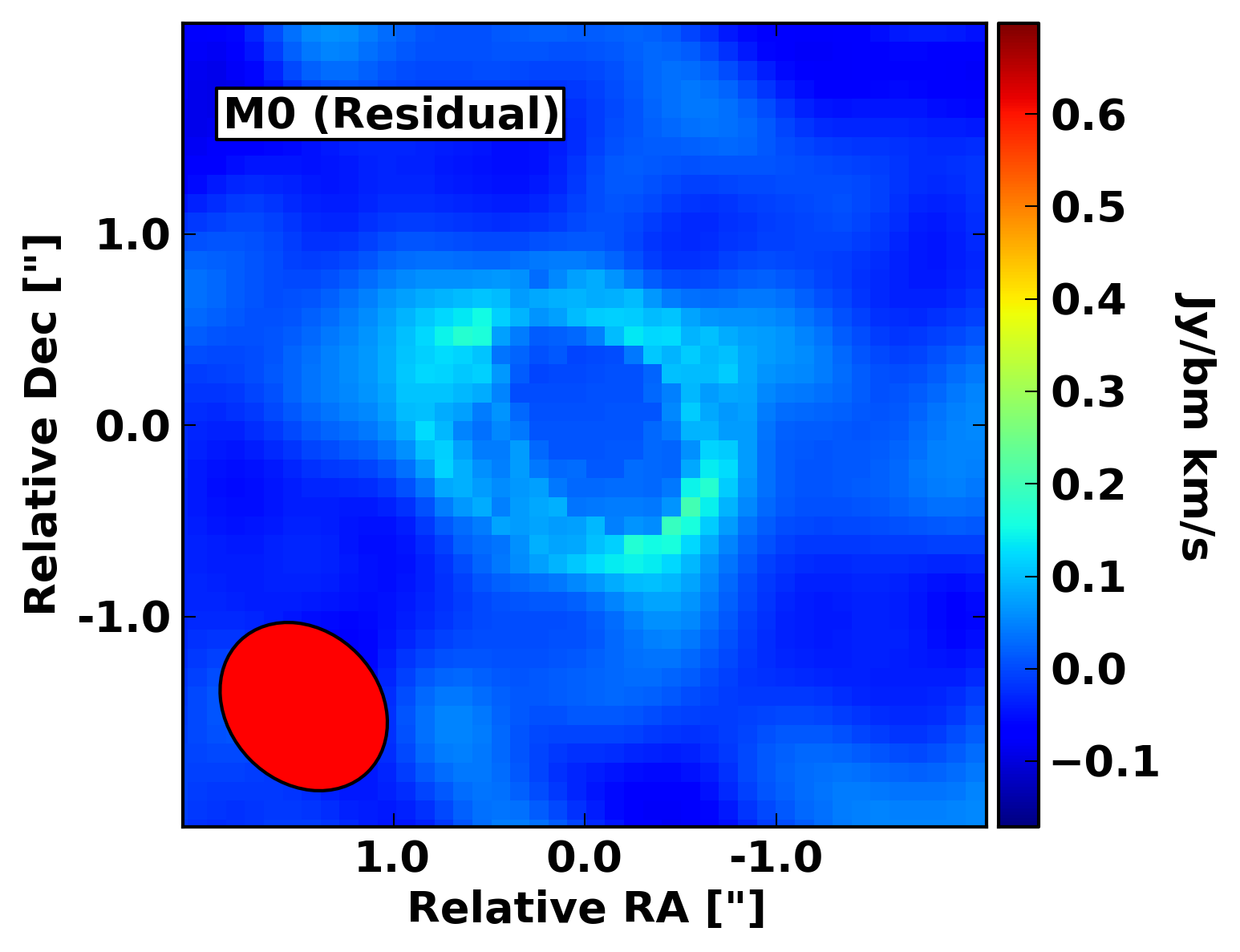} \\
\includegraphics[scale=0.5]{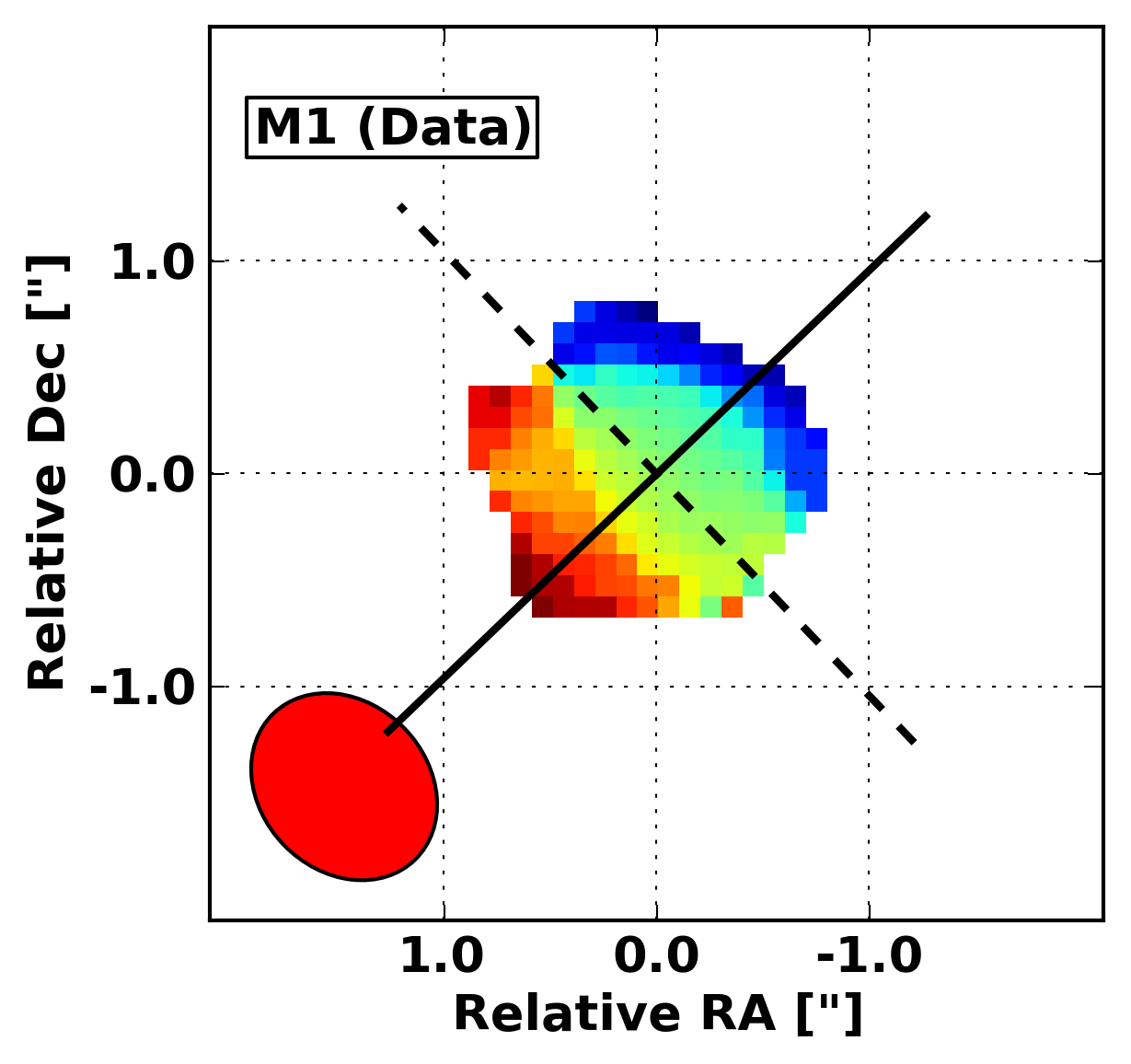} & \hspace{-0.5cm} \includegraphics[scale=0.5]{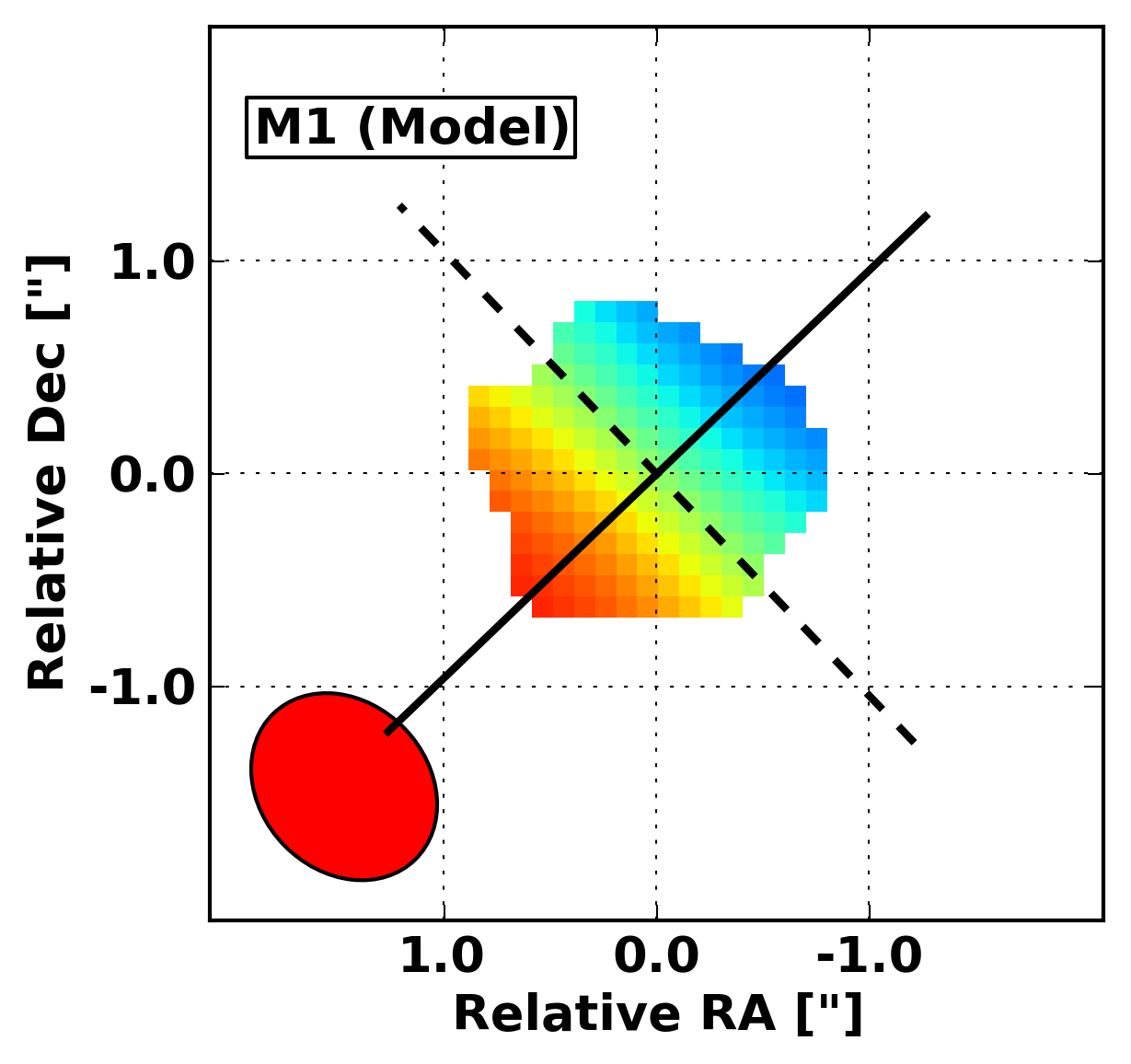} & \hspace{-0.5cm}
\includegraphics[scale=0.5]{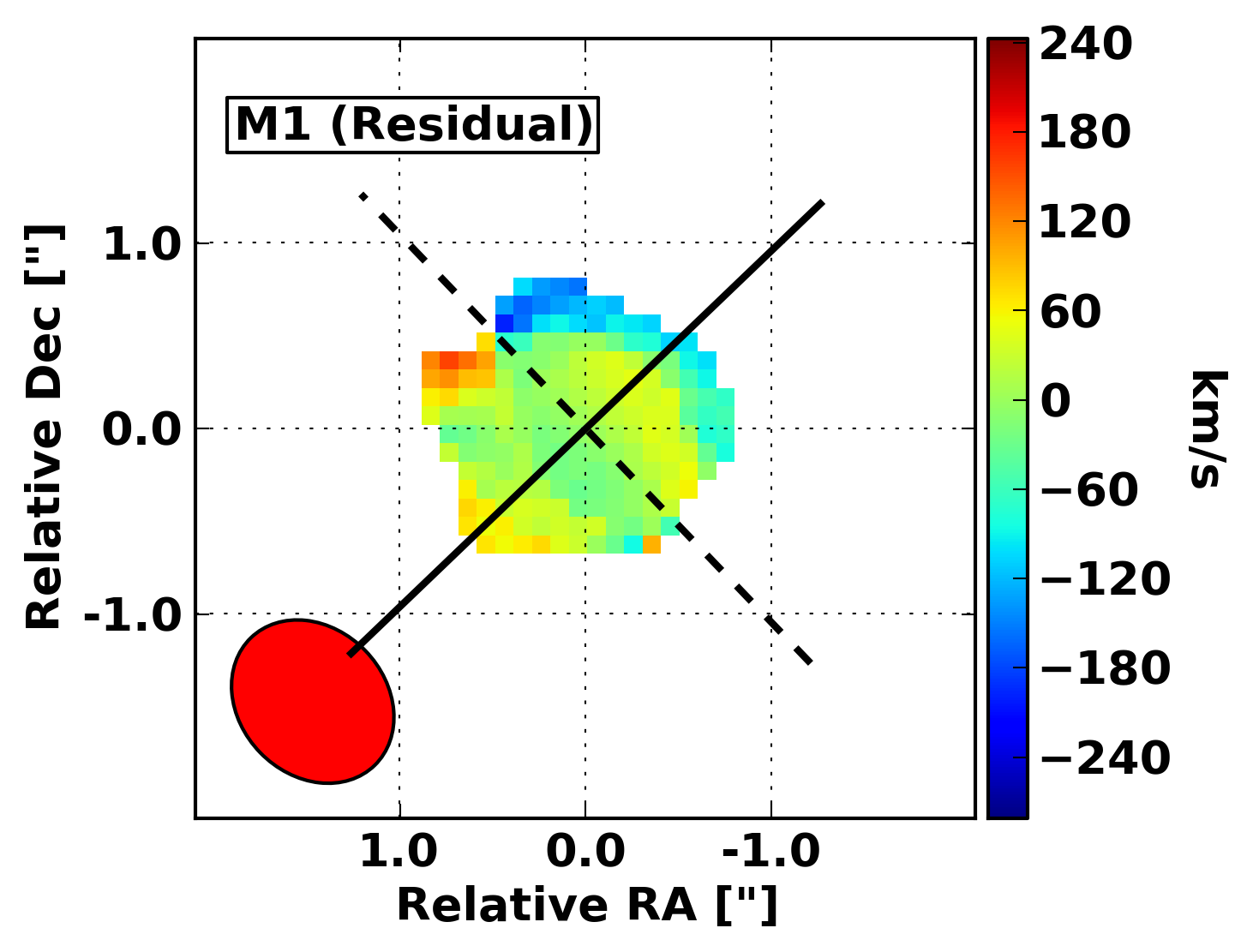} \\
\includegraphics[scale=0.5]{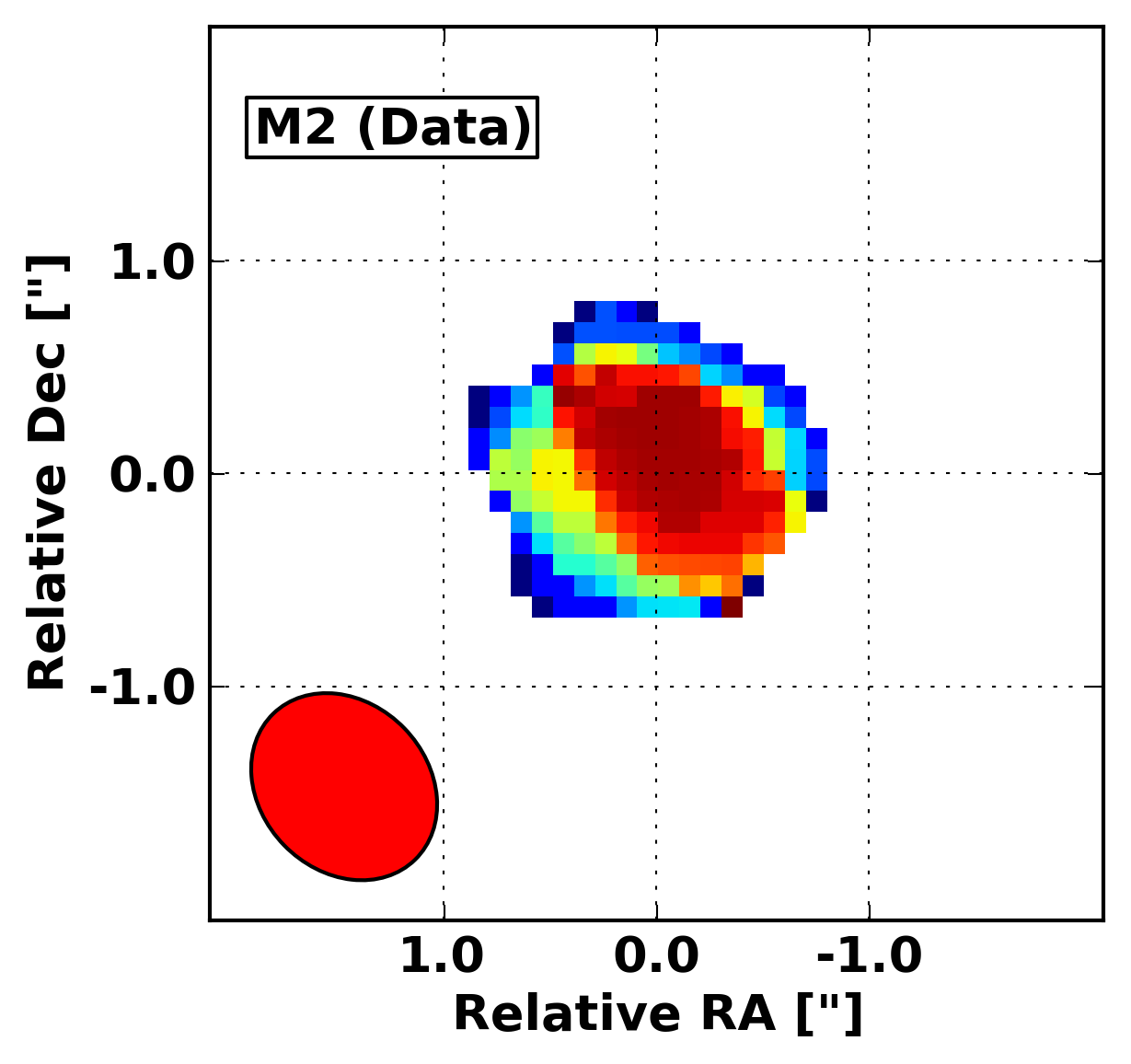} & \hspace{-0.5cm} \includegraphics[scale=0.5]{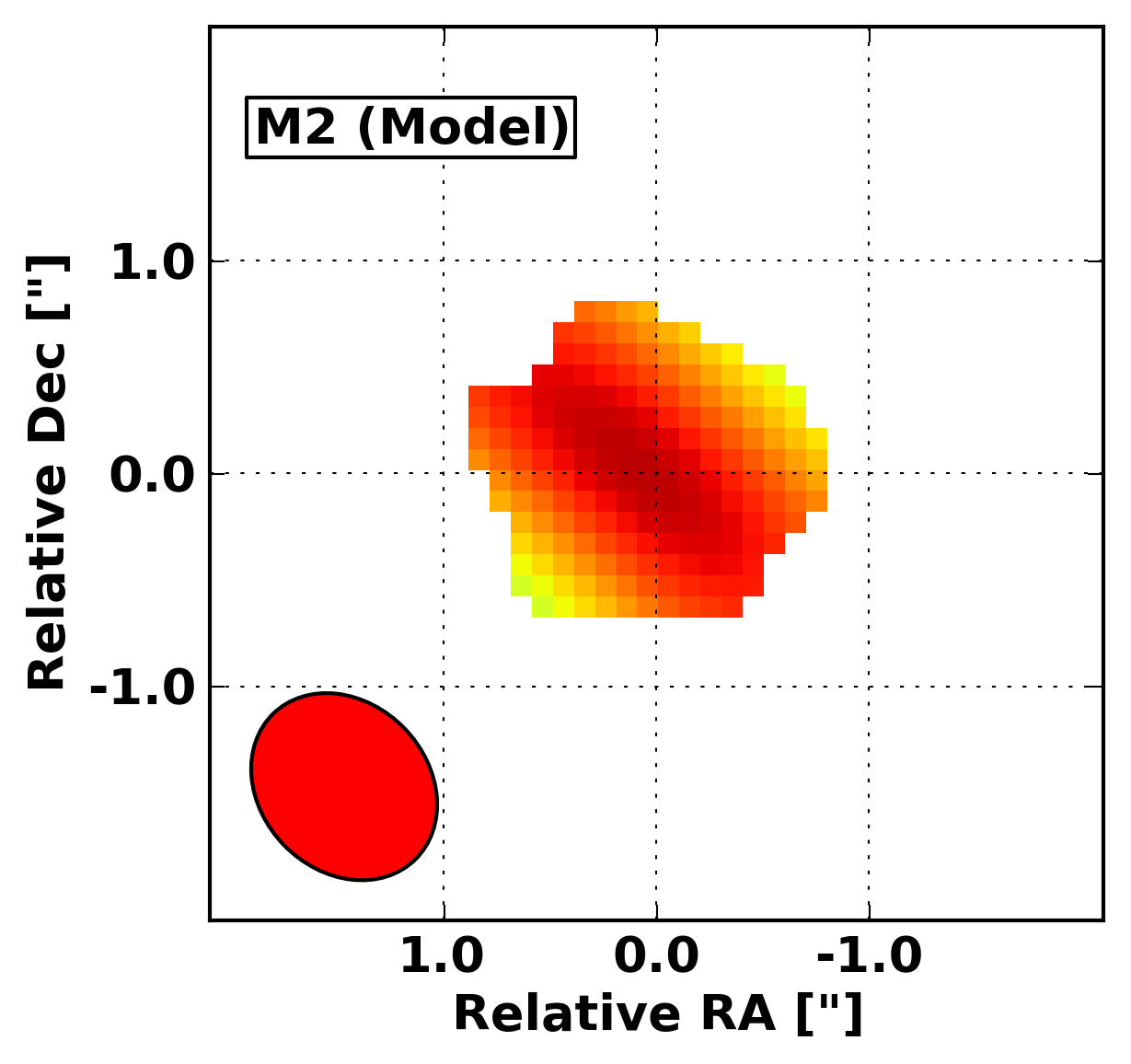} & \hspace{-0.5cm}
\includegraphics[scale=0.5]{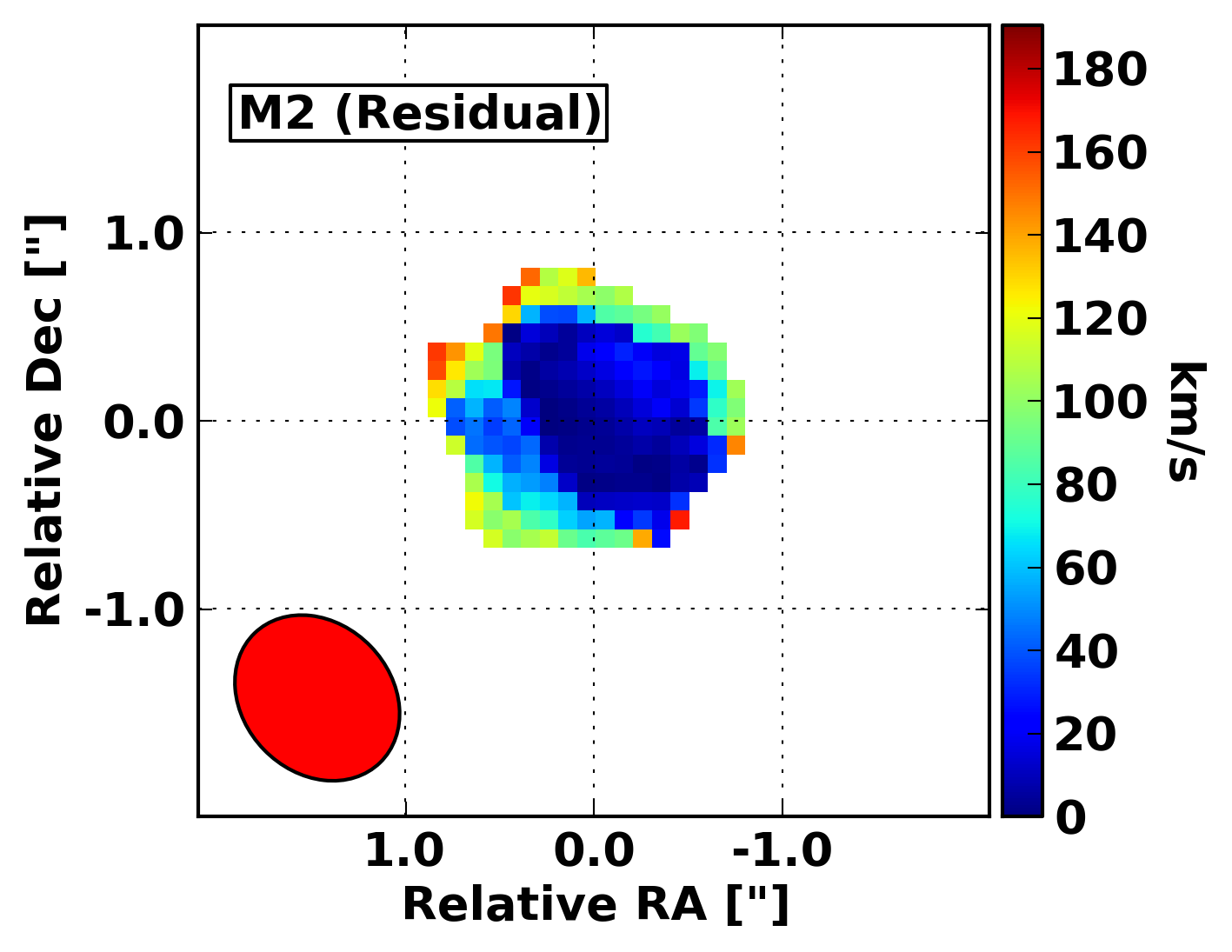} \\
\end{tabular}
\caption{Line intensity (top; M0), velocity (middle; M1) and velocity dispersion (bottom; M2) maps for G1234N. The observed data are shown in the left most panels and the best-fit models from $^{\rm 3D}$Barolo together with the fit residuals are shown in the middle and right panels respectively.} 
\label{fig:maps_G1234N}
\end{center}
\end{figure*}

\begin{figure*}
\begin{center}
\begin{tabular}{ccc}
\includegraphics[scale=0.5]{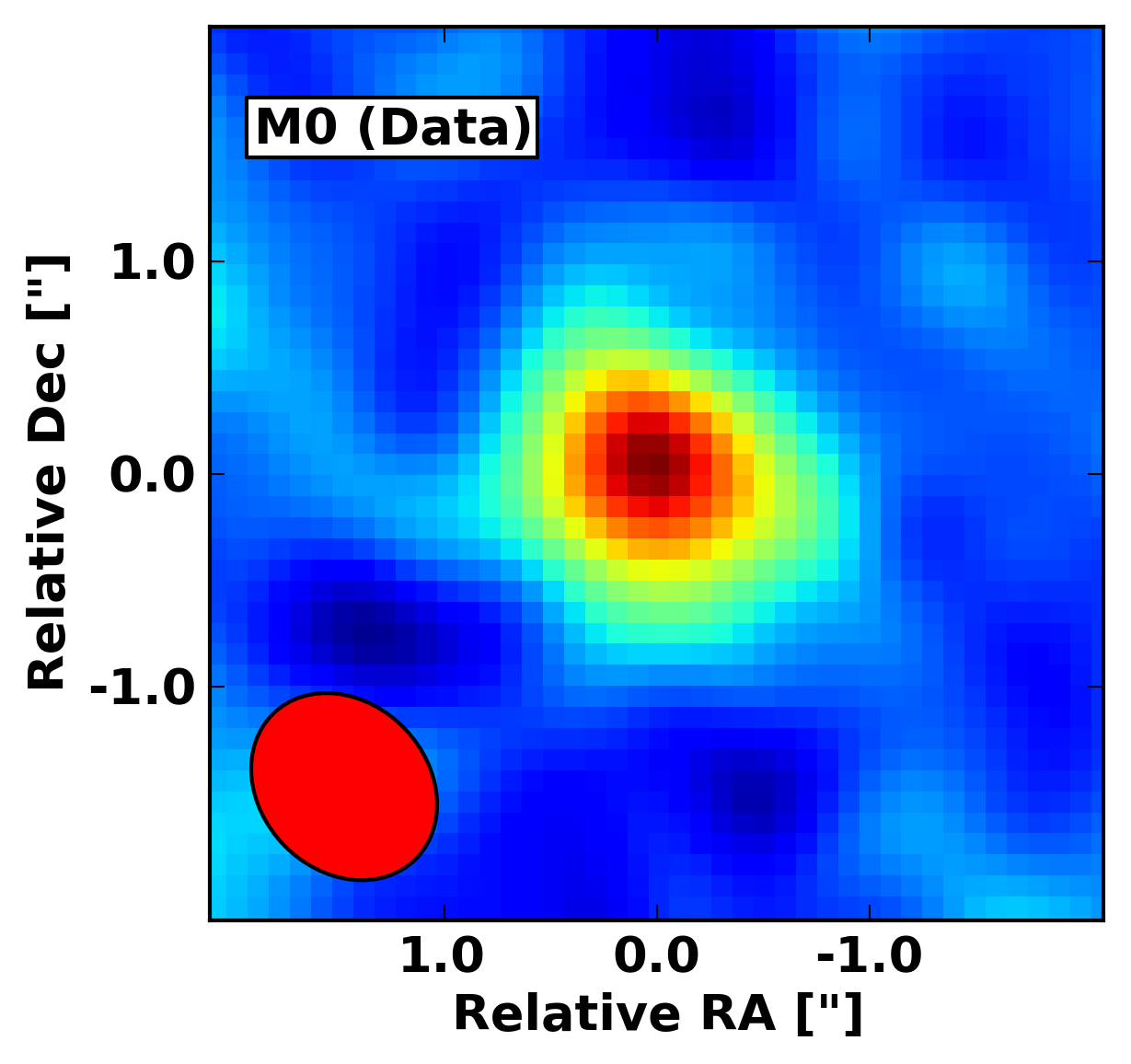} & \hspace{-0.5cm} \includegraphics[scale=0.5]{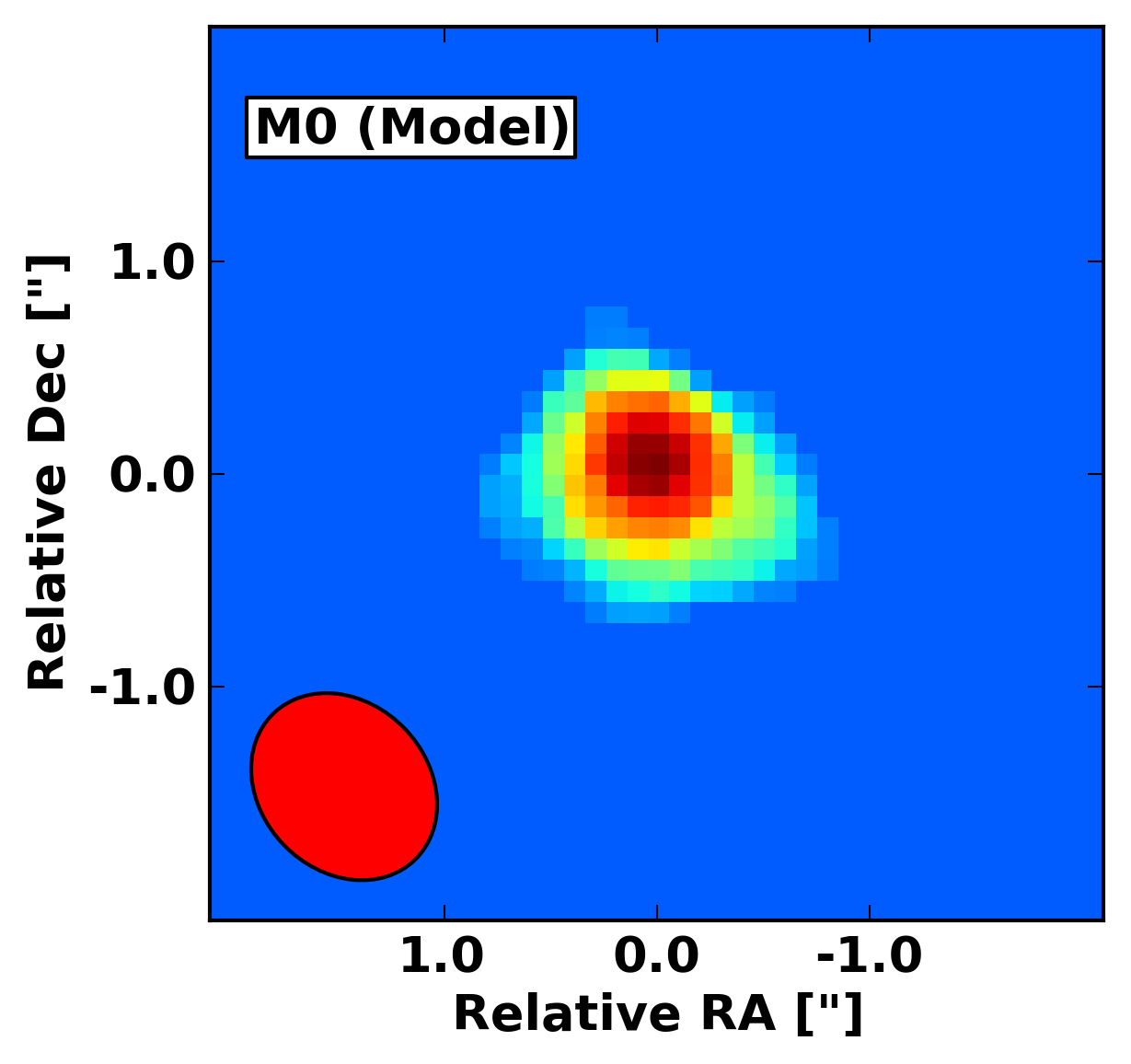} & \hspace{-0.5cm}
\includegraphics[scale=0.5]{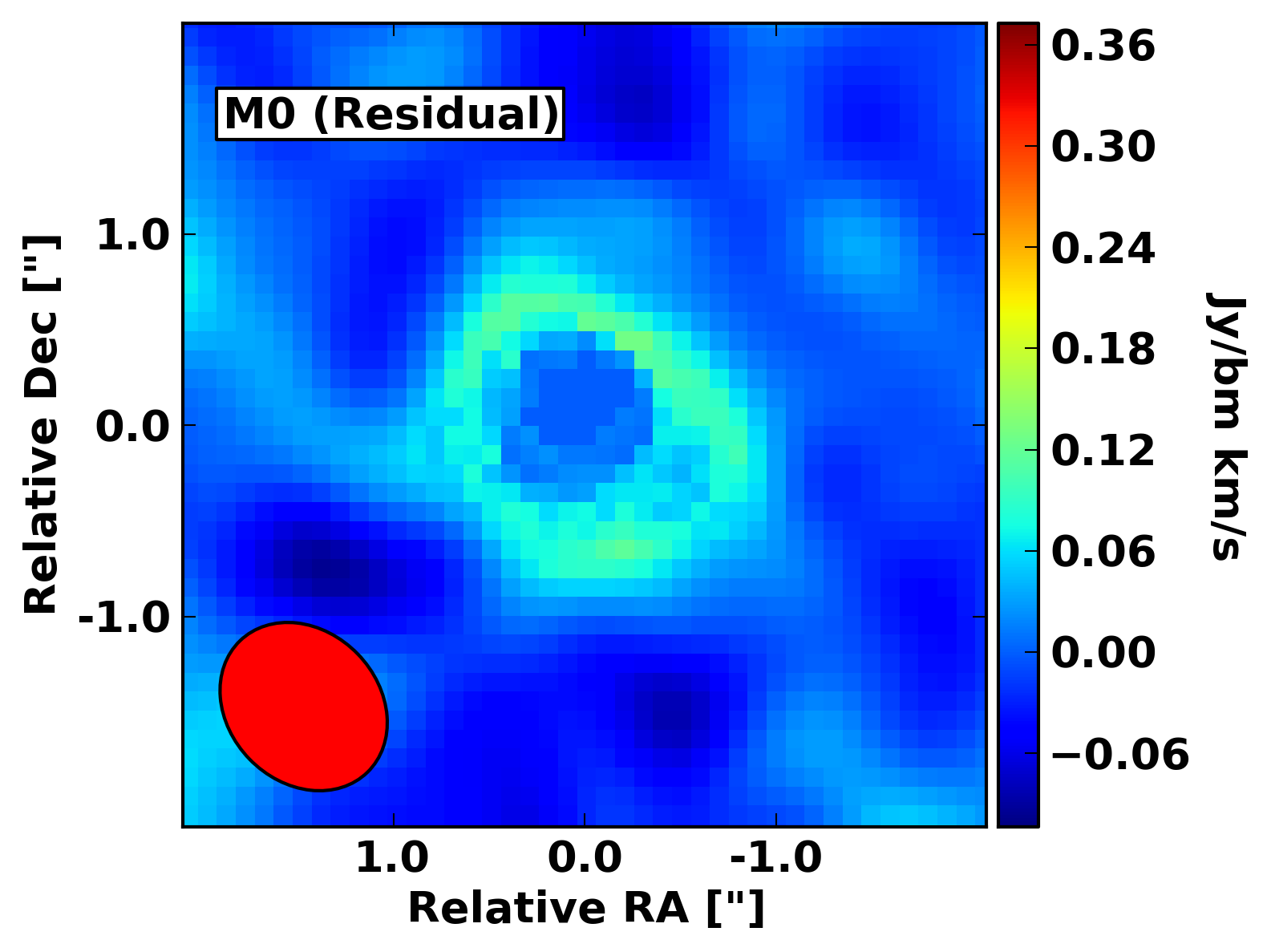} \\
\includegraphics[scale=0.5]{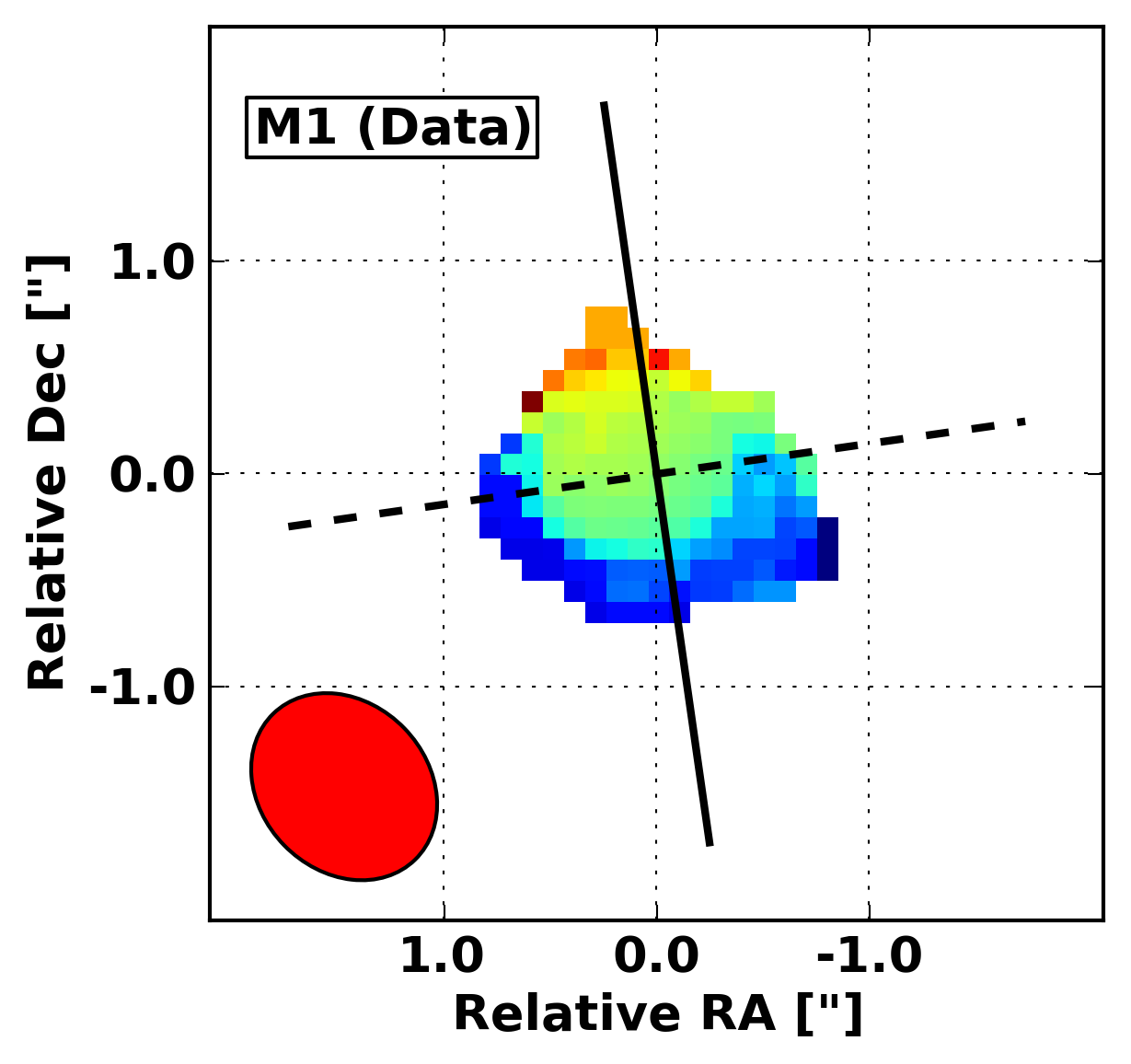} & \hspace{-0.5cm} \includegraphics[scale=0.5]{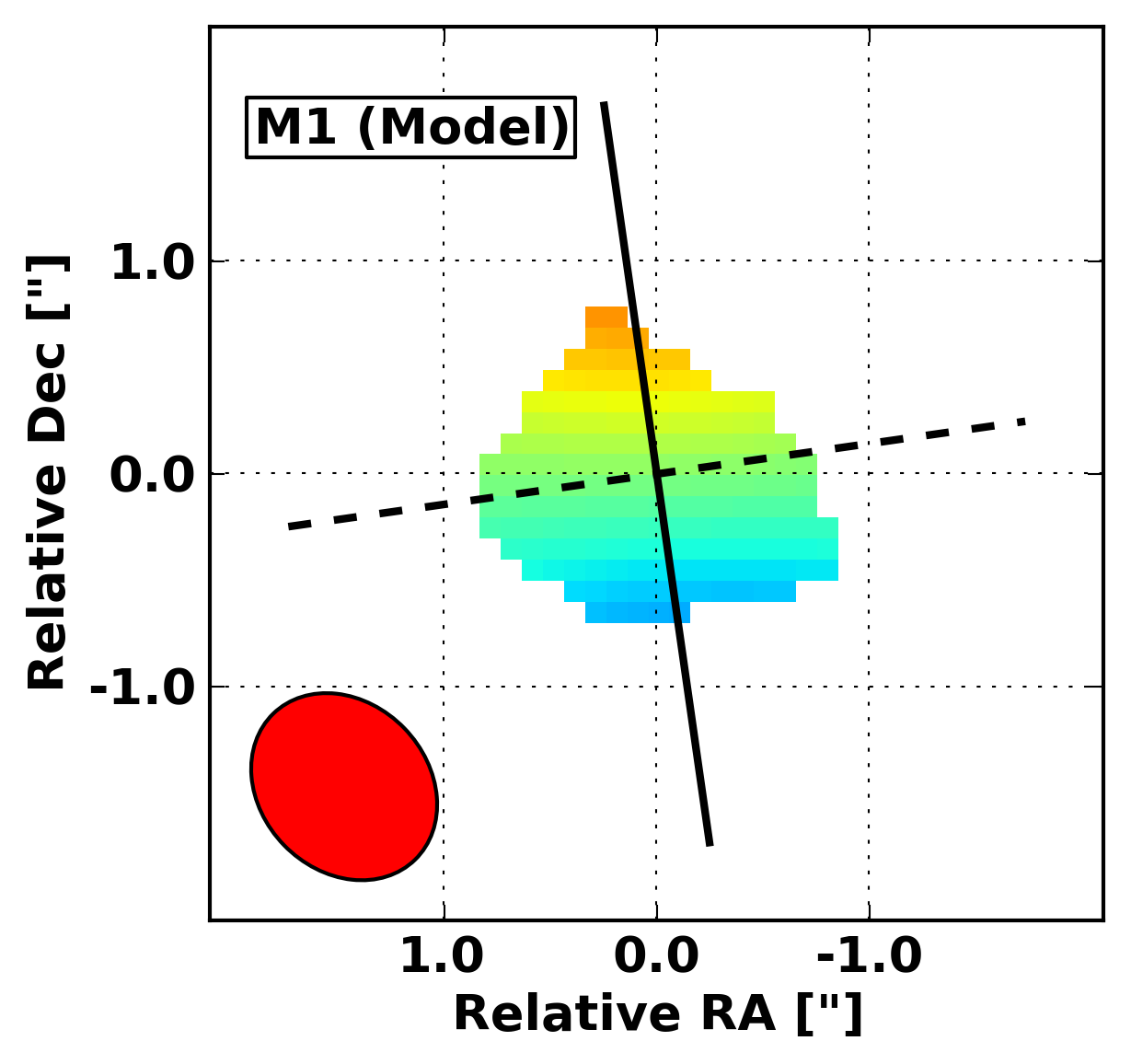} & \hspace{-0.5cm}
\includegraphics[scale=0.5]{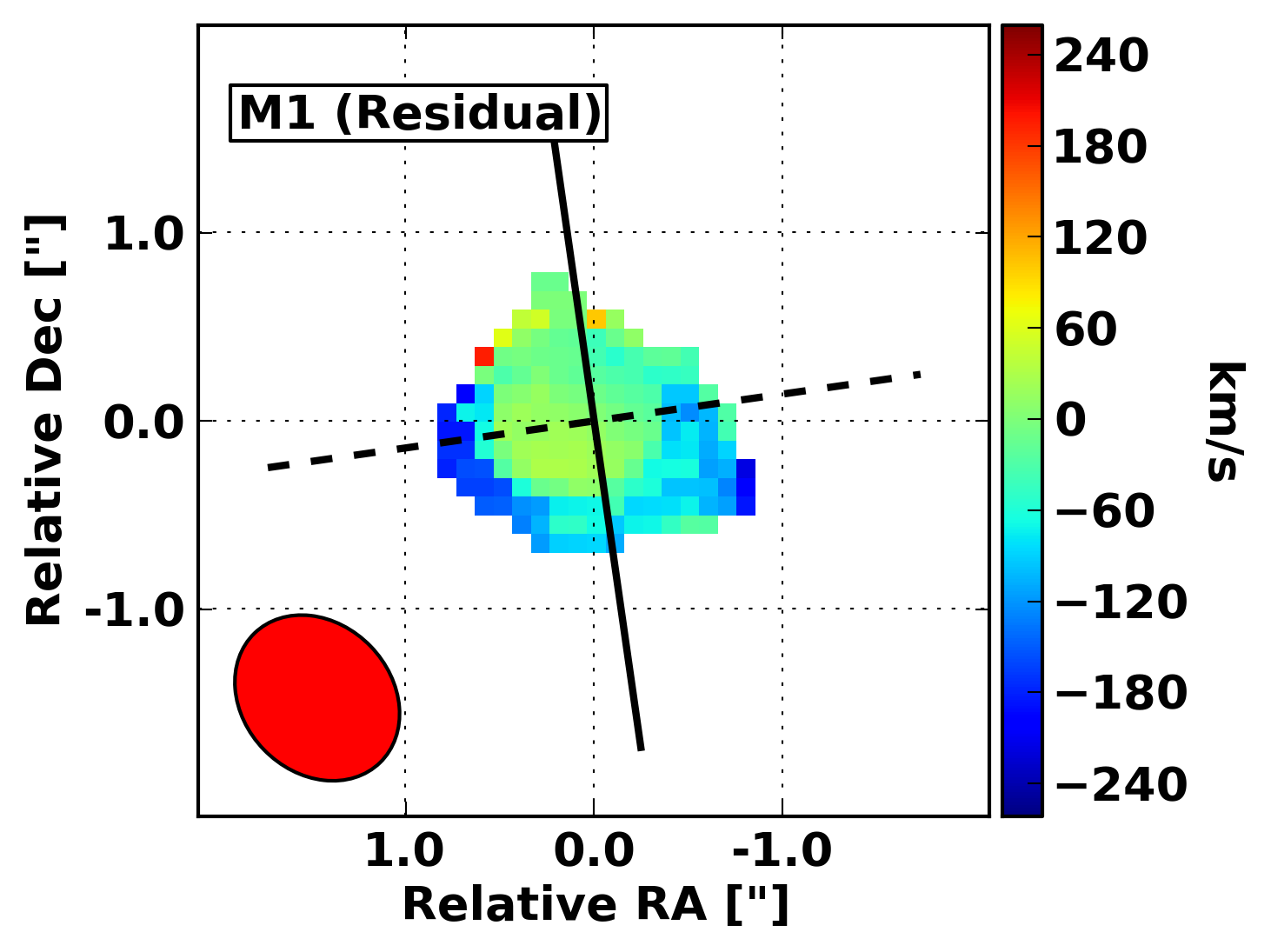} \\
\includegraphics[scale=0.5]{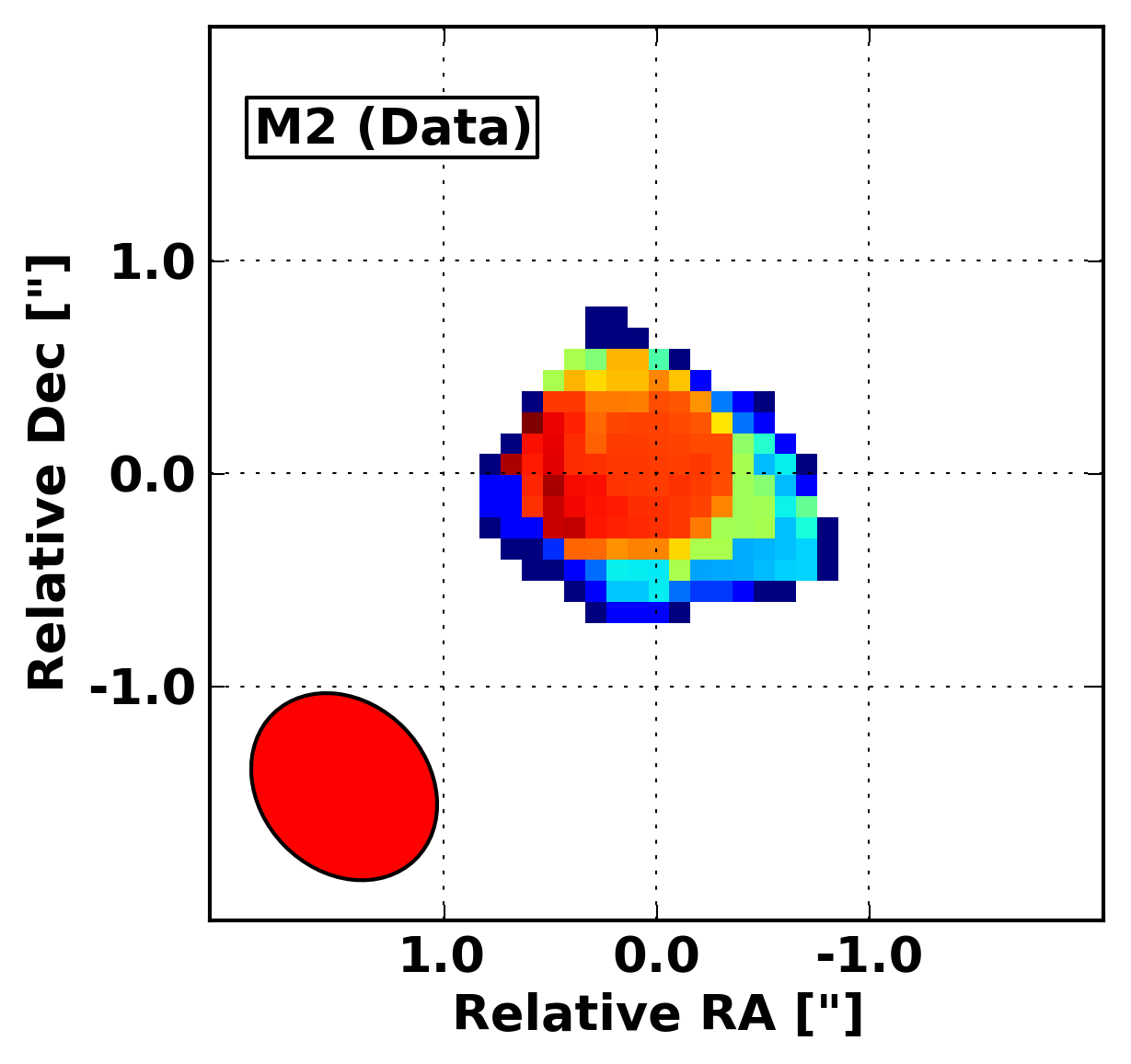} & \hspace{-0.5cm} \includegraphics[scale=0.5]{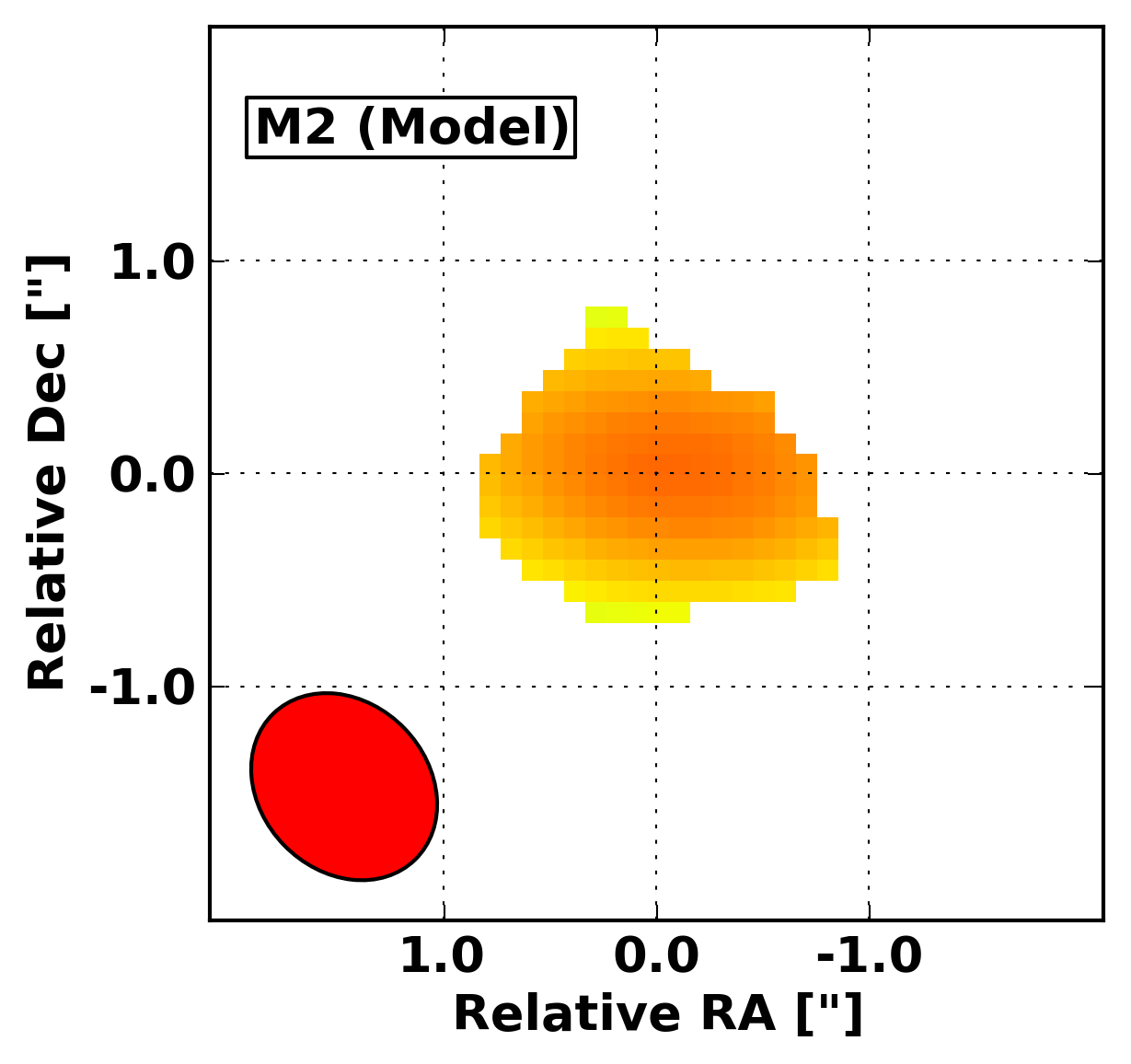} & \hspace{-0.5cm}
\includegraphics[scale=0.5]{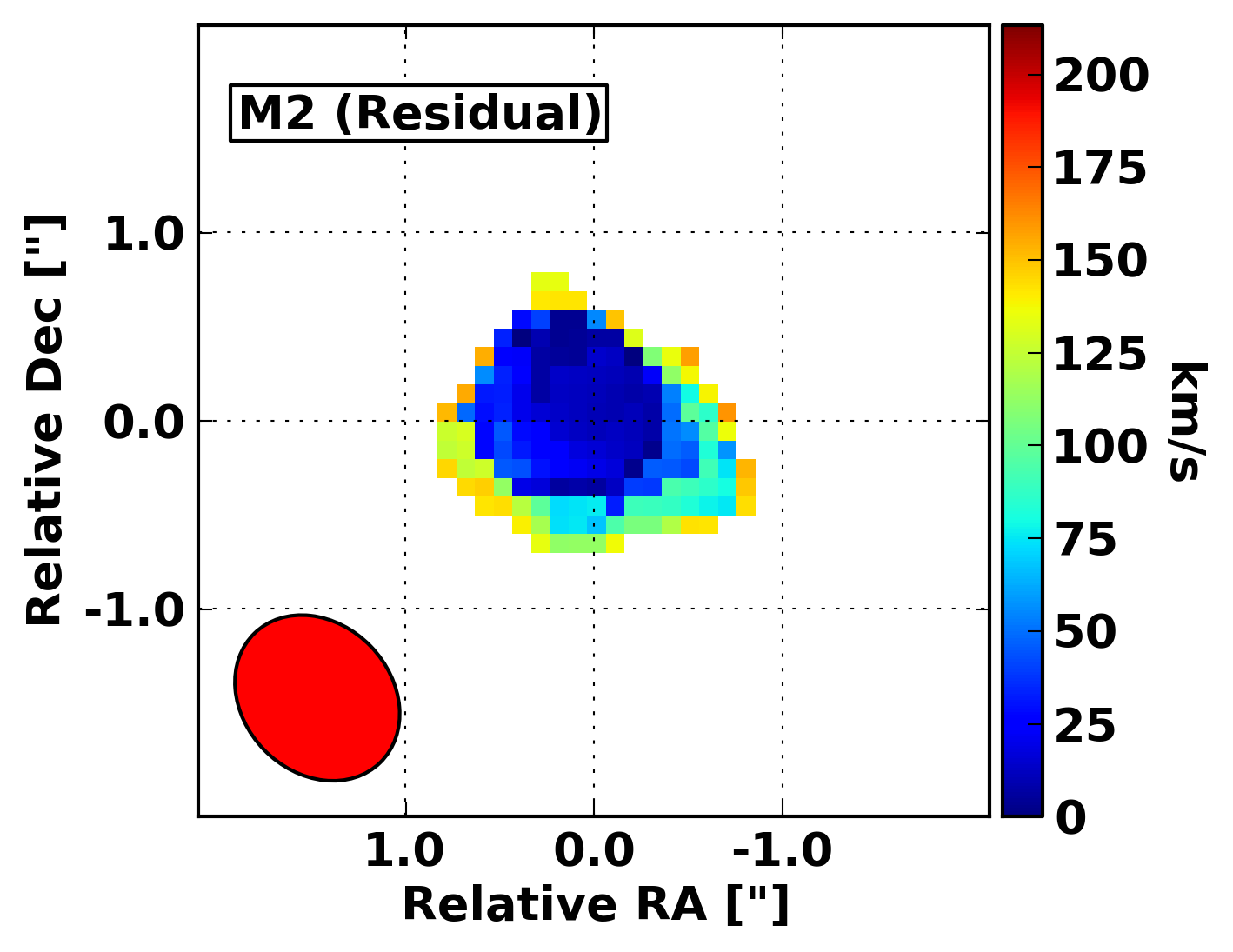} \\
\end{tabular}
\caption{Line intensity (top; M0), velocity (middle; M1) and velocity dispersion (bottom; M2) maps for G1234S. The observed data are shown in the left most panels and the best-fit models from $^{\rm 3D}$Barolo together with the fit residuals are shown in the middle and right panels respectively.}
\label{fig:maps_G1234S}
\end{center}
\end{figure*}

As can be seen in the moment 1 maps in Fig. \ref{fig:maps_J1234}, \ref{fig:maps_G1234N} and \ref{fig:maps_G1234S}, a velocity gradient is evident across the full spatial extent of all three galaxies that comprise the J1234 system. The best-fit $^{\rm 3D}$Barolo models together with the fit residuals can also be seen in the same figures. In general the fit residuals are small, at least in the central portions of these galaxies, suggesting that a smooth velocity gradient provides a reasonable approximation to these data. The resulting dynamical masses are summarised in Table \ref{tab:dynmod}. With dynamical masses of $\sim10^{11}$M$_\odot$, all three galaxies in the J1234 system can be considered among the most massive systems seen at these epochs. 

\begin{table*}
\begin{center}
\caption{Source size and dynamical mass measurements for all three galaxies in the J1234 system.}
\label{tab:dynmod}
\begin{tabular}{lccc}
\hline
& ULASJ1234 (QSO) & G1234N & G1234S \\
\hline
z$_{\rm{sys}}$ & 2.501 & 2.515 & 2.498 \\
r$_{1/2,\rm{Gauss}}$ / $\arcsec$ & 0.24$\pm$0.07 & 0.29$\pm$0.09 & 0.27$\pm$0.12 \\
r$_{1/2,\rm{Exp}}$ / $\arcsec$ & 0.32$\pm$0.02 & 0.45$\pm$0.02 & 0.51$\pm$0.02 \\
log$_{10}$(M$_{\rm{dyn}}$/M$_\odot$) - $^{\rm 3D}$Barolo & 10.84$\pm0.23$ & 11.11$\pm$0.16 & 11.17$\pm0.17$ \\
log$_{10}$(M$_{\rm{dyn}}$/M$_\odot$) - Exp Disk & 11.7$\pm^{0.4}_{0.2}$ & 11.2$\pm^{0.2}_{0.3}$ & 11.5$\pm^{0.9}_{0.4}$ \\
\hline
\end{tabular}
\end{center}
\end{table*}

We also derive dynamical masses for all three galaxies using the exponential disk modelling. The resulting dynamical masses summarised in Table \ref{tab:dynmod} are larger than those obtained from $^{\rm 3D}$Barolo by 0.9 dex, 0.1 dex and 0.3 dex for the three galaxies. As stated earlier, the exponential disk models are sensitive to mass over a much larger physical radius than the $^{\rm 3D}$Barolo models. Furthermore, the inclination of these galaxies is poorly constrained and adds significant additional uncertainty to the mass estimates. Given these factors, the discrepancies in the dynamical mass estimates from the two methods is not surprising. 

\begin{figure}
\begin{center}
\hspace{-1.8cm}
\includegraphics[scale=0.7]{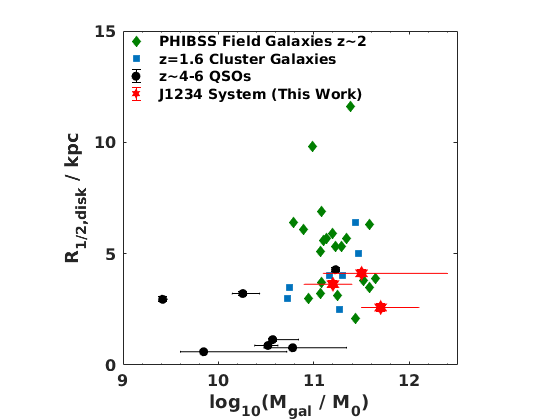} 
\caption{Host galaxy mass versus disk half-light radius for the three galaxies in the J1234 system and compared to various samples from the literature. The PHIBSS field galaxy sample is taken from \citet{Tacconi:13}, the cluster sample measurements are taken from \citet{Noble:18} and the high-redshift quasar sample is from \citet{Pensabene:20}.}
\label{fig:J1234_mass_size}
\end{center}
\end{figure}

 In Fig. \ref{fig:J1234_mass_size} we show the dynamical mass versus the half-light radius of the CO emission from the exponential disk models for all three galaxies in the J1234 system and compare to both field and cluster galaxy samples from the literature at $z\sim1-2.5$. The field galaxies from the PHIBSS sample correspond to CO(3-2) measurements \citep{Tacconi:13} whereas the $z\sim1.6$ cluster galaxy measurements are taken from the CO(2-1) observations presented in \citet{Noble:18}. We take the galaxy mass as the sum of the stellar mass and molecular gas mass in the case of the literature samples and as the dynamical mass in the case of our quasar host galaxy and its companions. In general, all three galaxies in the J1234 system are at the more massive and more compact end of the distribution of $z\sim2$ galaxies from the literature. We also show the disk half-light radii for $z\sim4-6$ quasars from \citet{Pensabene:20}, which uses the same disk modelling as we have employed in this work. The higher redshift quasars are generally found to be more compact in size but this likely reflects the differing spatial distribution of the [CII] gas in the case of the $z\sim4-6$ QSOs and the relatively low-J CO gas traced by the $z\sim2$ samples. 

In Paper I, we derived gas masses for all galaxies in the J1234 system using several different estimators, thus mitigating against the uncertainties implicit in using a single molecular tracer of the total gas mass. The resulting gas masses were found to be $\sim3\times10^{10}$M$_\odot$, $\sim2\times10^{10}$M$_\odot$ and $\sim1-7\times10^{10}$M$_\odot$ for ULASJ1234, G1234N and G1234S respectively. In the case of G1234S the large range results from a much larger CI-derived gas mass compared to the CO-derived gas mass (Paper I). Combining these gas mass estimates with the dynamical masses and assuming that the inner parts of the disks traced by our CO observations are indeed baryon-dominated, we can infer the molecular gas fractions. The gas fraction is traditionally defined as the ratio of the gas mass to the total stellar mass. In the absence of stellar mass estimates for our high-luminosity quasars, we infer the stellar masses from the dynamical masses via M$_\ast$=M$_{\rm{dyn}}-$M$_{\rm{gas}}-$M$_{\rm{BH}}$, as has been done in previous studies of high-luminosity quasars (e.g. \citealt{Bischetti:21}). In the case of the quasar, very different gas fractions are implied depending on whether we consider the dynamical mass from $^{\rm 3D}$Barolo, which represents the mass in the innermost regions of the galaxy, which are seen in CO emission, or whether we consider the exponential disk model mass, which extends out to 10kpc. In the former case, we obtain gas fractions of $\sim$85\% whereas the gas fraction relative to the extended disk model mass is more like $\sim$6\%. It is reasonable to assume that the galaxy is baryon dominated even out to 10kpc, in which case there must be a substantial stellar component already in place with M$_\ast \sim 10^{11}$M$_\odot$ in this quasar. G1234N has a gas fraction of $\sim$14-18\% whereas in G1234S the gas fraction is more uncertain owing to the discrepant gas masses from the CO and CI observations, and could be in the range 7-90\%. 

The star formation rates for the two companion galaxies are poorly constrained due to a lack of photometric data points sampling the peak of the galaxy dust SED. However, for the quasar host galaxy we find an SFR of $>$1000M$_\odot$ yr$^{-1}$ based on \textit{Herschel} and ALMA observations (Paper I; \citealt{Banerji:14}). The well-established star-forming ``main-sequence'' of galaxies at high-redshifts \citep{Daddi:07} can be used to predict how much star formation a typical star-forming galaxy of a certain stellar mass is expected to have at these redshifts. We find that a M$_\ast\sim$10$^{11}$M$_\odot$ galaxy at $z\sim2$ would have a star formation rate of $\sim$200M$_\odot$yr$^{-1}$ were it to lie on the main-sequence and the quasar host galaxy is therefore clearly experiencing a prolific starburst. Both starburst galaxies as well as the most massive main-sequence galaxies in the high redshift Universe have molecular gas fractions of $\sim$30-50\% \citet{Tacconi:13, Casey:14} with some suggestions that the gas fractions might be even higher in richer proto-cluster and cluster environments \citep{Noble:17, Hayashi:18}.

If we consider the inner region of the J1234 quasar, the gas fraction is very high and comparable to what is seen in the most gas-rich starburst galaxies. However, over more extended scales the galaxy would appear to be substantially depleted of gas. In the case of the quasar host galaxy, it is plausible that feedback from the AGN has begun to deplete the molecular gas reservoir e.g. as suggested by \citet{Brusa:18}. However the two companion galaxies also exhibit somewhat depleted gas reservoirs with currently no evidence for an actively accreting AGN within them. \citet{Tadaki:19} suggest that the most massive galaxies with masses of $>10^{11}$ M$_\odot$ in the highest density environments at these redshifts could have gas accretion suppressed due to inefficient cooling of massive haloes, which could explain the fact that both G1234N and G1234S have gas fractions that are at the lower end of what is seen in typical star-forming galaxies at these epochs. We caution however that there are considerable uncertainties in both the dynamical and gas masses as illustrated by the discrepancies in these estimates using different dynamical modelling methods and different gas mass tracers. 

\subsection{CO Line Profiles}

\begin{figure}
\begin{center}
\begin{tabular}{c}
\large{ULASJ1234}  \\
\includegraphics[scale=0.6]{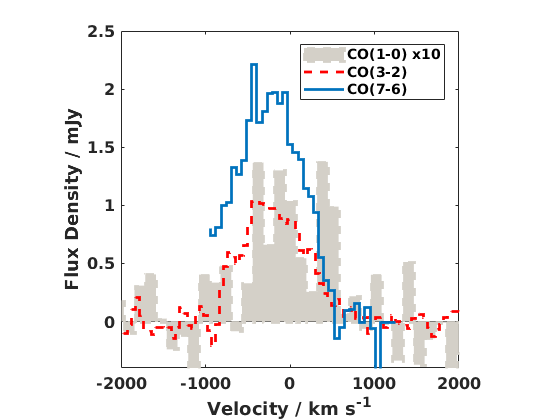} \\
\large{G1234N} \\
\includegraphics[scale=0.6]{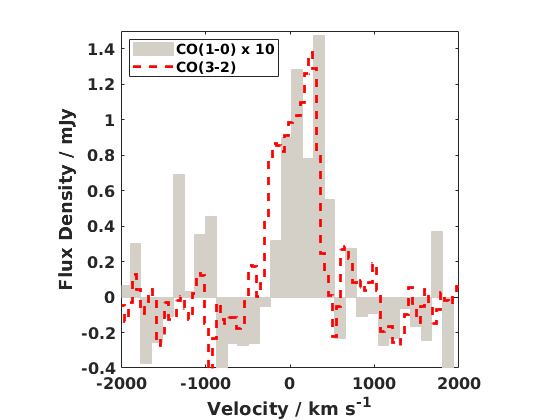} \\
\large{G1234S} \\
\includegraphics[scale=0.6]{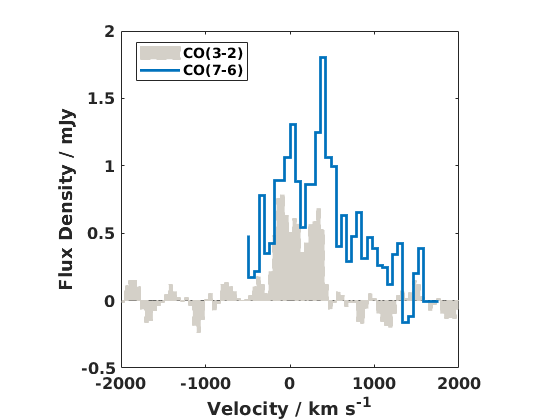} \\
\end{tabular}
\caption{Comparison of the CO line profiles in the the three galaxies in the J1234 system. The CO(1-0) line profiles have been scaled up by a factor of 10 for ease of visualisation.}
\label{fig:COvel_J1234}
\end{center}
\end{figure}

In Fig. \ref{fig:COvel_J1234} we compare the line profiles for the various different CO transitions presented here and in Paper I for all three galaxies in the J1234 system. Progressively higher J transitions of CO trace warmer, denser gas and a comparison of the line profiles therefore helps illustrate any potential differences in the kinematics of these different gas components. For the quasar J1234 and G1234N all the CO lines appear to be well matched in terms of their velocity and line shape suggesting that the cooler lower density, and warmer, denser gas are experiencing similar large scale motions in these systems. A double-peaked line profile is evident in the CO(7-6) emission from G1234S, which matches what is seen in the CO(3-2) emission. However for this particular galaxy we also see a very broad wing in the CO(7-6) emission extending out to velocities of $>$1000 km/s. Note - the blue side of the broad-wing is not traced by our observations as the CI(2-1) emission line affects this portion of the spectrum. Curiously, no such broad wings are seen in the CO(3-2) line profile of this galaxy. If the broad wing is ascribed to a molecular outflow, it appears to only be affecting the dense gas in the galaxy. Molecular outflows with similar observational properties to what is seen for G1234S, have previously been observed in luminous, high-redshift quasars \citep{Maiolino:12, Cicone:15}, where the AGN is responsible for driving the gas to velocities of $\sim$1000 kms$^{-1}$. Currently we have no evidence for an AGN being present in G1234S. However, no optical/infra-red counterpart to the galaxy exists in our imaging data of this field down to $i$-band magnitudes of $<25.1$ \citep{Banerji:14}, which suggests that the galaxy is highly obscured. We therefore speculate that G1234S with its broad wings seen in CO(7-6) emission and high excitation CO SLED (Paper I) may well host an obscured AGN although high-resolution, deep X-ray or mid infra-red data would be needed to confirm or refute this claim. It is also curious that only the dense gas is being affected by any putative outflow in this galaxy. Such an observation would be consistent with recent claims that star-formation can happen \textit{in-situ} within galactic outflows due to the dense outflowing gas cooling and fragmenting. Indeed recent observations of the prototypical local obscured quasar Mrk231 have shown that the fast outflowing gas in this galaxy shows an unusually high fraction of dense gas, which is conducive to forming stars \citep{Cicone:20}

\section{The J2315 System: A Major Merger}

\label{sec:J2315}

\subsection{Dynamical Mass \& Gas Fraction}

Our compact configuration ALMA observations for J2315 in B17 had already resolved the CO(3-2) emission in this quasar over physical scales of $>$20 kpc. In those observations we found that the velocity field could be satisfactorily modelled as a single (albeit very large) rotating gas disk. As can be seen in Fig. \ref{fig:CO_J2315}, the source breaks up into two distinct components in the higher spatial resolution observations presented in this paper, which clearly highlights that many high-redshift, massive galaxies that only have single component gas emission seen in low angular resolution observations, could well be merging systems. The two components are spatially separated by 15.0$\pm$0.5 kpc and very close together in velocity, with the northern companion galaxy blueshifted relative to the quasar host by only 170$\pm$20 km/s. We fit two 2-D Gaussians to the zeroth moment CO image shown in Fig. \ref{fig:CO_J2315}, finding major axis FWHM of (1.0$\pm$0.2)$\arcsec$ and (1.4$\pm$0.2)$\arcsec$ for the quasar host galaxy and the northern companion respectively, after deconvolving from the beam. This corresponds to half-light radii of $\sim$4-6 kpc at the redshifts of these galaxies, which is more than 2$\times$ the source sizes inferred for all three galaxies in the J1234 system based on the 2D Gaussian fitting (Table \ref{tab:dynmod}). 

\begin{figure*}
\begin{center}
\begin{tabular}{ccc}
\includegraphics[scale=0.5]{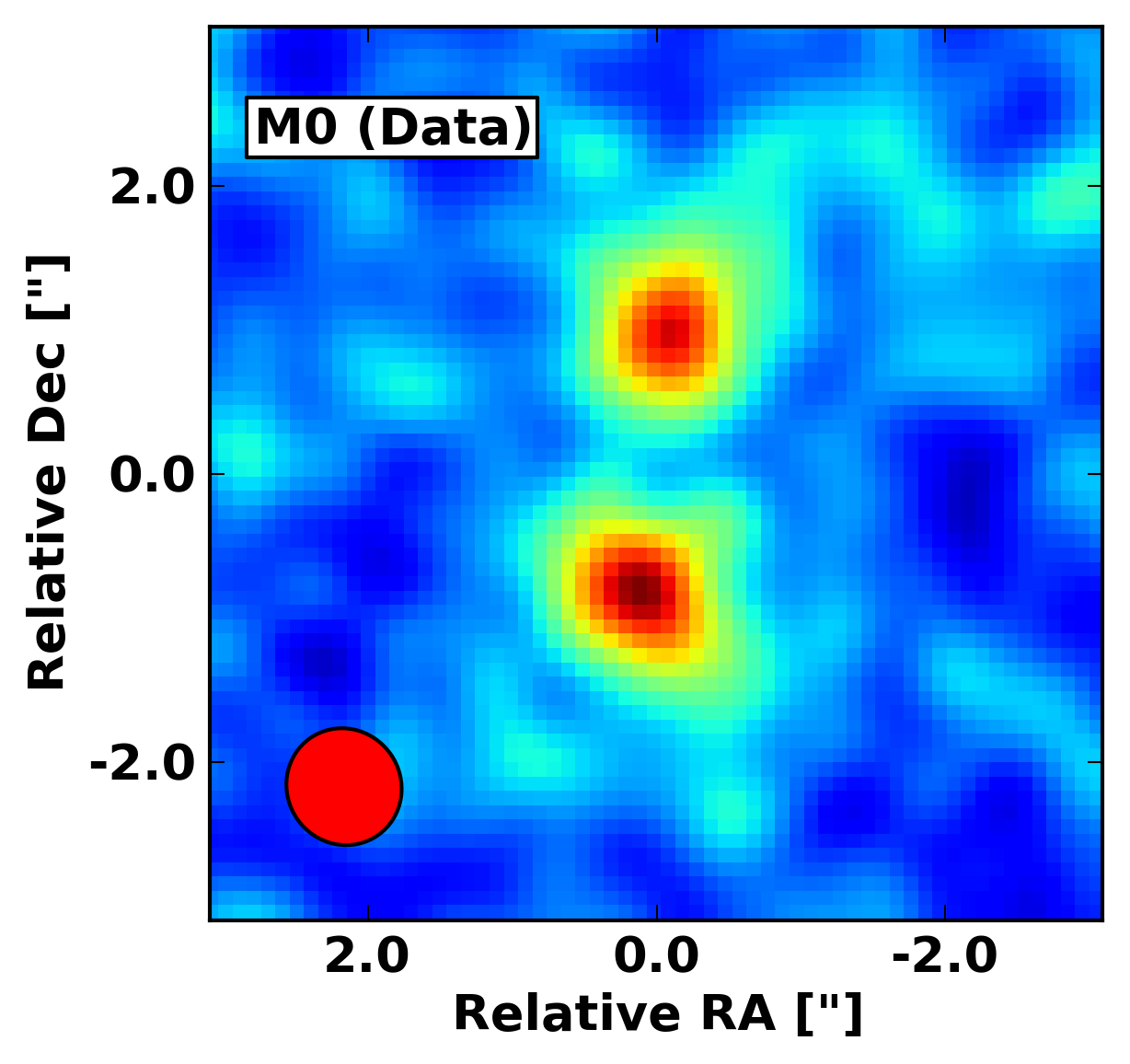} & \hspace{-0.5cm} \includegraphics[scale=0.5]{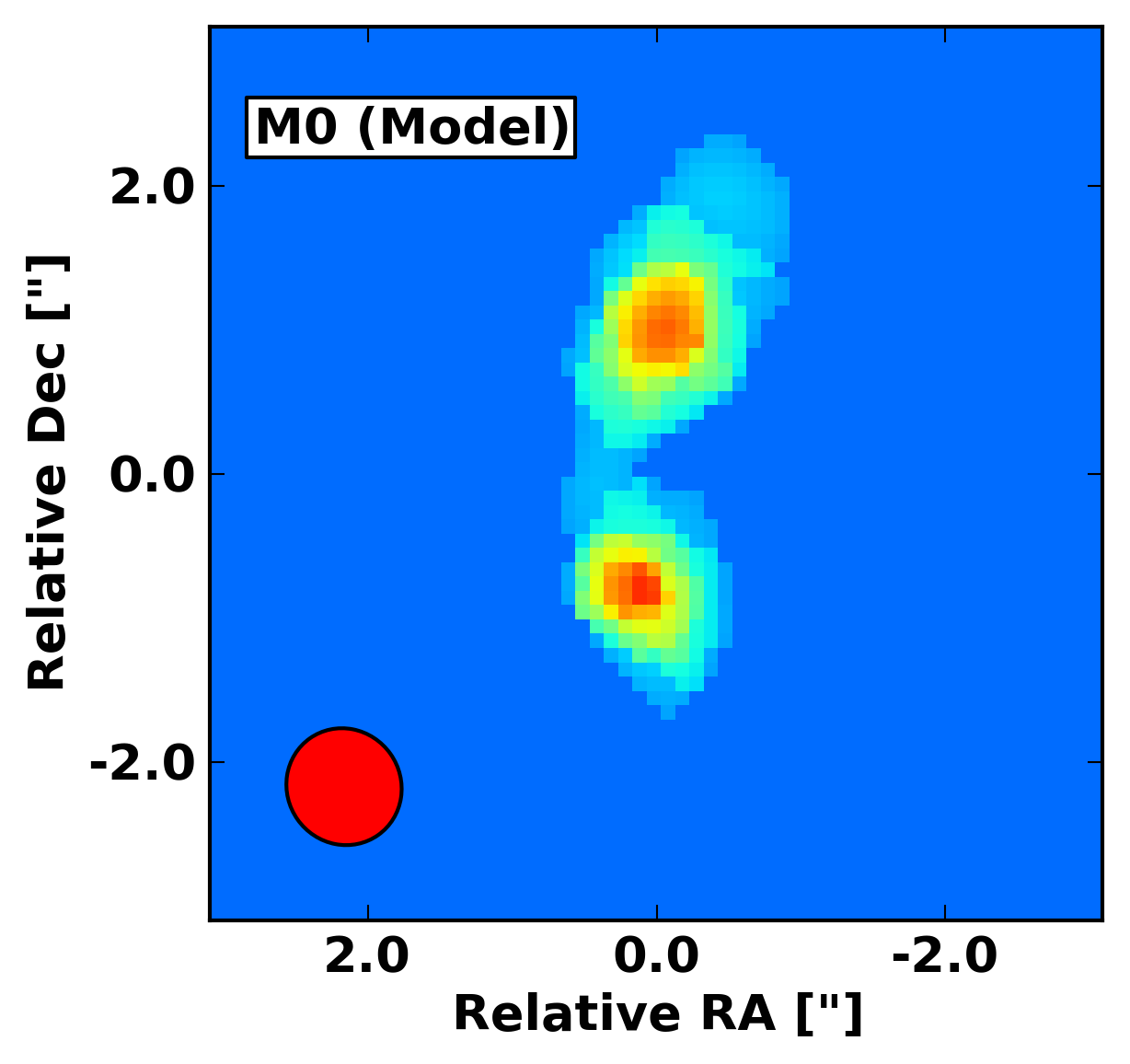} & \hspace{-0.5cm}
\includegraphics[scale=0.5]{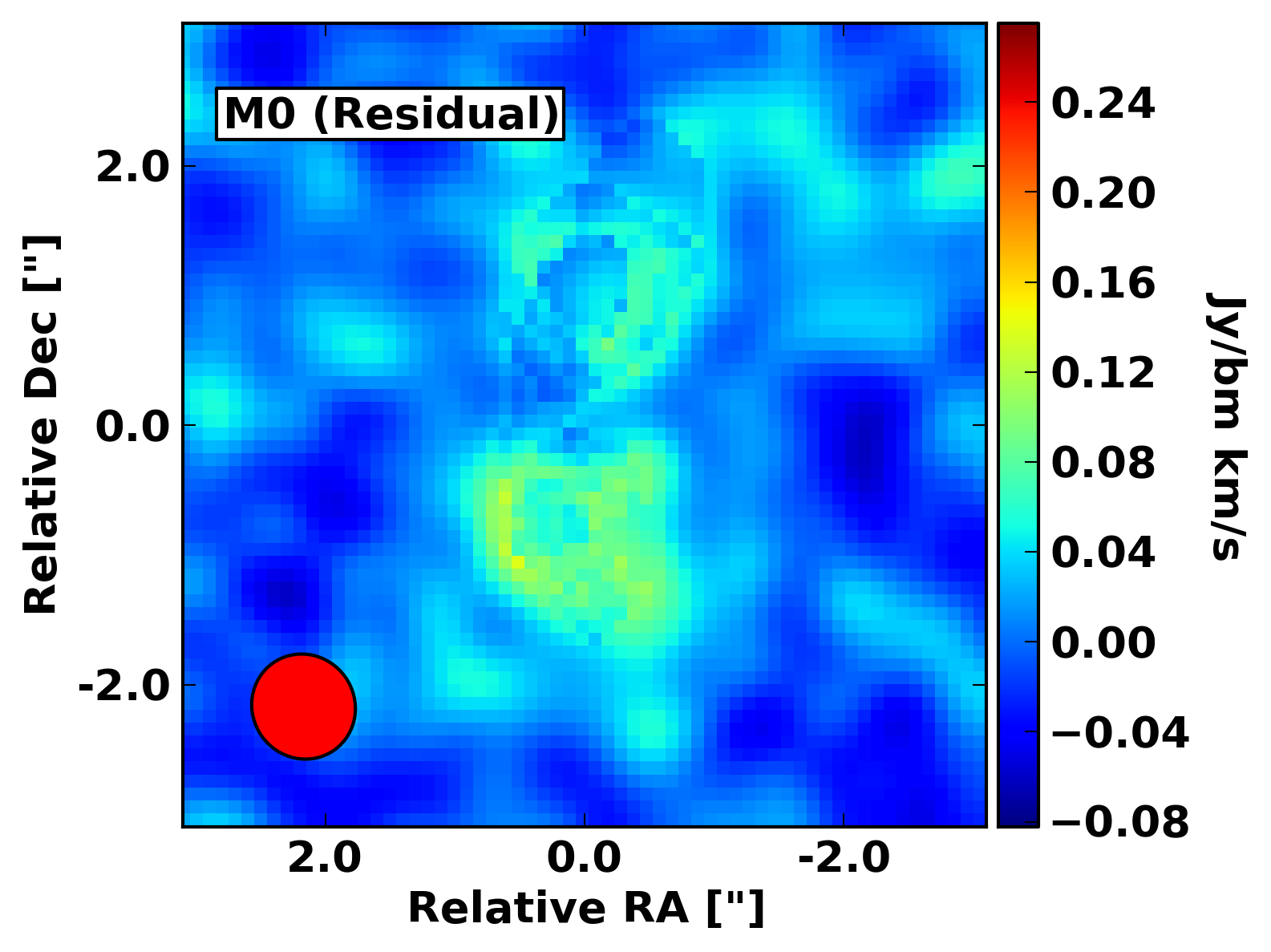} \\
\includegraphics[scale=0.5]{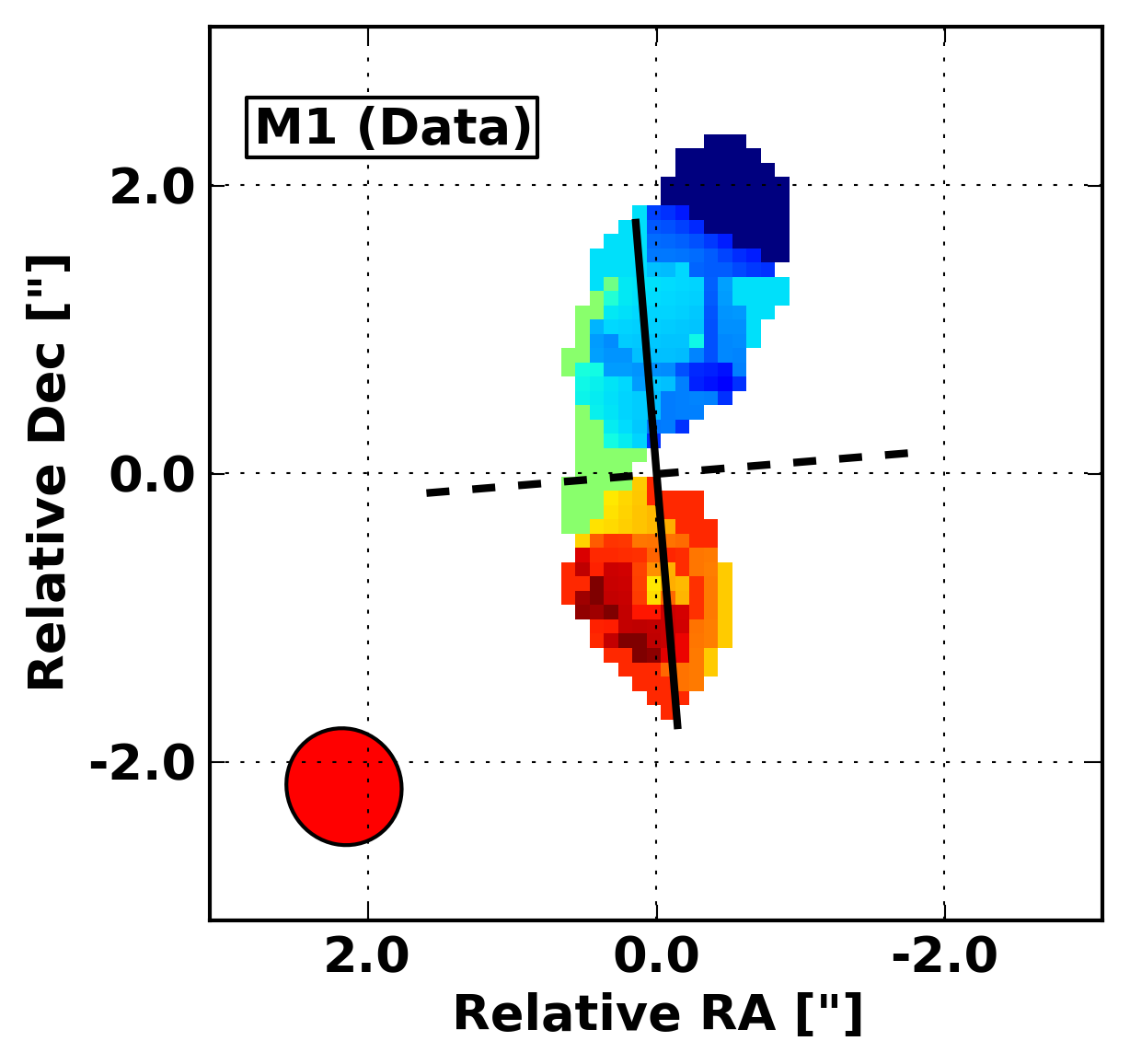} & \hspace{-0.5cm} \includegraphics[scale=0.5]{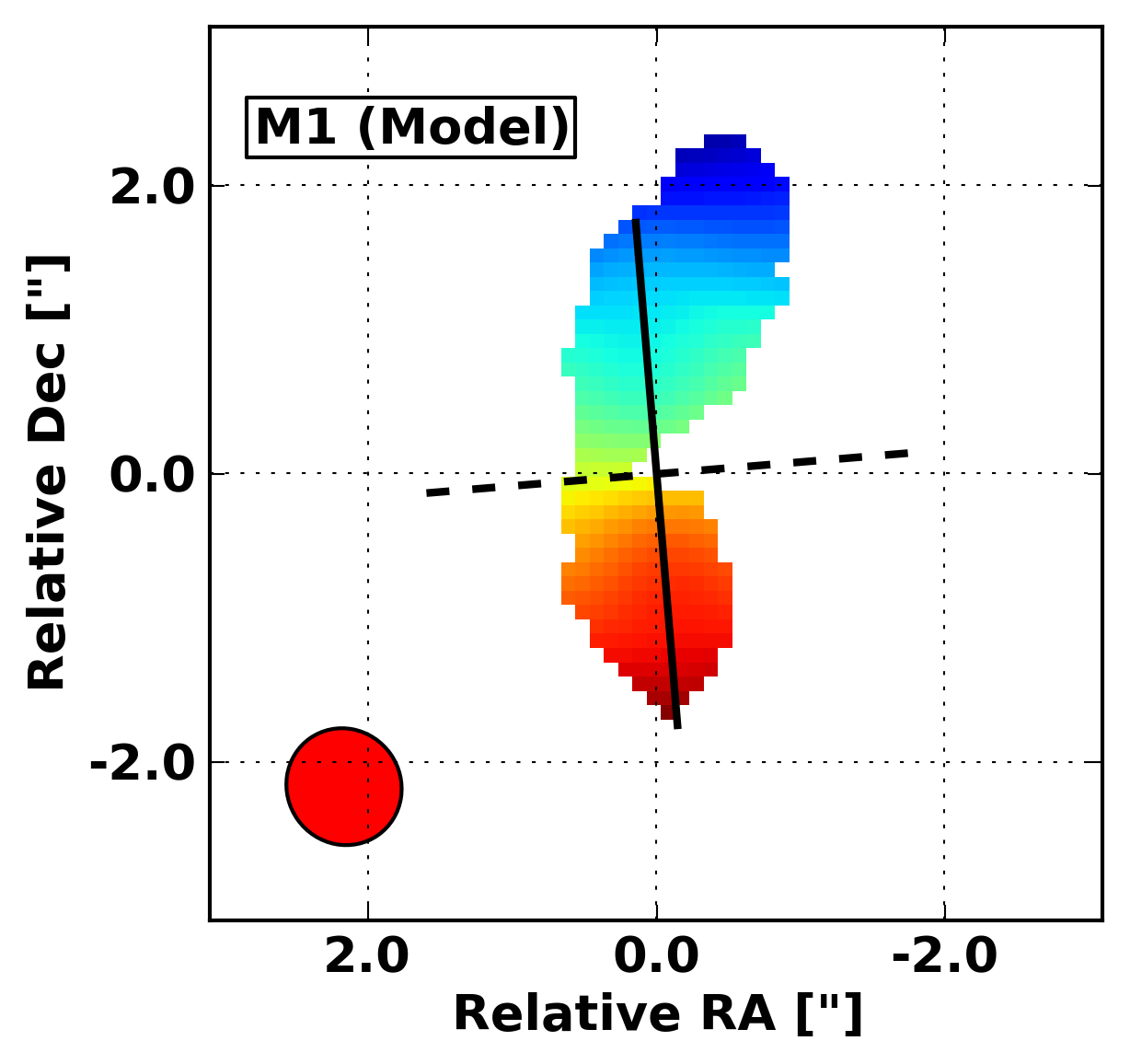} & \hspace{-0.5cm}
\includegraphics[scale=0.5]{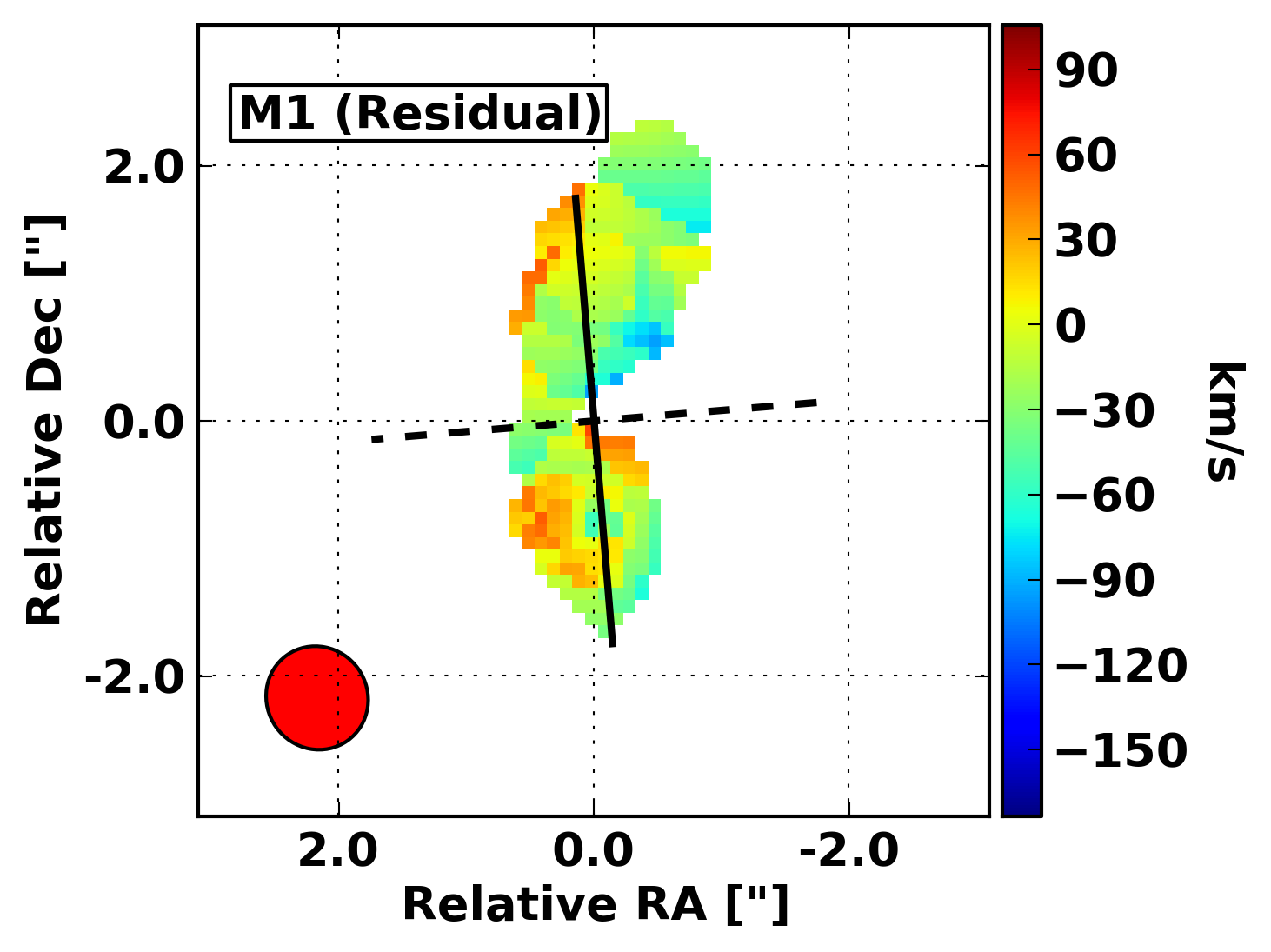} \\
\includegraphics[scale=0.5]{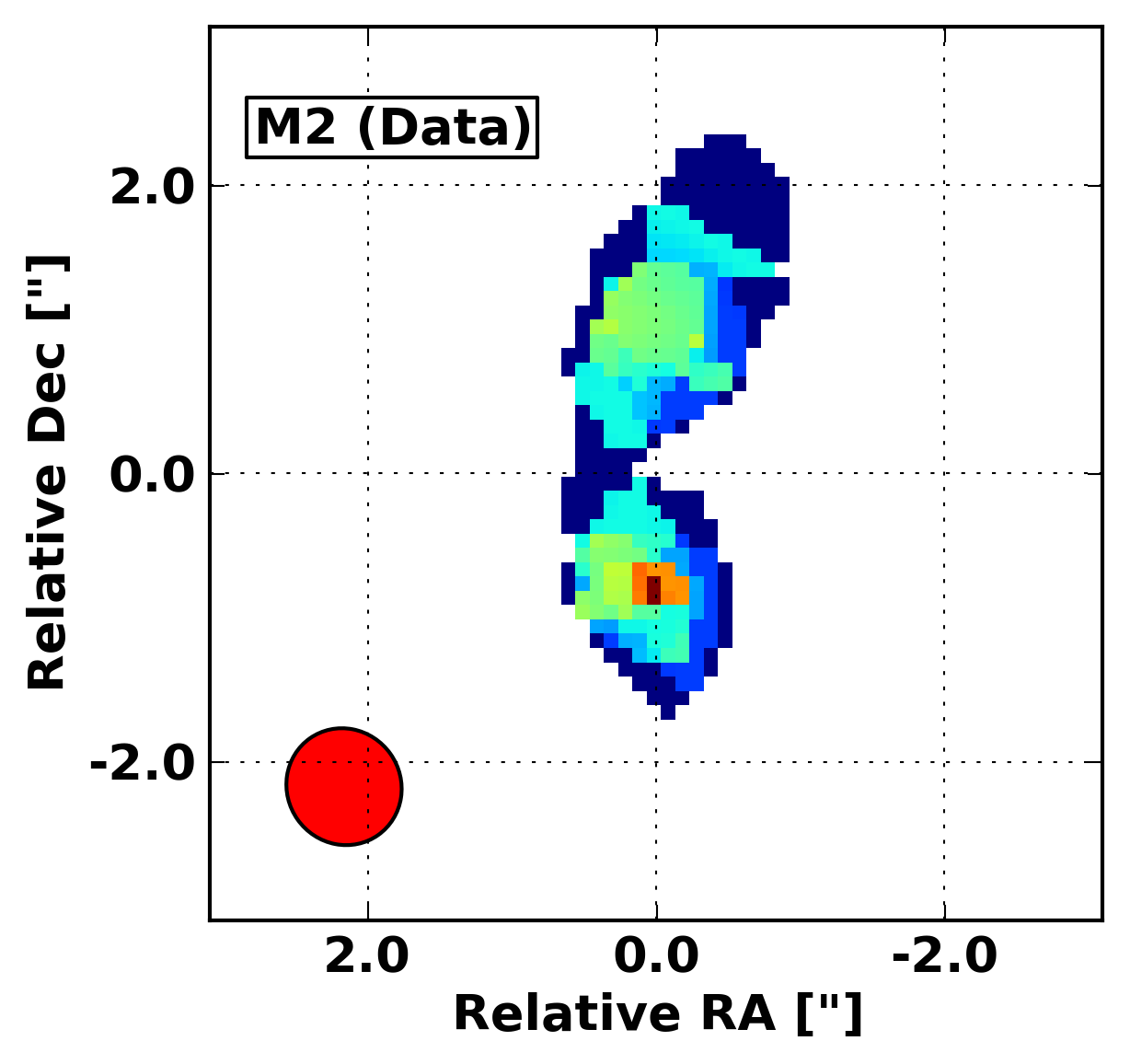} & \hspace{-0.5cm} \includegraphics[scale=0.5]{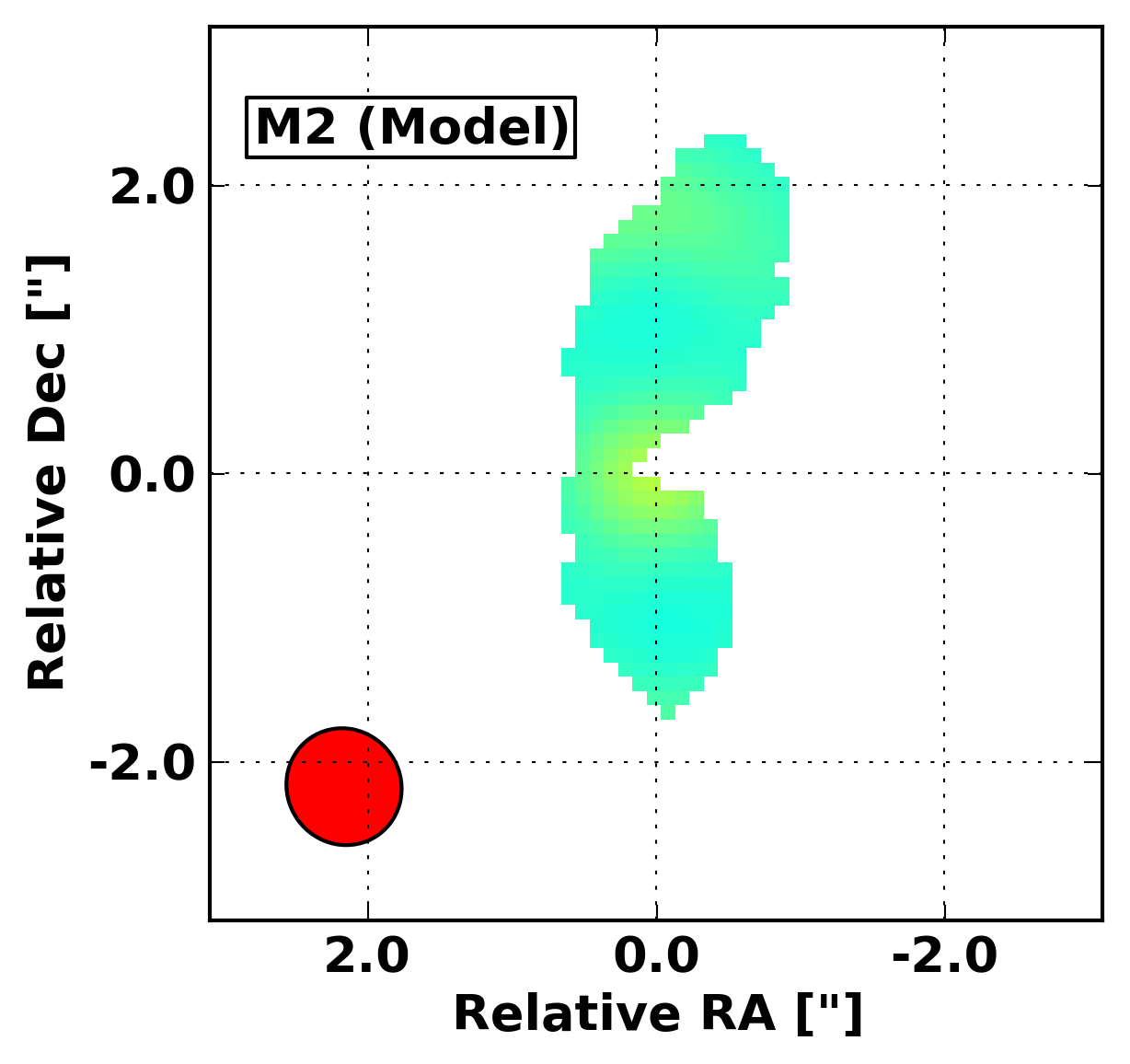} & \hspace{-0.5cm}
\includegraphics[scale=0.5]{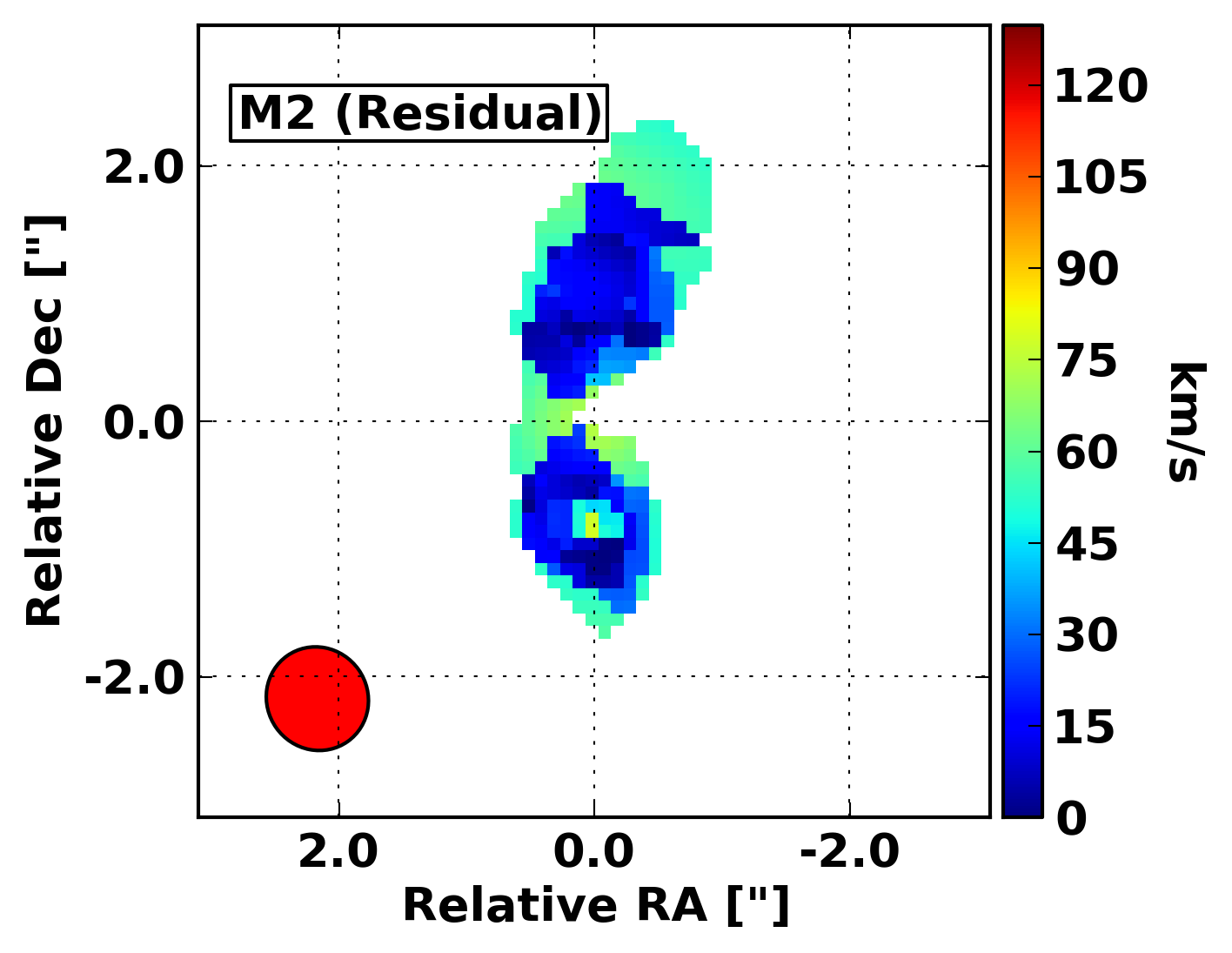} \\
\end{tabular}
\caption{Line intensity (top), velocity (middle) and line dispersion (bottom) maps for ULASJ2315. The source is clearly resolved into a major-merger of two gas-rich galaxies with comparable line intensities. The northern companion galaxy is blueshifted relative to the quasar host. The $^{\rm 3D}$Barolo best-fit dynamical models and residuals shown for completeness in the middle and right-most panels, do not provide a good description of these data as expected for a merging system.}
\label{fig:maps_J2315}
\end{center}
\end{figure*}

As was done for the galaxies in the J1234 system, we construct intensity, velocity and dispersion maps for both galaxies using the CASA task immoments. These maps can be seen in Fig. \ref{fig:maps_J2315}. The two galaxies of the merger are clearly resolved in the maps with comparable CO line intensities in each merging component. No velocity structure is evident in the individual galaxies at the current resolution, although we clearly see that the northern companion galaxy is blueshifted relative to the quasar host. 

For completeness, we also show the best-fit $^{\rm 3D}$Barolo models fit to the J2315 system as well as the fit residuals in the middle and right-most panels of Fig. \ref{fig:maps_J2315}. The fits are poor as expected as the system cannot be well modelled as a rotating disk and we do not consider the $^{\rm 3D}$Barolo fitting results in detail in the paper as we do not expect them to be robust. A simple estimate of the total mass of the system can instead be obtained using the virial theorem assuming a projected distance between the galaxies of 15 kpc and a velocity separation of 170 km/s. This leads to M$_{\rm{tot}}$=1.0$\times$10$^{11}$M$_\odot$ for both galaxies assuming that the galaxies are moving along the line-of-sight.


Alternatively, we can assume that both the galaxies in the J2315 system are individual rotating disks and use the FWHM of the CO line profiles shown in Fig. \ref{fig:CO_J2315} together with the rotating disk estimator \citep{Neri:03} to calculate dynamical masses for each galaxy. Inclination angles of 52 degrees and 61 degrees are estimated for the quasar host galaxy and the northern companion respectively by considering the inverse cosine of the minor-to-major axis ratio derived from the CO image. Correcting for these inclination values we find that the quasar host galaxy has M$_{\rm{dyn}}$=2.3$\times$10$^{10}$M$_\odot$ and the northern companion has M$_{\rm{dyn}}$=1.0$\times$10$^{10}$M$_\odot$, corresponding to a total mass of 3.3$\times$10$^{10}$M$_\odot$. Thus the two galaxies appear to be undergoing a major merger with a $\sim$1:2 mass ratio. 

In Paper I we used various different estimators to derive gas masses for this system. The total gas mass is $\sim$3-4$\times$10$^{10}$M$_\odot$. Combining this with the dynamical mass estimates above and estimating the stellar mass as M$_\ast$=M$_{\rm{dyn}}-$M$_{\rm{gas}}-$M$_{\rm{BH}}$ as was done in the case of J1234, we find very high gas fractions of $>$45\%. The lower limit assumes that the merging galaxies are moving along the line of sight and the gas fraction could be substantially higher if the inclination angle is sufficiently different from face-on.

\subsection{CO Line Profiles}

\label{sec:COvel_2315}

\begin{figure}
\begin{center}
\includegraphics[scale=0.6]{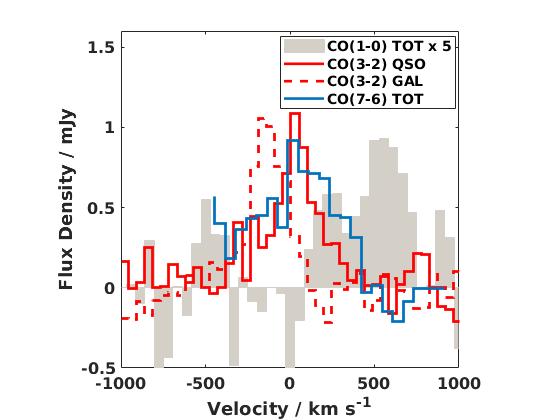} 
\caption{Comparison of the CO line profiles in J2315. The CO(1-0) and CO(7-6) line profiles are derived over the entire merging system as the individual components of the merger are not resolved in these data. Separate CO(3-2) line profiles for the QSO host and the northern companion galaxy are however shown. The CO(1-0) line profile has been scaled up by a factor of 5 for ease of visualisation.}
\label{fig:COvel_J2315}
\end{center}
\end{figure}

In Fig. \ref{fig:COvel_J2315} we compare the CO line profiles for the various different CO rotational transitions detected in the J2315 galaxies. The CO(1-0) and CO(7-6) profiles are integrated over the entire merging system as the individual components of the merger are not resolved in these data. We do however show separate CO(3-2) profiles extracted for the quasar host galaxy and the northern companion galaxy. These profiles are considerably less well matched compared to the J1234 quasar host galaxy (Fig. \ref{fig:COvel_J1234}) and the kinematics of the gas in this major-merger system is clearly messier as already indicated by the poor $^{\rm 3D}$Barolo fits. The CO(7-6) emission is significantly broader than the CO(3-2) emission with a FWHM of twice that of the CO(3-2) line (see Table 3 in Paper I). However, as can be seen in Fig. \ref{fig:COvel_J2315}, the broadening almost certainly arises because there isn't sufficient angular resolution in the ALMA Band 6 data to disentangle the companion galaxy from the quasar host. The separate CO(3-2) line profiles extracted for the quasar and the companion galaxy, straddle the broad CO(7-6) emission seen in the total system. As already noted in Paper I, the CO(1-0) emission is significantly redshifted by $\sim$ 500 kms$^{-1}$ relative to the other CO lines. Thus we have clear evidence for different gas dynamics affecting the cold/diffuse and warm/dense gas components in this major merger. In Paper I we also saw tentative evidence for the CO(1-0) emission originating from closer to the northern companion galaxy rather than the quasar, which could explain the velocity differences observed. However, it is interesting to note that the CO(3-2) emission in the companion galaxy is actually blueshifted rather than redshifted relative to the CO(3-2) in the quasar host. Higher angular resolution observations of all the CO lines are necessary to be able to locate co-spatial components of the gas emission and therefore disentangle the origin of the complicated velocity shifts observed in Fig. \ref{fig:COvel_J2315}. 

\section{MULTI-WAVELENGTH PROPERTIES OF THE HRQS}

\label{sec:merger}

In this Section we briefly review existing observational constraints on the multi-wavelength properties of the two HRQs and combine the information provided by these multi-wavelength data with the new constraints on the properties of the quasars from this work, in order to place the quasars and their host galaxies in context within an overall picture of galaxy black-hole co-evolution. Both sources were selected as red, infra-red luminous quasar candidates with broad emission lines present in the near infra-red spectra \citep{Banerji:12, Banerji:15}. X-ray observations \citep{Lansbury:20} place constraints on the accretion properties of the black hole and the line-of-sight gas column densities. Both quasars have similarly high X-ray luminosities - log$_{10}$(L$_{2-10 \rm{keV}}$)$>45$ - and Eddington ratios of $>$0.1, indicating that we are seeing the black holes during a phase of rapid accretion. The gas column densities are also significant - $\sim$10$^{22}$cm$^{-2}$, consistent with the quasars being in an obscured phase. In terms of the properties of the central engine, both HRQs are therefore very similar. 

\begin{figure*}
\begin{center}
\begin{tabular}{cc}
\hspace{-2cm}
\includegraphics[scale=0.7]{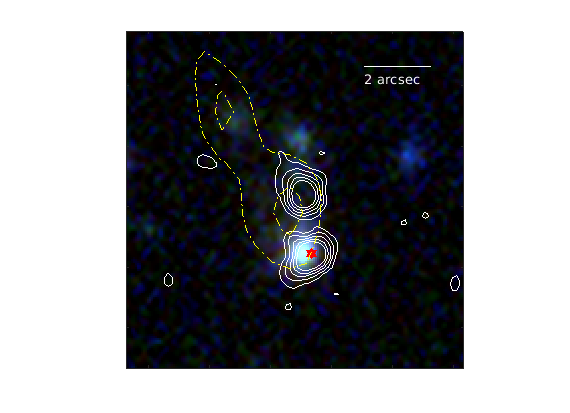} & \hspace{-1.5cm} \includegraphics[scale=0.7]{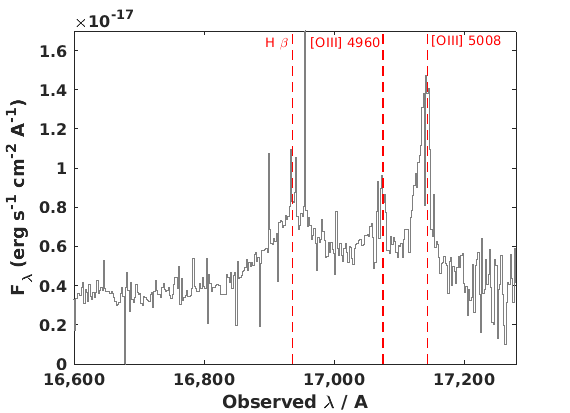}
\end{tabular}
\caption{\textit{Left:} A 10$\times$10 arcsec multi-colour $grz$ band composite image of the J2315 system from HSC Data Release 2. North is up and East is to the right. The white contours denote the CO(3-2) emission from this paper and the yellow dot-dashed contours denote the JVLA CO(1-0) emission from Paper I. The red star denotes the position of the quasar from near infra-red imaging. Morphologically ``fuzzy" blue emission is visible both at the location of the quasar and extending to the North co-incident with the location of one of the CO(1-0) peaks. \textit{Right:} A portion of the XShooter H-band spectrum (paper in preparation) centred on the position of the quasar marked with the red star on the left. The red vertical lines mark the expected positions of H$\beta$ and the [OIII] doublet based on the ALMA redshift from this paper. The [OIII] 5008.239 emission is clearly blueshifted relative to systemic signifying the presence of AGN-driven outflows.}
\label{fig:J2315_multiwave}
\end{center}
\end{figure*}

Deep optical photometry from current wide-field imaging surveys allows us to trace the rest-frame ultraviolet emission from the HRQs. As the quasar is highly obscured, in the majority of HRQs the rest-frame UV emission likely arises from unobscured star-forming regions in the quasar host galaxy \citep{Wethers:18}. J1234 and its companion galaxies are undetected in the $i$-band down to a 5$\sigma$ magnitude limit of $\sim$25.1 \citep{Banerji:14}. Far infra-red to millimeter observations \citep{Banerji:14, Banerji:18} however suggest significant levels of obscured star formation in the host galaxy. J2315 is detected in data from the Dark Energy Survey with a measured $i$-band magnitude of 22.5 \citep{Wethers:18}. The source is morphologically extended in the Dark Energy Survey imaging confirming that the UV emission arises from unobscured regions within the quasar host galaxy rather than the quasar itself. The UV-derived star formation rate for the galaxy is estimated to be $\sim$100-200 M$_\odot$yr$^{-1}$ \citep{Wethers:18}. Recent, deeper and higher spatial resolution observations from the HyperSuprimeCam Survey Data Release 2 \citep{Aihara:19} clearly shows resolved emission spatially co-incident with the location of the quasar (Fig. \ref{fig:J2315_multiwave}; left panel) as well as more extended emission in the direction of the northern companion galaxy.  The rest-frame UV observations of the quasar host galaxies therefore indicate very different levels of obscuration towards the star-forming regions despite similar levels of obscuration towards the black holes, with the host galaxy of J1234 being completely obscured in the UV whereas J2315 shows evidence for unobscured star-forming regions, albeit the main bulk of star formation activity in this galaxy is also likely obscured \citep{Banerji:18, Wethers:20}. 

In Paper I we presented constraints on the interstellar medium properties of the two HRQs using detections of multiple molecular emission lines of CO and CI. The excitation conditions in J1234 are consistent with those seen in other luminous AGN with the CO spectral line energy distribution (SLED) peaking at high J values. Photo-dissociation region modelling suggests high radiation field strengths in the interstellar medium of the galaxy. J2315 on the other hand appears to be dominated by lower excitation gas. The CO SLED peaks around J$_{\rm{upper}}$=3 and the radiation field strength is $\sim$1.5-2 dex lower than that inferred for J1234. The ISM conditions in J2315 are therefore more reminiscent of star-forming galaxies rather than AGN. 

\citet{Temple:19} study narrow-line region outflows in HRQs and find that [OIII] is undetected in J1234, which, even after correcting for the significant levels of obscuration in the galaxy, suggests that the narrow-line region has either been over-ionised or physically removed from the galaxy. This is also the case for $\sim$10 per cent of very high-luminosity optically selected unobscured quasars \citep{Coatman:19} and consistent with AGN feedback having had a significant effect on the narrow line region gas. In the case of J2315, broad blue wings are seen in the [OIII] emission line profile extending to velocities of $\sim$2000 km s$^{-1}$, providing direct evidence for high-velocity outflowing gas being powered by the luminous AGN. The [OIII] and H$\beta$ region from a recent XShooter spectrum of J2315 are shown in the right-hand panel of Fig. \ref{fig:J2315_multiwave} (paper in preparation). 

While we should be cautious about drawing conclusions from a sample of only two quasars studied here in detail, the comprehensive multi-wavelength observations that have been assembled for the two HRQs paint a consistent picture of them being seen in different phases during their evolution. J2315 is conceivably an earlier phase, where two galaxies are viewed in the process of undergoing a major merger. While the quasar itself and a significant amount of the ongoing star formation is obscured by dust, regions of unobscured star formation remain visible in the rest-frame UV with their location broadly consistent with the extended reservoir of molecular gas. Evidence for quasar-driven outflows is provided by the broad blue wings seen in the [OIII] emission line profiles. Nevertheless, the molecular gas fraction is high and the excitation conditions of the interstellar medium are more consistent with those of star-forming galaxies rather than luminous quasars, which might suggest that the effects of AGN feedback are yet to have a dramatic impact on the properties of the host galaxy. 

J1234 can putatively be assumed to be seen at a more evolved stage. The quasar together with both companion galaxies are completely obscured at rest-frame UV and optical wavelengths. A significant starburst appears to be occurring in the quasar host galaxy based on the ALMA dust continuum detections (Paper I), which may or may not have been triggered by a merger. While the velocity fields of the CO(3-2) molecular gas observed in this work are consistent with rotational dynamics, the relatively modest resolution of the observations do allow the possibility of a merging system seen close to or soon after final coalescence when the gas is perhaps settling into a rotation pattern. Unlike in the case of J2315, the star formation is entirely obscured in the rest-frame UV. No [OIII] emission is detected in the galaxy, suggesting that powerful AGN driven winds have already over-ionised or cleared the galaxy of its [OIII] gas. Further evidence for efficient feedback is provided by the low molecular gas fraction observed in this system. The ISM properties are very similar to those observed in other high-luminosity quasars with highly excited warm and dense phases of the molecular gas being dominant. 

\section{The M$_{\rm{BH}}$-M$_{\rm{gal}}$ relation}

\label{sec:mbh_mgal}

Our spatially resolved observations of the HRQ host galaxies have allowed us to estimate their dynamical masses. Such dynamical mass estimates allow luminous quasars to be placed on the M$_{\rm{BH}}$-M$_{\rm{gal}}$ relation \citep{Willott:15, Willott:17, Shao:17}. The majority of optically selected quasars with M$_{\rm{BH}}\gtrsim10^9$M$_\odot$ have black holes that are significantly over-massive relative to their host galaxies, and compared to the local relation \citep{Kormendy:13}. While this is largely a selection effect of optical surveys targeting the most massive black holes at the rapidly declining end of the mass function (e.g. \citealt{Lauer:07}), it is also symptomatic of a much larger scatter in the M$_{\rm{BH}}$-M$_{\rm{gal}}$ relation at high-redshifts relative to the local Universe. On the other hand, the population of SCUBA selected submillimeter galaxies at $z\sim2$ were inferred by \citet{Alexander:08} to have black holes that are under-massive relative to the local relation. Dust-obscured quasars are a plausible transition population between the starburst and optical quasar phases. The emergence of wide area surveys at mid and far-infrared wavelengths such as the \textit{WISE} All-Sky Survey and the \textit{Herschel} Extragalactic Surveys in the last decade, has allowed a wider population of obscured galaxies and AGN to be assembled, with higher dust temperatures than the initial SCUBA detected galaxies. At least some of these also have black holes that are over-massive relative to their hosts (e.g. \citealt{Assef:15, Matsuoka:18}) although the number of obscured galaxies with both host and black hole mass measurements remains limited. The location of our heavily reddened quasars on the M$_{\rm{BH}}$-M$_{\rm{gal}}$ plane could therefore place interesting constraints on the relative timescales for black-hole and galaxy growth as these massive galaxies are being assembled.

Before turning to the M$_{\rm{BH}}$-M$_{\rm{gal}}$ ratio in our quasars we first re-evaluate the black-hole mass estimates for our two quasars that were presented in \citet{Banerji:12} and \citet{Banerji:15}. As noted in those papers, there are large systematic uncertainties associated with the black-hole mass derivations, not least because the use of a continuum luminosity at 5100\AA\@ in the black-hole mass estimate requires significant dust-corrections given the typical extinction levels in the HRQs. These corrections are particularly large for the ALMA sample, which correspond to the reddest quasars in the population. Recent analysis of deep optical photometric data for these quasars also suggests that there could be a significant contribution from the host galaxy to the rest-frame 5100\AA\@ continuum fluxes for the reddest quasars \citep{Wethers:18}. We therefore take this opportunity to derive new, more conservative estimates of the black-hole masses for the two quasars in this paper. In particular for ULASJ2315 we utilise a higher signal-to-noise, more red-sensitive XShooter spectrum (paper in prep) compared to the SINFONI spectrum presented in \citet{Banerji:15} which allows a more robust estimate of the H$\alpha$ linewidth. We also make use of X-ray observations of both our quasars from \citet{Banerji:14} and \citet{Lansbury:20} to derive hard X-ray luminosities, which suffer significantly less from the effects of extinction compared to the optical continuum. These X-ray luminosities are L$_{\rm{2-10 keV}}=1.1\times10^{45}$ erg/s and $3.0\times10^{45}$ erg/s for J1234 and J2315 respectively. We then use the calibration in \citet{Bongiorno:14} to derive the new black-hole masses:

\begin{equation}
\begin{split}
\rm{log}_{10}(M_{\rm{BH}}) &=7.11 + 2.06 \rm{log}_{10} \left( \frac{\rm{FWHM}_{\rm{H}\alpha}}{1000 \rm{km s}^{-1}}\right) \\
& + 0.693 \rm{log}_{10} \left( \frac{\rm{L}_{\rm{2-10 keV}}}{10^{44} \rm{erg s}^{-1}}\right)
\end{split}
\label{eq:MBH}
\end{equation}

\begin{figure}
\begin{center}
\includegraphics[scale=0.6]{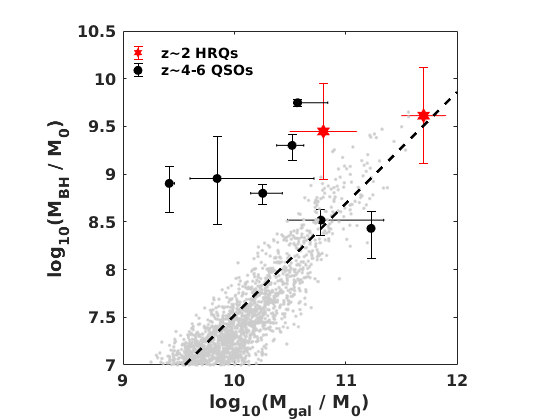} 
\caption{Black hole mass versus galaxy mass. In the case of observational data, the galaxy masses all refer to dynamical masses inferred using resolved observations of molecular emission lines arising from the ISM in the quasar host galaxy. In the case of our HRQs the uncertainties on the black hole masses are shown to be 0.5 dex, which encodes the level of systematic uncertainty expected in these masses due to various factors. The resolved dynamical mass estimates for the z$\sim$4-6 quasars are taken from \citet{Pensabene:20} and the black hole masses come from \citet{DeRosa:11, DeRosa:14, Trakhtenbrot:11, Venemans:15}. The grey points show the total baryonic mass contained within 2$\times$ the stellar half mass radius for Illustris galaxies \citep{Sijacki:15} at $z\sim2$. The solid line denotes the \citet{Kormendy:13} relation.}
\label{fig:mbh_mdyn}
\end{center}
\end{figure}

\noindent The resulting black-hole masses for J1234 and J2315 are 4.1$\times$10$^9$M$_\odot$ and 2.8$\times$10$^9$M$_\odot$ respectively, which can be considered to be conservative estimates and are lower than the original black-hole masses presented in \citet{Banerji:12} and \citet{Banerji:15} for these quasars, although still within the $\sim$0.5 dex systematic uncertainty expected in these black hole mass measurements \citep{Shen:11}. Using these new black-hole masses we show the location of our two $z\sim2.5$ dusty quasars on the M$_{\rm{BH}}$-M$_{\rm{gal}}$ relation in Fig. \ref{fig:mbh_mdyn} and compare to a higher redshift quasar population where host galaxy dynamical masses have been estimated in a similar way using resolved ALMA observations \citep{Pensabene:20}. We use the dynamical masses inferred from the exponential disk modelling as a proxy for the galaxy mass, as this is more representative of the total mass of the galaxy rather than the mass of the CO bright region (See Table \ref{tab:dynmod}). As a reminder, the exponential disk modelling evaluates the mass enclosed within a radius of 10 kpc by assuming a surface brightness profile that extends beyond the CO-bright regions traced by the observations. The \citet{Pensabene:20} high-z sample, which we use as our comparison sample has had galaxy dynamical masses inferred in exactly the same way as our sample - i.e. using an exponential disk model to model the dynamics of the spatially resolved [CII] gas in these quasars and evaluating the  mass enclosed within a radius of 10 kpc. While inferences regarding the locations of different populations of quasars on the M$_{\rm{BH}}$-M$_{\rm{gal}}$ plane are complicated by the large systematic uncertainties on single-epoch black-hole mass measurements as well as galaxy mass estimates, conclusions from the comparison between our sample and the z$\sim4-6$ sample should nevertheless be robust given that both black-hole masses and galaxy masses are evaluated in a very similar way for both samples and make the same underlying assumptions. Our dusty quasars are similar to the luminous $z\sim4-6$ quasars in that their black-hole masses put them on or above the local relation Thus, even if these sources are seen during a transition phase, their black-holes appear to already be fully grown. 

In Fig. \ref{fig:mbh_mdyn} we also show the total baryonic mass contained within 2$\times$ the stellar half mass radius, versus black hole mass for $z\sim2$ galaxies in the Illustris simulations \citep{Sijacki:15}, which is described in more detail in the next Section. It is interesting to note that J1234, which is the more ``evolved'' of our two systems appears to be broadly consistent with an extrapolation of the Illustris galaxies to the highest masses, whereas J2315, the major merger, lies further away from the relation suggesting some further evolution of the total galaxy mass is necessary in this system in order to bring it on to the M$_{\rm{BH}}$-M$_{\rm{gal}}$ relation predicted at these redshifts.

\subsection{Comparison to Illustris}

It is interesting to use the outputs of the Illustris simulations \citep{Vogelsberger:2014, Nelson:2015, Sijacki:15} to consider both the progenitors of our $z\sim2.5$ quasars as well as the kinds of systems they would evolve into at the present day. The Illustris-1 simulations used here simulate a total comoving volume of (106.5 cMpc)$^3$ and self-consistently evolve five different types of resolution elements from $z=127$ to the present day. The public data release paper \citep{Nelson:2015} contains more details regarding the simulation outputs that we make use of. At a $z\sim2.5$, corresponding to the redshift of our quasars, the simulated volume contains 22079 black holes, 345 of which have M$_{\rm{BH}}>10^8$M$_\odot$ and  29 of which have M$_{\rm{BH}}>10^9$M$_\odot$. While the simulation volume is not large enough to contain $>10^8$M$_\odot$ black holes at the highest redshifts of $z\sim6$, the simulations can nevertheless be used to track back the $z \sim 2.5$ population to understand the progenitors of these massive black holes at higher redshifts.  

\begin{figure*}
\begin{center}
\begin{tabular}{cc}
\includegraphics[width=8.5cm]{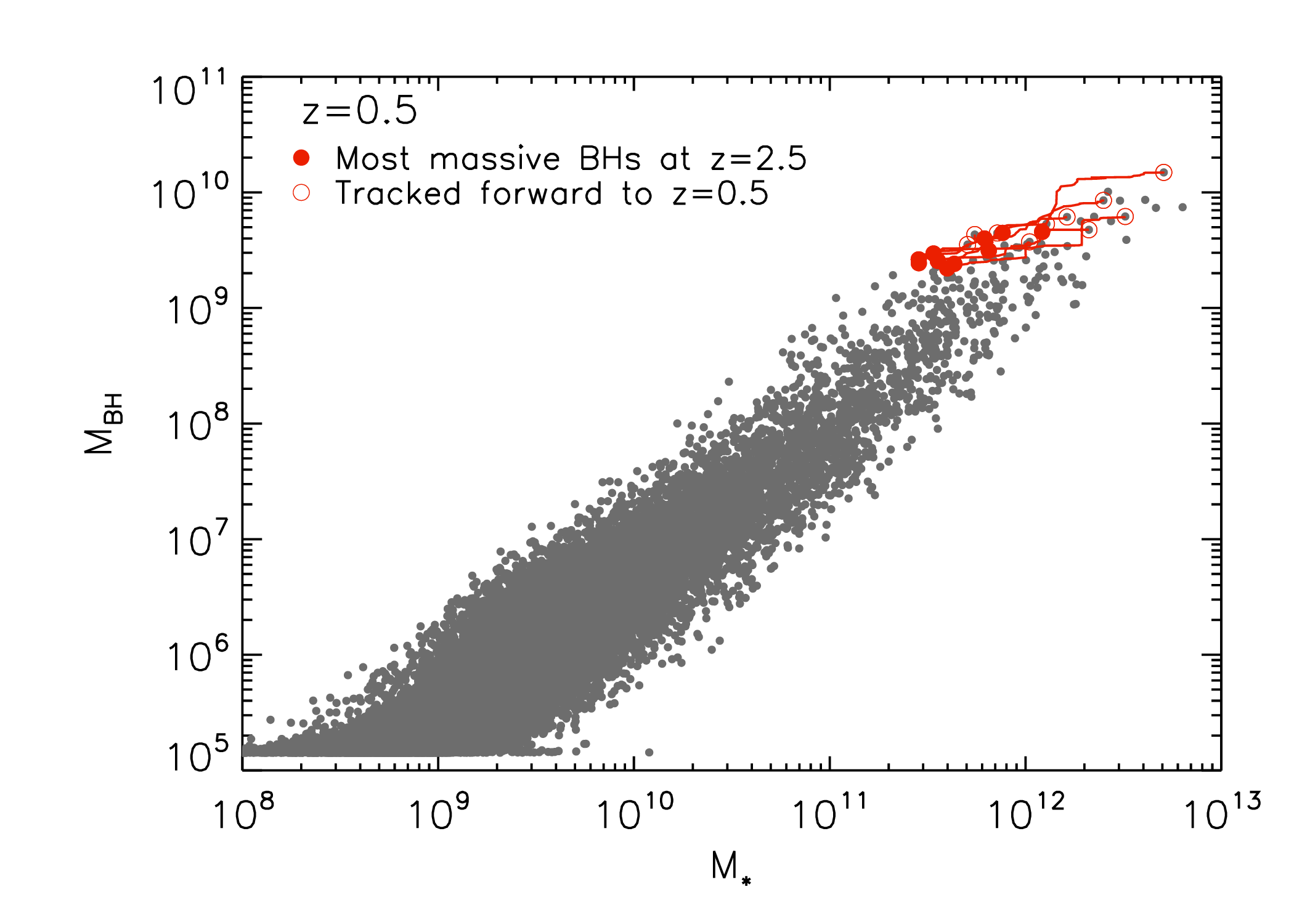} & \hspace{-0.5cm} \includegraphics[width=8.5cm]{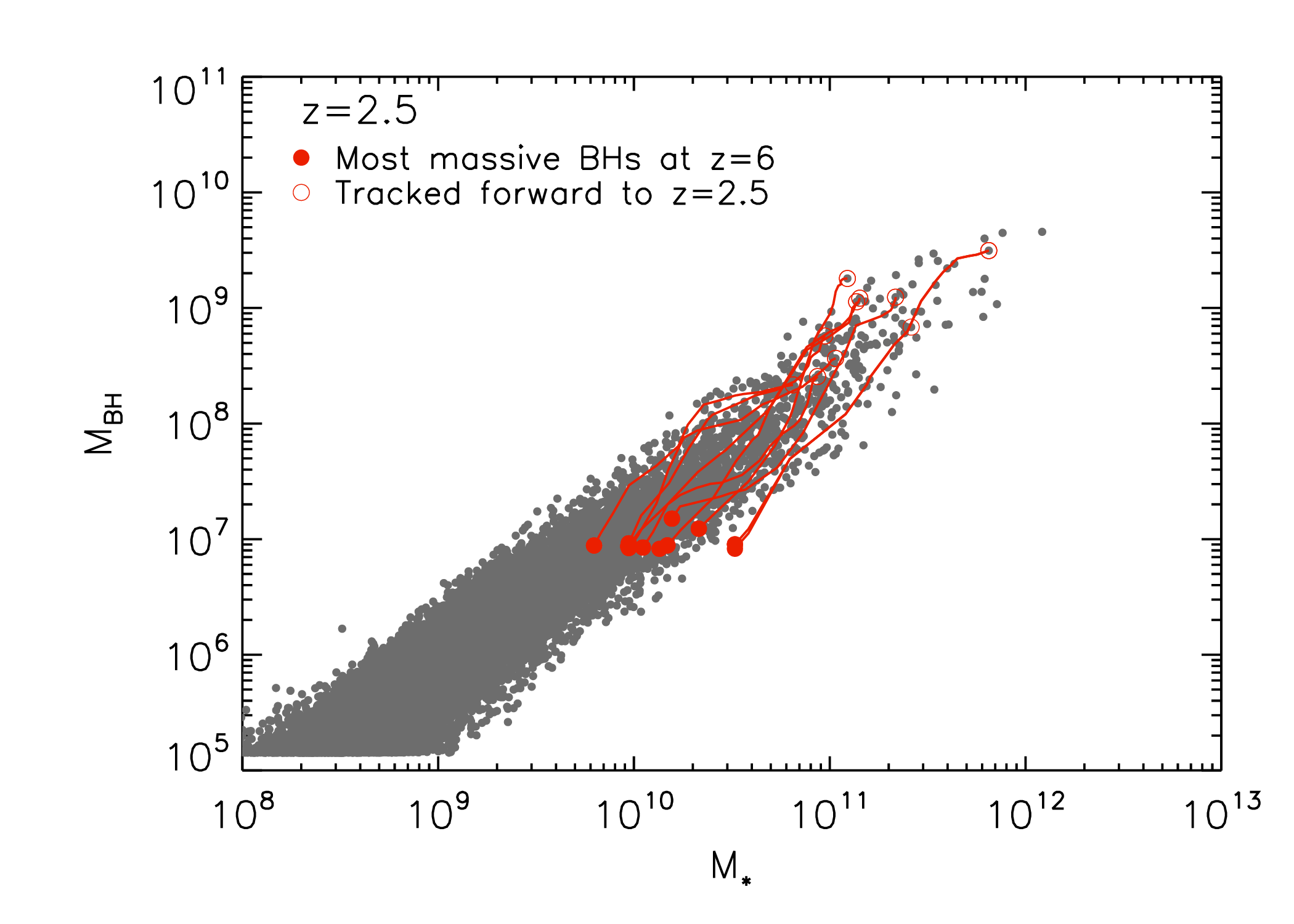} \\
\end{tabular}
\caption{The red points and lines track the evolution of galaxy and black hole mass in the Illustris simulation for the most massive black holes at $z=2.5$ tracked forward to $z=0.5$ (left) and for the most massive black holes at $z=6$ tracked forward to $z=2.5$ (right). The background set of grey points trace the $z=0.5$ (left) and $z=2.5$ (right) galaxy populations in Illustris.}
\label{fig:Illustris}
\end{center}
\end{figure*}

\citet{DeGraf:15} and \citet{Ricarte:17} both found that the most massive black holes with M$_{\rm{BH}} \sim 10^9$ M$_\odot$ assembled a larger fraction of their black hole mass at earlier times. In the left panel of Fig. \ref{fig:Illustris} we show the evolution of the most massive black holes in Illustris at $z\sim2.5$ tracked forward to $z\sim0.5$. This confirms that the bulk of the black hole mass is already assembled at $z\sim2.5$ and the evolution happens in a mostly horizontal direction from this point onwards with the galaxy stellar mass growing significantly more than the black hole mass. To quantify the direction of the evolution, we define the mean slope, $\alpha$ of the evolutionary tracks in the M$_{\rm{BH}}$-M$_{\rm{gal}}$ relation as: 

\begin{equation}
  \alpha=\frac{\rm{log}_{10}(M_{\rm{BH}, z2})-\rm{log}_{10}(M_{\rm{BH}, z1})}{\rm{log}_{10}(M_{\rm{\ast}, z2})-\rm{log}_{10}(M_{\rm{\ast}, z1})} 
\end{equation}

\noindent Selecting the 10 most massive black holes in Illustris at $z1=2.5$ and tracking them forward to $z2=0.5$ results in a slope of $0.64 \pm 0.28$. This is also broadly consistent with the putative direction of evolution from the J2315 major merger system to the J1234 system discussed in Section \ref{sec:merger} and seen in Fig. \ref{fig:mbh_mdyn}.  

Given the location of the $z\sim4-6$ quasars in Fig. \ref{fig:mbh_mdyn} relative to our $z\sim2.5$ quasars, it is also interesting to ask whether the evolution from $z\sim6$ to $z\sim2.5$ also occurs in a predominantly horizontal direction in the M$_{\rm{BH}}$-M$_{\rm{gal}}$ plane. As can be seen in the right-hand panel of Fig. \ref{fig:Illustris}, this is not the case. The Illustris volume is too small to contain a typical high-luminosity $z\sim6$ quasar, but the most massive black holes in Illustris at $z\sim6$ with M$_{\rm{BH}} \sim 10^7$ M$_\odot$ could putatively evolve to  M$_{\rm{BH}} \sim 10^9$ M$_\odot$ quasars at $z\sim2.5$. The evolution at these epochs happens in a diagonal direction - i.e. between $z\sim6$ and $z\sim2.5$ black holes tend to grow more rapidly than their host galaxies. Selecting the 10 most massive black holes at $z1=6$ and tracking them forward to $z2=2.5$ results in a mean slope in the M$_{\rm{BH}}$-M$_{\rm{gal}}$ plane of $1.88 \pm 0.4$, considerably steeper than the slope of the evolution from $z=2.5$ to $z=0.5$. The highest redshift quasars shown in Fig. \ref{fig:mbh_mdyn} are therefore unlikely to be the progenitors of our $z\sim2.5$ quasar population. Indeed, in Illustris the z$\sim$6 progenitors of the most massive z$\sim$2.5 black holes lie $\sim 0.13 \pm 0.18$ dex below the best-fitting scaling relation, suggesting that although the z$\sim$6 quasars will continue to grow along with their host galaxies, the z$\sim$2.5 quasars likely originated from z$\sim$6 black holes which were under-massive relative to their host galaxies. 

\section{Conclusions}

We have presented spatially resolved observations of the CO(3-2) emission from ALMA in two high-luminosity, heavily reddened quasar systems seen at the peak epoch of galaxy formation at $z\sim2.5$. Both quasars have very similar luminosities, black-hole masses and dust obscurations and the new ALMA observations clearly demonstrate that there is a range in gas morphologies, gas fractions and dynamical masses seen in these systems. 

\begin{itemize}
    \item The quasar J1234 has two massive, CO-luminous companion galaxies. The CO(3-2) emission shows a clear velocity gradient in the quasar host as well as the two companion galaxies located $\sim$ 170 kpc and $\sim$ 90 kpc (projected distance) from the quasar. By modelling these velocity gradients using the tilted-ring model $^{\rm{3D}}$Barolo, we infer dynamical masses of $\sim$10$^{11}$M$_\odot$ in all three galaxies. Dynamical masses estimated using an exponential disk model are larger in all cases, presumably because they trace mass over a much larger physical extent of $\sim$10 kpc.  
    
    \item The gas fractions in all three galaxies in the J1234 system are somewhat lower than expected from typical star-forming galaxies at these redshifts, particularly when deriving these gas fractions using the full dynamical mass encompassed within $\sim$ 10 kpc. This is consistent with the process of AGN feedback and inefficient cooling in the most massive halos suppressing the accretion of gas into the system. 
    
    \item All three galaxies in the J1234 system are at the more massive and more compact end of the distribution of both field and cluster galaxies at $z\sim2$ on a mass-size relation. 
    
    \item We compare the velocity profiles of the CO(3-2) emission to the line profiles of other CO lines detected in the J1234 system galaxies. For the galaxy G1234S, we uncover very broad wings in the CO(7-6) gas emission that are not seen in the CO(3-2) line profiles. This could indicate the presence of a molecular outflow with a much higher fraction of warm, dense gas in the outflow. The implied outflow velocities are high and lead us to speculate about the presence of a ``hidden", highly obscured AGN within G1234S, although this needs to be confirmed with deep, high resolution X-ray and/or mid infra-red data. 
    
    \item We resolve the quasar J2315 into a close separation major merger with a $\sim$1:2 mass ratio. The two galaxies in the merger have a physical separation of $\sim$15 kpc and a velocity separation of $\sim$170 km/s providing evidence that major mergers are associated with luminous accretion and prolific star formation in high-redshift, obscured quasar systems. We also present new rest-frame UV and optical images of the system from HyperSuprimeCam where we see unobscured UV emitting regions spatially co-incident with the extended molecular gas reservoir.
    
    \item The gas fraction in J2315 is $\gtrsim$45\%, which is consistent with or higher than other star-forming galaxies and much higher than in the J1234 system. This could suggest that AGN feedback is yet to fully kick in and that this system is viewed at an earlier evolutionary stage compared to J1234.

    \item The various CO line profiles in the J2315 merging system show significant velocity shifts relative to one another suggesting that the warm/dense and cool/diffuse gas phases are undergoing different motions and are likely not co-spatial. In particular, the CO(7-6) emission is much broader than the CO(3-2) emission most likely as the two merging galaxies are not resolved in these data, and the CO(1-0) emission is redshifted by $\sim$500 kms$^{-1}$ relative to the other CO lines.  
    
    \item We use the dynamical mass estimates from our resolved observations to place our heavily reddened quasars on the M$_{\rm{BH}}$-M$_{\rm{gal}}$ relation. J1234 is consistent with the predictions from the Illustris simulations for $z\sim2$ galaxies. J2315 on the other hand, has a black hole mass that is larger than would be implied by the host galaxy mass. This is consistent with J1234 representing a later phase in the evolution of these massive galaxies compared to J2315. 
    
    \item We use the outputs from the Illustris simulations to understand both the progenitors and likely descendants of our $z\sim2.5$ quasars. We find that most of the mass in the black hole for such massive quasars is expected to be assembled by $z\sim2.5$ consistent with very massive black holes already being in place at these redshifts. From this epoch the galaxies mainly grow in stellar mass until the present day. We also demonstrate that the high luminosity $z\sim6$ quasar population is unlikely to evolve into our $z\sim2.5$ quasars. Instead our massive $z\sim2.5$ quasars likely evolved from black-holes that were under-massive relative to their host galaxies at $z\sim6$. 
    
\end{itemize}

Overall these observations highlight the diversity in gas morphologies, environments and gas fractions even within a uniformly selected quasar sample with very similar black-hole properties and AGN obscuration. The differences could be consistent with the two quasars being seen in different evolutionary stages along a major-merger sequence. During final approach, when the galaxies are separated by a few tens of kpc (as is the case for the J2315 major-merger system), the gas fractions are still elevated as the effects of AGN feedback are yet to fully kick in. These feedback processes likely rapidly become more efficient following final coalescence, when the AGN begins to deplete the gas supply leading to the low gas fractions we observe in J1234. It is also important to point out however that there are numerous examples of unobscured, UV-luminous quasars with similar host galaxy properties to both J1234 (e.g. \citealt{Coppin:08}) and J2315 (e.g. \citealt{Wagg:14}) suggesting that AGN continuum obscuration is not necessarily a pre-requisite for gas-rich, starburst/merger host galaxies. With a sample of only two obscured quasars analysed in the present work, it is difficult to draw any firm conclusions regarding how they fit into a standard, merger-driven paradigm of massive galaxy formation. Crucially, matched control samples of unobscured, UV-luminous, high-redshift quasars with similar resolved observations of the gas emission are still missing. Assembling these large samples is costly, but nevertheless essential if we are to put the obscured quasar population in context within galaxy formation scenarios.

\section*{Acknowledgements}

We thank the MNRAS referee for their comments, which have helped improve the paper. MB acknowledges support from The Royal Society via a University Research Fellowship. GCJ acknowledges ERC Advanced Grant 695671 ``QUENCH'' and support by the Science and Technology Facilities Council (STFC). CD acknowledges support from ERC starting grant 638707 and support from the STFC. MB would like to thank Matt Auger, Chris Carilli, Paul Hewett, Debora Sijacki and Matthew Temple for many useful discussions. This paper makes use of the following ALMA data: ADS/JAO.ALMA 2016.1.01308.S. ALMA is a partnership of ESO (representing its member states), NSF (USA) and NINS (Japan), together with NRC (Canada), NSC and ASIAA (Taiwan), and KASI (Republic of Korea), in cooperation with the Republic of Chile. The Joint ALMA Observatory is operated by ESO, AUI/NRAO and NAOJ.

\section*{Data Availability}

The ALMA data presented in this paper can be downloaded from the ALMA archive using the Program ID 2016.1.01308.S (PI: Manda Banerji). 




\bibliographystyle{mnras}
\bibliography{alma_ref2} 



\bsp	
\label{lastpage}
\end{document}